\documentclass[12pt,showkeys]{book}

\usepackage[francais]{babel}
\usepackage[latin1]{inputenc} 
\usepackage[T1]{fontenc} 
\usepackage{fancyhdr}
\usepackage{graphicx}
\usepackage{latexsym}
\usepackage{amsfonts}
\usepackage{amssymb}
\usepackage{amsmath}
\usepackage{comment}
\usepackage{mathrsfs}

\newcommand{\be}{\begin{equation}}
\newcommand{\ee}{\end{equation}}
\newcommand{\ba}{\begin{eqnarray}}
\newcommand{\ea}{\end{eqnarray}}
\newcommand{\pl}{\left\{}
\newcommand{\pr}{\right\}}
\newcommand{\al}{\left|}
\newcommand{\ar}{\right|}
\newcommand{\rr}{\right)}
\newcommand{\rl}{\left(}
\newcommand{\ccl}{\left[}
\newcommand{\ccr}{\right]}
\newcommand{\tr}{{\rm Tr}}
\newcommand{\dd}{\partial}
\newcommand{\ffitil}{\widetilde \varphi}
\newcommand{\ffi}{\varphi}
\newcommand{\xib}{\bar \xi}
\newcommand{\DD}{{\widetilde {\mathscr D}}}
\newcommand{\thetab}{\bar \theta}
\newcommand{\zb}{\bar z}
\newcommand{\chib}{\bar \chi}
\newcommand{\Phib}{\overline{\Phi}}
\newcommand{\psib}{\bar \psi}
\newcommand{\lambdab}{\bar \lambda}
\newcommand{\sigmab}{\bar \sigma}
\newcommand{\Fb}{\overline F}
\newcommand{\Db}{\overline D}
\newcommand{\nablab}{\overline \nabla}
\newcommand{\Wb}{\overline W}
\newcommand{\Qb}{\overline Q}
\def\unit{\relax{\rm 1\kern-.26em I}}

\fancypagestyle{plain}{
	\fancyhead{}
	
}

\begin{document}
\begin{titlepage}

\vspace*{-2cm}
\includegraphics[width=30mm]{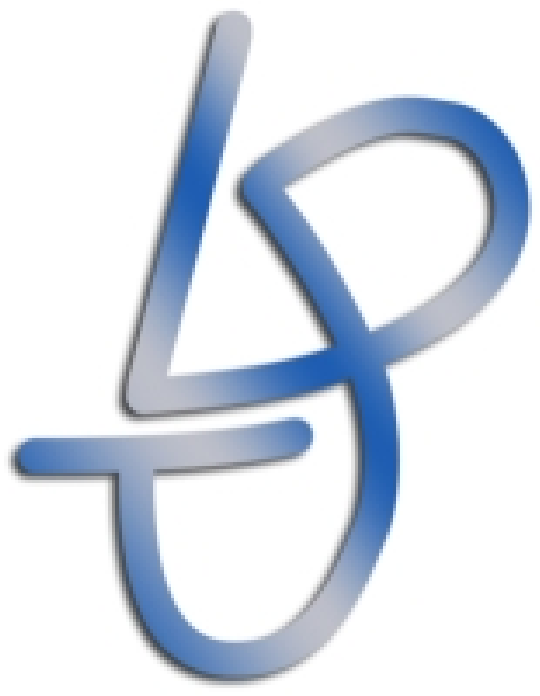}
\hfill
\includegraphics[width=20mm]{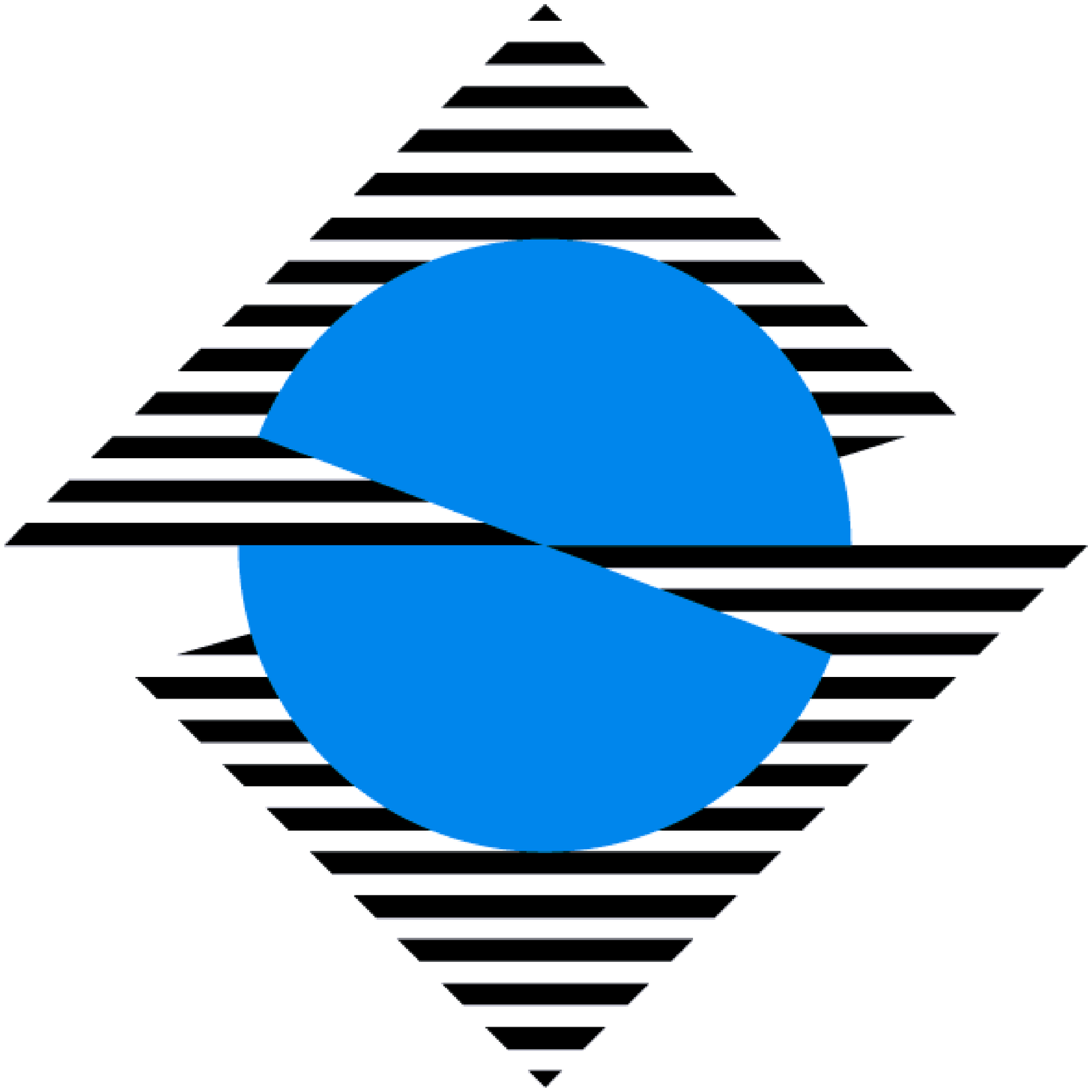}

\begin{center}
\fbox{{\bf UNIVERSIT\'E PARIS-SUD XI}}

\vskip1cm

{\bf TH\`ESE}
\vskip1cm

Sp\'ecialit\'e: {\bf  PHYSIQUE TH\'EORIQUE}\\
\vskip0.3cm
Pr\'esentée\\
 pour obtenir le grade de
\vskip0.75cm
\large {\bf Docteur de l'Université Paris XI}

\vskip0.75cm

par

\vskip0.5cm

{\sc \bf Chloé Papineau}

\vskip1cm

Sujet: 

\vskip0.2cm

\Large{\bf Conséquences cosmologiques et phénoménologiques des théories des cordes}

\end{center}
\vskip0.75cm
\begin{flushleft}

Soutenue le 13 juillet 2007 devant la commission d'examen:\\
\vskip0.75cm
$\begin{array}{lll}
\mbox{MM}. &    \mbox{Ignatios Antoniadis}, & \mbox{rapporteur},\\
  & \mbox{Pierre Bin\'etruy}, & \mbox{président},\\
  &  \mbox{Emilian Dudas}, & \mbox{directeur de th\`ese},\\
  & \mbox{Ulrich Ellwanger}, & \mbox{rapporteur},\\
  &  \mbox{Christophe Grojean}, & \\
  & \mbox{Mariano Quir\'os}. &
   \end{array}$

\end{flushleft}
\end{titlepage}

\thispagestyle{empty}
\vspace*{2.5cm}
~\\
\thispagestyle{empty}
\thispagestyle{empty}
\vspace*{2.5cm}
~\\
\newpage
\thispagestyle{empty}

\vspace*{-2cm}
\includegraphics[width=30mm]{logo-LPT}
\hfill
\includegraphics[width=20mm]{logoP11}

\begin{center}
\fbox{{\bf UNIVERSIT\'E PARIS-SUD XI}}

\vskip1cm

{\bf TH\`ESE}
\vskip1cm

Sp\'ecialit\'e: {\bf  PHYSIQUE TH\'EORIQUE}\\
\vskip0.3cm
Pr\'esentée\\
 pour obtenir le grade de
\vskip0.75cm
\large {\bf Docteur de l'Université Paris XI}

\vskip0.75cm

par

\vskip0.5cm

{\sc \bf Chloé Papineau}

\vskip1cm

Sujet: 

\vskip0.2cm

\Large{\bf Conséquences cosmologiques et phénoménologiques des théories des cordes}

\end{center}
\vskip0.75cm
\begin{flushleft}

Soutenue le 13 juillet 2007 devant la commission d'examen:\\
\vskip0.75cm
$\begin{array}{lll}
\mbox{MM}. &    \mbox{Ignatios Antoniadis}, & \mbox{rapporteur},\\
  & \mbox{Pierre Bin\'etruy}, & \mbox{président},\\
  &  \mbox{Emilian Dudas}, & \mbox{directeur de th\`ese},\\
  & \mbox{Ulrich Ellwanger}, & \mbox{rapporteur},\\
  &  \mbox{Christophe Grojean}, & \\
  & \mbox{Mariano Quir\'os}. &
   \end{array}$

\end{flushleft}

\newpage
\thispagestyle{empty}
\vspace*{2.5cm}
~\\
\thispagestyle{empty}
\thispagestyle{empty}
\vspace*{2.5cm}
~\\

\chapter*{Dans l'ordre chronologique ... ou presque}

La primeur en la matière revient à mes parents sans lesquels, et c'est peu dire, je ne serais certainement pas là. Par leur amour inconditionnel, même lorsqu'ils ne comprenaient pas mes choix, ils m'ont montré la voie du bonheur et de la réussite. Je remercie aussi le reste de ma famille, Camille, Fabien, et Séverine, pour leur soutien et leurs encouragements. Je vous dois tant, à tous...

Ma gratitude va également à Asmâa Abada, Corinne Augier, Pierre Binétruy, Christos Charmousis, Ulrich Ellwanger, ainsi qu'aux autres professeurs rencontrés lors de mes études, qui ont su m'aiguiller et répondre à mes nombreuses questions, particulièrement l'année du DEA. C'est en partie grâce à leurs conseils avisés et leur appui que cette thèse a vu le jour.

Je remercie chaleureusement Dominique Schiff et Hendrik Hilhorst qui m'ont accueillie au sein du LPT, se sont montrés attentifs et attentionnés, et m'ont permis de communiquer sur mes travaux, ici et ailleurs. J'exprime mes remerciements à Patrick Mora qui m'a offert un cadre de thèse exceptionnel au sein du CPHT à l'\'Ecole Polytechnique. Merci également à tout le personnel administratif des deux laboratoires pour leur patience face à mon incapacité à gérer la ``paperasse''.

Au coeur de ces pages, au coeur de cette thèse et au coeur de mon infinie gratitude, je ne sais trouver les mots pour remercier Emilian Dudas pour son soutien indéfectible, sa patience, sa gentillesse et sa disponibilité. \c{C}a a été un honneur et un plaisir d'apprendre de lui pendant trois ans. Au delà des leçons de physique, pour lesquelles je ne dirai jamais assez merci, je retiens aussi les leçons de vie : l'humilité et la capacité à se remettre constamment en question comptent parmi les qualités que j'espère avoir acquises, au moins un peu...

Je tiens également à remercier les membres du groupe de Théorie des Cordes du CPHT, Ignatios Antoniadis, Elias Kiritsis, Hervé Partouche et Marios Petropoulos, coordinateur du groupe, ainsi que les membres du groupe de Cosmologie du LPT, Martin Bucher, Christos Charmousis, Renaud Parentani et Bartjan Van Tent, et du groupe de Physique au-delà du Modèle Standard, Asmâa Abada, Abdelhak Djouadi, Ulrich Ellwanger, Yann Mambrini, et Grégory Moreau.

Pour tous ces petits moments de la vie de tous les jours, les discussions, les cafés, les fous-rires, les débats voire les disputes, un grand merci à mes camarades thésards et post-docs : Benjamin Basso, Nicol\'as Bernal, Florian Bonnet, Charles Bouchart, Luca Carlevalo, Pablo Camara, Roberto Casero, Emmanuel Chang, Cédric Delaunay, Florian Domingo, Pascal Grange, Razvan Gurau, Umut Gursoy, Benjamin Haas, Irene Hidalgo, Liguori Jego, François-Xavier Josse-Michaux, Euihun Joung, Balazs Kozma, Xavier Lacroze, Vincenzo Laporta, Alexey Lokhov, Jean Macher, Tristan Maillard, Axel Marcillaud de Goursac, Liuba Mazzanti, Yacine Mehtar-Tani, Vassilis Niarchos, Francesco Nitti, Domenico Orlando, Angel Paredes, Mathieu Remazeilles, Alberto Romagnoni, Mathieu Segond, Emmanuel Sérié, Shweta Sharma, Adrian Tanasa, Ana Teixeira, Maria-Cristina Timirgaziu, Sudhir Vempati, Fabien Vignes-Tourneret, et Robin Zegers.

Merci à mes amis, Kamal Bouhouch, Antonin Bourgeois, Matthias Delescluse, Aurore Franco, Pierre Hosteins, François Kiyak, Stéphane Pruvot, Astrid Quillien, et tous ceux que j'oublie....

Je me dois d'exprimer ma reconnaissance à Christophe Grojean, Stéphane Lavignac et Géraldine Servant qui m'ont accueillie à certaines des nombreuses conférences ou écoles auxquelles j'ai eu la chance de participer. Je ne saurais énumérer tout ce que m'ont appris ces expériences.

Je remercie également mes collaborateurs Valéry A. Rubakov et Stefan Pokorski.

Ce manuscrit doit beaucoup à l'aide et la relecture de Charles Bouchart, Pablo Camara, Christos Charmousis, Yann Mambrini, Grégory Moreau, et Alberto Romagnoni.

J'exprime enfin ma reconnaissance à Ignatios Antoniadis et Ulrich Ellwanger qui ont gentiment accepté de lire ce manuscrit et d'écrire un rapport en un temps record. Un grand merci à Pierre Binétruy, Christophe Grojean et Mariano Quir\'os pour m'avoir fait l'honneur d'être membres du jury de soutenance.

\newpage
\thispagestyle{empty}
\vspace*{2.5cm}
~\\
\thispagestyle{empty}
\thispagestyle{empty}
\vspace*{2.5cm}
~\\

\newpage
\thispagestyle{empty}
\vspace*{2.5cm}
~\\
\thispagestyle{empty}
\thispagestyle{empty}
\vspace*{2.5cm}
~\\

\tableofcontents

\thispagestyle{empty}
\cleardoublepage
\thispagestyle{empty}
\thispagestyle{empty}
\vspace*{2.5cm}
~\\
\newpage
\thispagestyle{empty}
\cleardoublepage
\thispagestyle{empty}
\thispagestyle{empty}


\chapter[Introduction]{De l'importance des symétries}


Un point majeur dans la compréhension des lois de la Nature réside dans la notion de symétrie. La conviction que les symétries jouent un rôle essentiel en physique est devenue de plus en plus forte au cours des siècles. En revanche, l'analyse des propriétés d'une symétrie (ou invariance) en termes de groupe, de ses générateurs et de ses représentations est beaucoup plus récente et occupe désormais une place centrale dans l'approche des théoriciens aux lois fondamentales de la physique. En effet, les transformations associées à une symétrie forment un groupe dont les représentations permettent de décrire les objets physiques intéressants vis-à-vis de cette symétrie. Avec le théorème de Noether (1918) vient la correspondance entre symétrie et conservation d'une grandeur. Par exemple, l'invariance par translation et par rotation conserve l'énergie, l'impulsion et le moment angulaire. Le groupe associé à ces transformations est le groupe de Poincaré, dont les représentations sont définies par le spin et la masse (Wigner, 1939).

Parmi les symétries accessibles, on distingue les symétries d'espace-temps, qui agissent sur les coordonnées $x^{m}$ et dont le groupe de Poincaré fait partie, et les symétries internes, qui sont définies en chaque point de l'espace-temps et agissent sur les objets physiques (les champs). Dans l'ensemble des symétries internes, on retrouve les symétries chirales, le nombre baryonique, le nombre leptonique, ou encore les symétries de saveur et de couleur. Les symétries internes sont classées en deux catégories : les symétries globales et les symétries de jauge. Les premières s'appliquent aux objets indépendamment de leur localisation tandis que pour les symétries de jauge, les paramètres de transformations sont des fonctions dépendant du point où s'applique la transformation. Enfin, la structure du groupe qui décrit une symétrie permet de différencier les symétries continues, pour lesquelles les paramètres des transformations peuvent prendre un nombre continu de valeurs, et les symétries discrètes, pour lesquelles les paramètres appartiennent à un ensemble discret.

Connaissant le groupe d'une symétrie, le lagrangien le plus général préservant cette symétrie est alors clairement déterminé. Tous les champs de la théorie sont dans des représentations du groupe, et tous les termes d'interactions (entre ces champs) invariants sous les transformations du groupe sont présents.

La remarquable confirmation expérimentale des prédictions du Modèle Standard de la Physique des particules, qui a pour groupe de jauge $SU(3)\times SU(2)_L\times U(1)$, est une forte indication que cette approche est la bonne. Les sensibilités actuelles vont au delà du pourcent de précision et aucune déviation ferme n'a été observée. Néanmoins, il existe de nombreux signes nous montrant que ce modèle n'est pas la théorie ultime. Les résultats expérimentaux allant dans ce sens sont la preuve que les neutrinos sont massifs puisqu'ils oscillent, la mesure de la constante cosmologique qui est trop faible, la mise en évidence de la matière noire, l'inflation, et la baryogénèse. D'un point de vue théorique, le secteur électrofaible souffre d'un problème de hiérarchie dramatique qui amène à reconsidérer la question de la naturalité, les couplages de jauge ne s'unifient pas, le nombre de générations et les hiérarchies entre les différentes masses des fermions, ou de manière équivalente des couplages de Yukawa, ne sont pas expliquées. Il faut donc considérer le Modèle Standard comme une limite à basse énergie d'une théorie plus fondamentale. Pourtant, une chose est sûre, quelle que soit la nouvelle physique apparaissant au-delà du Modèle Standard, sa limite à des énergies aux alentours de l'échelle électrofaible \textit{doit} reproduire les données que nous avons. La précision avec laquelle le Modèle Standard a été testé contraint donc de manière drastique le panel de théories pouvant englober le Modèle Standard.

Pour aller au-delà du Modèle Standard et trouver une explication à ces nombreux problèmes, une des démarches adoptées est d'élargir le groupe de symétrie de la théorie. Il faut alors trouver des mécanismes brisant une partie du groupe élargi de sorte que les symétries résiduelles soient exactement celles du Modèle Standard. Diverses voies ont été engagées ces dernières décennies : l'élargissement du groupe de jauge, du groupe de symétries globales ou du groupe de symétries d'espace-temps. Les théories de Grande Unification, dont les plus connues sont $SU(5)$ et $SO(10)$, s'incrivent dans le premier cas. Bien que très élégantes et permettant d'expliquer en partie les saveurs et le nombre de générations, l'échelle d'unification se situe autour de $10^{16}$ GeV et le problème de hiérarchie n'est pas résolu. La supersymétrie et les dimensions supplémentaires, outils que nous développons dans cette thèse, s'inscrivent dans les deux autres cas.

Les théories de type Kaluza-Klein incorporent des dimensions supplémentaires en supposant que celles-ci sont compactes. Le rayon de compactification, qui représente la taille de ces dimensions, doit être suffisamment petit pour expliquer que nous n'y soyons pas sensibles. Le contenu en particule est en effet élargi, et les particules supplémentaires que nous percevons à quatre dimensions ont une masse proportionnelle à l'inverse du rayon. Donc si le rayon est trop grand, la masse est accessible à LEP, par exemple. On peut également supposer que les dimensions supplémentaires sont de rayon très large, voire infini (dimensions plates), mais que la matière ordinaire à quatre dimensions est en quelque sorte piégée dans une sous-variété, la brane, d'un espace-temps plus large. Ces deux approches peuvent être combinées : c'est le cas dans des modèles ``warped extra-dimensions'' qui constituent un moyen de résoudre le problème de hiérarchie.

La supersymétrie, par ailleurs, élargit le groupe de symétries internes grâce à des générateurs qui anti-commutent entre eux et commutent avec les générateurs du groupe de Poincaré. Les fermions et les bosons sont alors des composantes de multiplets supersymétriques. L'argument majeur en faveur de la supersymétrie est qu'elle résout naturellement le problème de hiérarchie. Elle garantit également l'unification des couplages de jauge, et peut fournir des candidats à la matière noire. Mais la supersymétrie n'est pas réalisée dans la Nature à des échelles d'énergie observables, et il faut la briser. Une brisure spontanée de cette symétrie globale entraîne la présence d'une particule de Goldstone non-massive dans le spectre, ce qui est exclu. Nous jaugeons la supersymétrie en rendant le paramètre des transformations local. Fait remarquable, ce processus force la théorie à inclure la gravitation. À basse énergie, une brisure spontanée de supersymétrie dans une théorie de supergravité fait apparaître des termes non-supersymétriques dans le lagrangien. Néanmoins, ces termes ne rajoutent pas de divergences quadratiques et donc le problème de hiérarchie reste résolu. C'eût été trop beau : les théories de supergravité contiennent des termes non-renormalisables qui nous poussent inévitablement à aller encore au-delà.

La supergravité est naturellement présente dans les théories de cordes. Ces théories postulent que les objets fondamentaux ne sont plus ponctuels mais ont une extension spatiale. Les particules élémentaires sont vues comme des excitations de la corde. Celle-ci vit donc dans deux dimensions (le temps et la dimension de la corde) plongées dans un espace-temps à dix dimensions, ce nombre étant une condition d'auto-cohérence de la théorie. Les théories de cordes sont à l'heure actuelle les meilleures candidates à l'unification des quatre interactions fondamentales. En outre, elles comportent un seul paramètre : la tension de la corde. Leur action effective à basse énergie s'écrit comme une théorie des champs. Les théories de cordes contiennent donc tous les ingrédients que nous avons évoqués pour aller au-delà du Modèle Standard : groupe de jauge élargi, dimensions suplémentaires compactes, branes, supergravités (une ou plusieurs selon les théories). La réduction de dix à quatre dimensions d'espace-temps donne lieu à une phénoménologie très riche. Selon le mode de compactification employé, on obtient des théories très diverses. Un défi majeur dans ce sens est non seulement de retrouver le groupe de jauge et le contenu en champs du Modèle Standard, mais aussi de vérifier les contraintes cosmologiques comme l'inflation et la constante cosmologique.

Le sujet central de cette thèse est l'étude des mécanismes de brisure d'une symétrie. Ce manuscrit s'organise comme suit :
\begin{itemize}
\item[$\bullet$] Le deuxième chapitre introduit les dimensions supplémentaires et présente un modèle à deux dimensions supplémentaires intéressant dans le cadre du problème de hiérarchie.
\item[$\bullet$] Le troisième chapitre est une introduction à la supersymétrie et la supergravité.
\item[$\bullet$] Le quatrième chapitre présente deux mécanismes de brisure dynamique de symétrie par des phénomènes non-perturbatifs : les modèles de Nambu-Jona-Lasinio (NJL) et les modèles avec condensation de jauginos. Le modèle à deux dimensions supplémentaires est repris et l'on montre sa dualité avec des modèles de NJL à quatre dimensions.
\item[$\bullet$] Le cinquième chapitre traite des dualités de Seiberg dans les théories de Chromodynamique Quantique (QCD) supersymétrique, et détaille un modèle de brisure dynamique de supersymétrie.
\item[$\bullet$] Après avoir expliqué ce que sont les champs de modules issus de la réduction dimensionnelle des théories de cordes, le sixième chapitre présente un modèle permettant leur stabilisation. Nous proposons également une alternative combinant la stabilisation des modules avec les phénomènes non-perturbatifs brisant dynamiquement la supersymétrie.
\end{itemize}


\chapter{Dimensions supplémentaires\label{dimsup}}


Une manière d'étendre le Modèle Standard est de considérer que l'espace-temps fondamental contient plus de quatre dimensions. Cette idée est renforcée par le fait que les supercordes vivent naturellement dans un espace-temps dix-dimensionnel (c'est en fait une condition de consistance de la théorie).

La façon dont les dimensions supplémentaires se manifestent à nous doit être contrainte puisque nous n'avons jusqu'à présent détecté aucun signal que nous aurions pu interpréter comme la preuve que d'autres dimensions existent. À ce jour, deux grandes classes de modèles se dégagent pour expliquer comment les dimensions supplémentaires sont cachées \cite{Rubakov:2001kp}.

Le modèle originel proposé par les mathématiciens T. Kaluza et O. Klein \cite{Kaluza:1921tu} dans les années 1920 tentait d'unifier les deux forces connues à l'époque, la gravitation et l'électromagnétisme, en ajoutant une dimension d'espace. En supposant que cette cinquième dimension est enroulée sur elle-même en un cercle de rayon $R$ très petit, par exemple de l'ordre de la longueur de Planck ($\sim 10^{-33}$ cm), on explique de manière très simple pourquoi notre monde n'y est pas sensible. On peut de la même façon supposer que plusieurs dimensions supplémentaires sont présentes, et qu'elles ont toutes un rayon très petit. Les modèles se basant sur ce principe sont dits de type Kaluza-Klein.

Depuis plus récemment, l'approche des univers branaires connaît un très grand engouement. Dans ces constructions, c'est notre Modèle Standard (ou une partie de ses secteurs) qui est piégé dans un sous-espace à quatre dimensions, appelé ``brane''. Les dimensions supplémentaires peuvent alors être larges voire infinies, plates ou courbées (``warped extra dimensions'').


\section{Modèles de type Kaluza-Klein\label{xdim}}


Considérons pour commencer une seule dimension supplémentaire, de coordonnée $y$. Dire que cette dimension est compacte et prend la forme d'un cercle de rayon $R$, c'est dire que $y$ appartient à l'ensemble $\ccl 0, 2\pi R\ccr$ et que l'on identifie les deux points extrêmes. Un champ scalaire $\Phi$ se propageant dans cet espace-temps se décompose donc en modes de Fourier
\be
\Phi(x^{\mu},y) \ = \ \frac{1}{\sqrt{2\pi R}} \sum_{n \, \in \, \mathbb Z} \exp \rl \frac{iny}{R} \rr \phi_n (x^{\mu}) \quad ,
\label{decompscalxdim}
\ee
avec $\mu = 0, \ldots, 3$ décrivant nos quatre dimensions. La base des $e^{iny/R}$ correspond aux fonctions propres de l'impulsion dans la direction compacte, et le champ vérifie donc les conditions de périodicité $\Phi(x^{\mu},0)=\Phi(x^{\mu},2\pi R)$.

L'action de ce champ respecte l'invariance de Lorentz à cinq dimensions
\be
\mathcal S_5 \ = \ \int d^4 x \int_0^{2\pi R} dy \pl \dd_M \Phi^{\dagger} \dd^M \Phi - m_0^2 \Phi^{\dagger} \Phi \pr \quad ,
\label{action5Dscal}
\ee
avec $M = 0, \ldots, 3,5$ où $M=5$ correspond à la dimension supplémentaire\footnote{La métrique de Minkowski est $\eta_{MN} = \text{diag}\rl +, -, \ldots, - \rr$.}.

Le champ $\Phi$ a une dimension de masse $3/2$. Le facteur $1/\sqrt{2\pi R}$ dans la décomposition (\ref{decompscalxdim}) assure que les modes de Fourier $\phi_n$, appelés modes de Kaluza-Klein (KK), ont une dimension canonique à quatre dimensions.

En injectant la décomposition (\ref{decompscalxdim}) dans l'action (\ref{action5Dscal}), on trouve
\be
\mathcal S_5 = \frac{1}{2\pi R} \int d^4 x \int_0^{2\pi R} dy \sum_{m, \, n \, \in \, \mathbb Z} \pl \dd_{\mu} \phi_m^* \dd^{\mu} \phi_n - \rl m_0^2 + \frac{mn}{R^2} \rr \phi_m^* \phi_n \pr e^{i\frac{\rl n - m \rr y}{R}} \ . \nonumber
\ee
L'intégration sur $y$ mène\footnote{On utilise $\int_0^{2 \pi R} dy e^{i\frac{\rl n - m \rr y}{R}}  = 2 \pi R \delta_{mn}$.} à l'action effective à quatre dimensions
\be
\mathcal S_4 \ = \ \int d^4 x \sum_{n\, \in \, \mathbb Z} \pl \dd_{\mu} \phi_n^* \dd^{\mu} \phi_n - \rl m_0^2 + \frac{n^2}{R^2} \rr \phi_n^* \phi_n \rr \quad . \label{actioneff5Dscal}
\ee

Nous voyons donc qu'un état de Klein-Gordon à cinq dimensions $\dd^M \dd_M \Phi = m_0^2 \Phi$ donne lieu à une tour infinie discrète d'états de Klein-Gordon à quatre dimensions $\dd^{\mu} \dd_{\mu} \phi_n = m_n^2 \phi_n$ avec
\be
m_n^2 \ = \ m_0^2 \ + \ \frac{n^2}{R^2} \quad .
\label{massKKscal}
\ee
L'état $n=0$, appelé mode zéro, possède la masse $m_0^2$.

La procédure se généralise facilement à un nombre $d$ de dimensions supplémentaires de rayons $R_i$, avec $i=1, \ldots, d$. En particulier, l'expression des masses (\ref{massKKscal}) devient
\be
m_{n_1, \ldots, n_d}^2 \ = \ m_0^2 \ + \ \sum_{i = 1}^d \frac{n_i^2}{R_i^2} \quad . \nonumber
\ee

Faisons la même construction pour un fermion à quatre composantes $\Psi$. En s'inspirant de (\ref{decompscalxdim}), on le décompose en
\be
\Psi(x^{\mu},y) \ = \ \frac{1}{\sqrt{2\pi R}} \sum_{n \, \in \, \mathbb Z} \exp \rl \frac{iny}{R} \rr \psi_n (x^{\mu}) \quad .
\label{decompfermionxdim}
\ee
Supposons pour plus de simplicité que le champ est libre,
\be
\mathcal S_5 \ = \ \int d^4 x \int_0^{2\pi R} dy \rl i \overline{\Psi} \gamma^M \dd_M \Psi \rr \quad ,
\label{action5Dfermion}
\ee
où les matrices de Dirac $\gamma_M = \rl \gamma_{\mu}, i \gamma_5 \rr$ forment une représentation de l'algèbre de Clifford $\pl \gamma_M , \gamma_N \pr = 2 \eta_{MN}$. En effectuant l'intégration sur $y$ comme dans le cas du champ scalaire, on trouve une action
\be
\mathcal S_4 \ = \ \int d^4 x \sum_{n\, \in \, \mathbb Z} \pl i  \psib_n \gamma^{\mu} \dd_{\mu} \psi_n  - \frac{in}{R}  \psib_n \gamma_5 \psi_n \pr \quad . \label{actioneff5Dfermion}
\ee
On a donc ici aussi une tour d'états KK de masses\footnote{Pour les fermions, seuls les modules au carré des masses ont une sens.} $\frac{n^2}{R^2}$ mais les termes de masse sont du type Dirac $\psib_R \psi_L + \psib_L \psi_R$. Par conséquent, tous les modes $n \neq 0$ sont non-chiraux. Ceci constitue un réel problème car notre but est de construire des modèles avec dimensions supplémentaires tels que le Modèle Standard en soit une théorie effective.


\subsection{Orbifolds\label{Orbifolds}}

\begin{figure}[ht!]
\begin{center}
\includegraphics[scale=0.4]{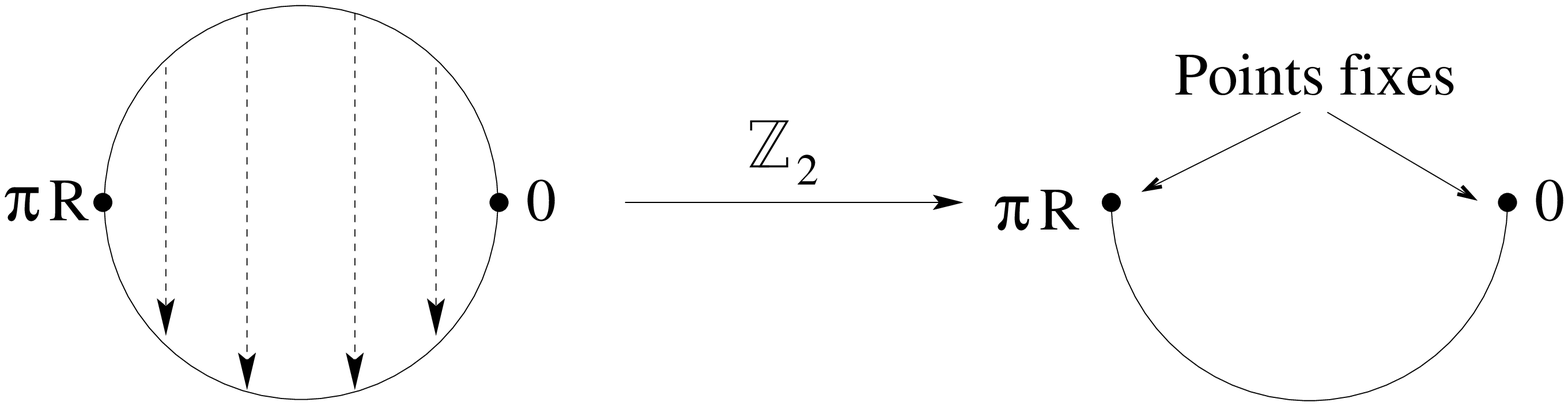}
\caption{Sous l'orbifolding $\mathbb Z_2 : \, y \rightarrow -y\,$, le cercle devient un segment avec deux points fixes à ses extrêmités.}
\label{orbifold}
\end{center}
\end{figure}

Un moyen de retrouver la chiralité à cinq dimensions est d'imposer une symétrie supplémentaire $y \rightarrow -y$. Sous cette dernière, le cercle de rayon $R$ devient un segment de longueur $\pi R$. Les points $y=0$ et $y = \pi R$ sont des points fixes puisqu'ils sont transformés en eux-mêmes, comme illustré sur la Fig. \ref{orbifold}.

De manière générale, si les dimensions supplémentaires forment une variété $K$, imposer une symétrie $\Gamma$ sur $K$ donne lieu à un groupe quotient $K/\Gamma$ appelé \textit{orbifold}, qui n'est en général plus une variété ; le processus s'appelle ``orbifolding''. L'orbifold contient toujours des points fixes, qui sont les points de $K$ invariants sous $\Gamma$.

Pour mieux comprendre ce que cela implique, reprenons le champ scalaire $\Phi$. On peut réécrire le développement (\ref{decompscalxdim}) de la façon suivante
\be
\Phi (y) = \frac{1}{\sqrt{\pi R}} \pl \frac{1}{\sqrt{2}}\phi_0 + \sum_{n \geqslant 1} \rl \phi^{(+)}_n  \cos \frac{ny}{R} + \phi^{(-)}_n  \sin \frac{ny}{R} \rr \pr \quad , \label{decompscalcossin}
\ee
avec
\begin{eqnarray}
&&\phi^{(+)}_n \ = \ \frac{1}{\sqrt{2}} \rl \phi_n + \phi_{-n} \rr \quad , \nonumber \\
&&\phi^{(-)}_n \ = \ \frac{i}{\sqrt{2}} \rl \phi_n - \phi_{-n} \rr \quad . \nonumber
\end{eqnarray}

La symétrie de l'orbifold autorise les deux parités pour $\Phi$. Si l'on impose $\Phi (-y) = - \Phi(y)$, alors les fonctions $\phi^{(+)}_n$ sont toutes identiquement nulles, et le mode zéro disparaît. Dans le cas où il n'y aurait pas de terme de masse $m_0$ dans l'action (\ref{action5Dscal}), cela permet d'éviter la présence d'un champ non-massif à quatre dimensions.

En quoi l'orbifold restaure-t'il la chiralité ? Sous la symétrie $\mathbb Z_2$, le terme cinétique $\overline{\Psi} \gamma^M \dd_M \Psi$ doit être invariant. Or il est clair que le terme $\dd_5$, lui, se transforme en $-\dd_5$. Il faut donc que
\be
\overline{\Psi} \gamma_5 \Psi \ = \ \overline{\Psi}_L \Psi_R - \overline{\Psi}_R \Psi_L  \nonumber
\ee
soit impair, ce qui implique que l'une des deux chiralités de $\Psi$ doit être impaire tandis que l'autre doit être paire. Il s'ensuit, en comparant avec le cas du champ scalaire (\ref{decompscalcossin}), que la chiralité paire se décompose sur la base des $\cos \frac{ny}{R}$, tandis que celle qui est impaire se décompose sur la base $\sin \frac{ny}{R}$ et n'a donc pas de mode zéro. On a bien une théorie chirale et il devient possible de construire des théories consistantes grâce aux orbifolds \cite{Dixon:1985jw}.


\subsection{Compactification à la Scherk-Schwarz}


En étudiant l'action (\ref{action5Dscal}) (respectivement (\ref{action5Dfermion})), on se rend compte que la décomposition (\ref{decompscalxdim}) (resp. (\ref{decompfermionxdim})) n'est pas la plus générale autorisée par la périodicité de la dimension supplémentaire. En effet, rien ne nous empêche d'imposer plutôt
\be
\Phi (y + 2 \pi R) \ = \ e^{i \beta} \Phi (y) \quad , \nonumber
\ee
avec $\beta$ une constante, si ce choix est une symétrie de l'action. De telles conditions, qui se généralisent très facilement au cas de plusieurs dimensions supplémentaires, sont dites \textit{twistées} et correspondent à une compactification de Scherk-Schwarz \cite{Scherk:1978ta}. Le champ se décompose selon
\be
\Phi(x^{\mu},y) \ = \ \frac{1}{\sqrt{2\pi R}} \exp \rl \frac{i\beta y}{2\pi R} \rr \, \sum_{n \, \in \, \mathbb Z} \exp \rl \frac{iny}{R} \rr \phi_n (x^{\mu}) \quad .
\label{decompscalxdimSS}
\ee
En particulier, pour $\beta = \pi$, le champ est anti-périodique $\Phi(y + 2 \pi R) = - \Phi(y)$.

En effectuant la compactification à partir de l'action (\ref{action5Dscal}), on trouve que les masses KK sont
\be
m_n^2 \ = \ m_0^2 \ + \ \frac{\rl n+\beta/ 2 \pi \rr^2}{R^2} \quad . \nonumber
\ee
Le compactification à la Scherk-Schwarz est donc un autre moyen d'obtenir un mode zéro massif dans le cas où $m_0 = 0$.

\vspace*{0.3cm}

Enfin, il est courant d'effectuer une compactification à la Scherk-Schwarz sur un orbifold. Par exemple,
\be
\Phi(y+2\pi R) \ = \ e^{i\beta} \Phi(y)  \quad \quad \text{et} \quad \quad \Phi(-y) \ = \ - \Phi(y) \nonumber
\ee
impliquent, pour $\beta$ générique, que $\Phi$ est nul aux points fixes, $\Phi(\pi R) = 0 = \Phi(0)$, ce qui correspond à des conditions de type Dirichlet.

Le choix des conditions aux bords des dimensions compactes et le choix de la transformation (parité) des champs sous l'action de l'orbifold sont soumis à contrainte. De manière un peu plus formelle, si l'action $Z_O$ de l'orbifold sur le champ est
\be
\Phi(\zeta_O(y)\,) \ = \ Z_O\, \Phi(y) \quad , \nonumber
\ee
et que la compactification à la Scherk-Schwarz impose un twist $T$ donné par
\be
\Phi(\tau(y)\,) \ = \ T \, \Phi(y) \quad , \nonumber
\ee
alors, en général, les deux transformations ne commutent pas : $Z_O \cdot T \neq T \cdot Z_O\,$.

On peut explicitement trouver les conditions de compatibilité entre les deux mécanismes. Concrètement, nous ne nous intéresserons qu'aux cas où les deux actions $Z_O$ et $T$ commutent. L'association de conditions de Scherk-Schwarz et des orbifolds peut être très intéressante pour engendrer des brisures de symétrie \cite{Quiros:2003gg}. Nous reviendrons sur ceci dans le paragraphe \ref{6Dmodel}.


\section{Les univers branaires\label{braneworld}}


Une autre classe de modèles avec dimensions supplémentaires existe, dans lesquels l'invariance de Lorentz à $d+4$ dimensions est explicitement brisée par la présence de branes, également appelés ``domain walls'' dans certaines réalisations de théorie des champs.

Le modèle le plus simple \cite{Rubakov:1983bb} contient un champ scalaire $\Phi$ à cinq dimensions, décrit par l'action
\begin{eqnarray}
&&\mathcal S \ = \ \int d^4 x dy \ccl \frac{1}{2} \rl \dd_M \Phi \rr^2 - V(\Phi) \ccr \quad ,  \nonumber \\
\text{avec} &&V(\Phi) \ = \ -\frac{1}{2}m^2 \Phi^2 + \frac{1}{4}\lambda \Phi^4 \quad . \label{RubShaposh}
\end{eqnarray}

\begin{figure}[ht!]
\begin{center}
\includegraphics[scale=0.8]{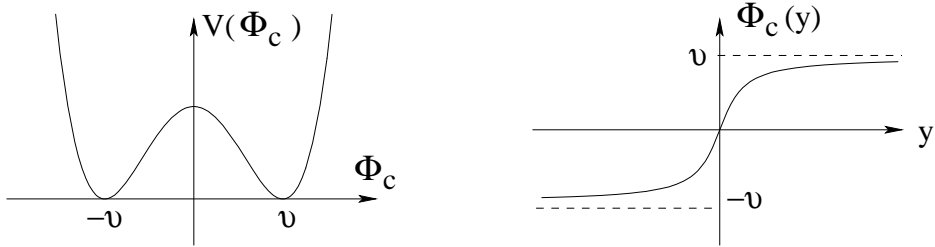}
\caption{Le potentiel (\ref{RubShaposh}) et la solution classique (\ref{soliton}) des équations du mouvement. Le soliton $\Phi_c$ décrit un domain wall qui sépare les deux vides $+v$ et $-v$.}
\label{domainwall}
\end{center}
\end{figure}

L'équation du mouvement $\dd_M \dd^M \Phi=- \dd V(\Phi)/\dd \Phi$ admet une solution classique
\be
\Phi_c(x^{\mu},y) =  \rl \frac{m}{\sqrt{\lambda}} \rr \text{th} \rl \frac{my}{\sqrt{2}} \rr \quad  \label{soliton}
\ee
dépendant de $y$, et qui prend les valeurs asymptotiques $\pm v= \pm \frac{m}{\sqrt{\lambda}}$. Son profil a été tracé sur la Fig. \ref{domainwall}. Cette solution brise explicitement l'invariance de Lorentz à cinq dimensions, mais elle préserve celle à quatre dimensions.

Si l'on couple un fermion de Dirac $\Psi$ au champ scalaire $\Phi$ par une interaction de Yukawa $\Phi \overline{\Psi} \Psi$, alors au niveau classique, autour de la solution solitonique $\Phi = \Phi_c$, les fermions sont localisés au voisinage de $y=0$. Leur profil dans la direction $y$ décroît exponentiellement quand on s'éloigne de la brane. De plus, le mode zéro $\psi$ de ce fermion est naturellement chiral $\gamma_5 \psi = - \psi$. Il est également possible de localiser des champs de jauge (et donc tout le Modèle Standard) mais cela est plus difficile à réaliser que dans l'exemple ci-dessus. Nous verrons au chapitre \ref{modules} que ces localisations apparaissent naturellement dans des théories issues de théories des cordes.


\subsection{Grandes dimensions supplémentaires}


La possibilité de localiser la matière sur une brane quadri-dimensionnelle explique pourquoi nous ne ressentons pas les dimensions supplémentaires de manière directe. Notamment, celles-ci ne sont plus contraintes à être très petites, ou même plates. Un des enjeux est d'apporter de nouvelles solutions aux problèmes de hiérarchie de jauge, de petitesse de la constante cosmologique, d'unification des couplages de jauge, etc., en utilisant la géométrie des dimensions supplémentaires. De ce fait, nous sommes amenés à inclure la gravitation dans nos modèles, ce qui a des conséquences sur la loi de Newton effective à quatre dimensions. Néanmoins, celle-ci a été testée par des expériences du type Cavendish jusqu'à des distances de l'ordre de $\sim 0.1$ cm. La taille $R$ des dimensions supplémentaires peut donc être aussi large que cette valeur\footnote{En termes d'énergie, cela signifie $R^{-1} \gtrsim \mathcal O(10^{-4})$ eV.}.

Les modèles dits ADD \cite{Arkani-Hamed:1998rs} s'appuient sur cette idée et reformulent le problème de hiérarchie en termes de volume de l'espace compact. En effet, la masse de Planck $M_P \sim 10^{19}$ GeV n'est pas l'échelle fondamentale d'une théorie vivant à $D=4+d$ dimensions. Celle-ci, que l'on notera $M_*$, intervient dans l'action d'Einstein-Hilbert
\be
\mathcal S \ = \ - \frac{M_*^{D-2}}{2} \int d^D x \sqrt{g^{(D)}} R^{(D)} \quad , \label{LEHDdim}
\ee
où $g^{(D)}$ et $R^{(D)}$ sont la métrique et le scalaire de Ricci de l'espace-temps $D$-dimensionnel. La constante de Newton $G_{(D)}$ est donnée par
\be
G_{(D)} \ = \ \frac{1}{8 \pi M_*^{D-2}} \quad . \label{constNewton}
\ee

L'idée des modèles ADD est de localiser le Modèle Standard sur une brane de sorte que seule la gravitation se propage dans les dimensions supplémentaires. Si la gravitation est homogène le long des coordonnées tranverses à la brane, alors on peut effectuer facilement l'intégration sur $d^d y$ dans l'action (\ref{LEHDdim}). On trouve
\be
\mathcal S \ = \ \frac{V_d M_*^{D-2}}{2} \int d^4 x \sqrt{g^{(4)}} R^{(4)} \quad , \nonumber
\ee
où $V_d$ est le volume des dimensions supplémentaires, $V_d \sim R^d$ en supposant que tous les rayons sont égaux et valent $R$.

La masse de Planck à quatre dimensions est donnée par
\be
M_P^2 \ \sim \ M_*^{D-2} \, R^{\, d} \ = \ M_*^{d+2} \, R^{\, d} \quad . \label{MPlanckDdim}
\ee
On peut donc expliquer le problème de hiérarchie en abaissant $M_*$ jusqu'à l'ordre de grandeur de l'échelle électrofaible : $M_* \sim 1$ TeV, ce qui implique
\be
R \ \sim \ M_*^{-1} \rl \frac{M_P}{M_*} \rr^{2/d} \ \sim \ 10^{\rl 32/d \rr - 17} \ \text{cm} \quad . \nonumber
\ee
Le cas d'une dimension supplémentaire a déjà été exclu expérimentalement. Le cas $d=6$ naturel dans les théories de supercordes impliquerait $R \sim 10^{-12}$ cm, ce que nous ne pourrons probablement jamais tester. Finalement, le problème de hiérarchie a été transféré en problème de nombre de dimensions, ou de taille ``non-naturellement'' large des dimensions supplémentaires.

Notons que le problème de hiérarchie n'est pas la seule motivation pour introduire les dimensions supplémentaires. Dans d'autres classes de modèles \cite{Antoniadis:1990ew}, ayant par exemple $R^{-1} \sim 1$ TeV, ou contenant des dimensions supplémentaires longitudinales à la brane, c'est l'unification des couplages et la brisure de supersymétrie (voir le chapitre \ref{SUSY-SUGRA}) qui sont étudiés.


\subsection{Dimensions supplémentaires courbées\label{warpedxdim}}


Jusqu'ici, nous avons négligé l'effet de la brane sur les dimensions supplémentaires. Puisque celle-ci porte le Modèle Standard, ou du moins une partie des champs de matière, elle génère un champ gravitationnel directement reliée à sa densité d'énergie, ce qui a pour effet de courber les directions transverses.

Considérons le cas d'une seule dimension supplémentaire, et ignorons l'épaisseur de la brane (fonction $\delta$). La densité d'énergie par unité de volume de la brane, aussi appelée \textit{tension}, est notée $\sigma$. L'action gravitationnelle complète est
\be
\mathcal S = \frac{1}{16 \pi G_{(5)}} \int d^4x dy \sqrt{g^{(5)}} \rl R^{(5)} + 16 \pi \Lambda_{(5)} G_{(5)} \rr + \sigma \int d^4 x \sqrt{g^{(4)}} \quad , \nonumber
\ee
où $\Lambda_{(5)}$ est la constante cosmologique à cinq dimensions, et $G_{(5)}$ la constante de Newton (\ref{constNewton}). On peut montrer que la constante cosmologique effective à quatre dimensions est nulle pour
\be
\Lambda_{(5)} \ = \ - \frac{4\pi}{3} G_{(5)} \sigma^2 \ \leqslant \ 0 \quad . \nonumber
\ee

L'espace-temps fondamental est alors de type anti-de Sitter, et la métrique correspondante a la forme
\be
ds^2 \ = \ e^{ - k\al y \ar}\, \eta_{\mu \nu} dx^{\mu} dx^{\nu} \ - \ dy^2 \quad , \nonumber
\ee
où $k = - \frac{\Lambda_{(5)}}{\sigma}$ est l'inverse du rayon de courbure de la dimension supplémentaire ($e^{-k\al y \ar}$ est appelé ``warp factor''). Ces modèles sont parfois dits non-factorisables à cause du facteur de courbure.

On peut exploiter ce résultat en plaçant le Modèle Standard sur une brane de tension négative, $\sigma < 0$, appelée brane infra-rouge (ou brane TeV), et une autre brane de tension positive, appelée brane ultraviolette (ou brane Planck), aux deux points fixes $0$ et $\pi R$ d'un orbifold \cite{Randall:1999ee}. Du fait de la courbure des dimensions, l'échelle induite sur la TeV brane est
\be
\Lambda \ \simeq \ e^{- \pi k R} M_P \quad , \nonumber
\ee
et l'échelle fondamentale à cinq dimensions est $M_* \sim M_P$. Pour obtenir une échelle $\Lambda \sim 1$ TeV, on a typiquement besoin de $k R \simeq 10$. Ici aussi, on peut expliquer le problème de hiérarchie, mais cette grâce à la courbure des dimensions supplémentaires.


\section{Brisure dynamique de symétrie en six dimensions\label{6Dmodel}}


Dans la suite, nous considérons un champ scalaire dans un espace-temps avec $d=2$ dimensions supplémentaires infinies et plates, et nous plaçons une brane\footnote{Par brane, nous entendons fonction delta bidimensionnelle $\delta^{2} (y)$.} à l'origine $y_1 = y_2 = 0$. Le modèle que nous étudions est
\be
\mathcal S \ = \  \int d^4 x d^2 y \, \frac{1}{2} \rl \dd_M \Phi \rr^2 - \int d^4 x  \pl - \frac{\mu^2}{2} \Phi^2 + \frac{\lambda}{4} \Phi^4 \pr_{y_1 = y_2 = 0} \quad , \label{SGW}
\ee
le second terme étant un potentiel de type Higgs localisé sur la brane. Puisque $\Phi$ a une dimension $2$, le paramètre de masse $\mu$ est adimensionné, tandis que $\ccl \lambda \ccr = -4$.

\begin{figure}[ht!]
\begin{center}
\includegraphics[scale=0.4]{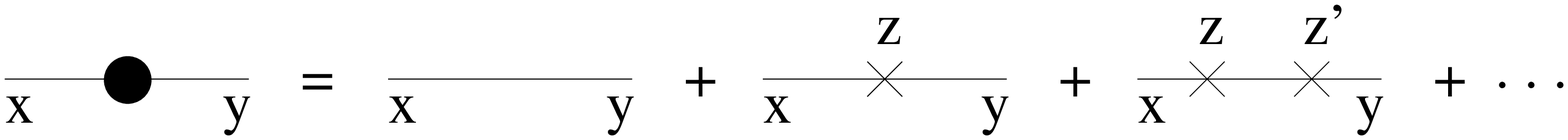}
\caption{Corrections du propagateur $G^{(2)}(x,y)$ de $\Phi$ au niveau classique.}
\label{goldbergerwise}
\end{center}
\end{figure}

Cette théorie contient des divergences au niveau classique \cite{Goldberger:2001tn}. Le propagateur de $\Phi$ reçoit des corrections venant de l'insertion de masse localisée $\mu^2$, comme représenté sur la Fig. \ref{goldbergerwise}.

En sommant toutes les contributions,
\begin{eqnarray}
G^{(2)}(x,y) &=& D(x,y) + \mu^2 \int d^2 z D(x,z) \delta^{2}(z) D(z,y) \nonumber \\
&+& \mu^4 \int d^2 z d^2 z' D(x,z) \delta^{2}(z) D(z,z') \delta^{2}(z') D(z',y) + \ldots \quad , \nonumber
\end{eqnarray}
avec $D(x,y)$ le propagateur nu. Cette expression se réécrit plus simplement dans une représentation mixte
\begin{eqnarray}
G^{(2)}_k(x,y) &=& D_k(x,y) + \mu^2 D_k(x,0) D_k(0,y) \nonumber \\
&+& \mu^4 D_k(x,0) D_k(0,0) D_k(0,y) + \ldots \label{resomGW} \\
&=& D_k(x,y) + \frac{\mu^2}{1 - \mu^2 D_k(0,0)} D_k(x,0) D_k(0,y) \quad , \nonumber
\end{eqnarray}
où $k$ est la quadri-impulsion le long de la brane et $x,y$ sont les coordonnées dans l'espace interne à deux dimensions ; $k$ est conservée à chacun des vertex des diagrammes de la Fig. \ref{goldbergerwise}. La dernière égalité correspond à une resommation de la série.

La quantité $D_k(0,0)$ est particulièrement simple à calculer en coordonnées polaires dans les directions transverses. On trouve
\be
D_k(0,0) \ = \ \frac{1}{4 \pi} \ln \rl \frac{\Lambda^2}{ k^2} \rr \quad , \label{Dk00}
\ee
où $\Lambda$ est une échelle de régularisation ultraviolette.

L'expression (\ref{resomGW}) se réécrit alors
\be
G^{(2)}_k(x,y) \ = \ D_k(x,y) + \mu^2(k) \, D_k(x,0) D_k(0,y) \quad . \nonumber
\ee
Dans cette expression, $\mu^2(k)$ désigne le couplage renormalisé déduit de (\ref{resomGW}) et (\ref{Dk00})
\be
\mu^2 (Q) \ = \ \frac{\mu^2(\Lambda)}{1 + \frac{\mu^2(\Lambda)}{4 \pi} \ln \frac{Q^2}{\Lambda^2}} \quad , \label{running}
\ee
où $\mu^2 (\Lambda)$ est le paramètre de masse intervenant dans l'action (\ref{SGW}). Le couplage $\mu^2(Q)$ décroît avec $Q$, c'est-à-dire qu'il devient fort dans l'infra-rouge.

Nous interprétons cette évolution des couplages de la façon suivante : les divergences qui apparaissent au niveau classique ne sont dûes qu'à l'épaisseur infiniment fine de la brane. En régularisant nos intégrales divergentes grâce au cutoff $\Lambda$, nous avons en fait régularisé l'épaisseur de la brane en $\epsilon \equiv 1/ \Lambda \neq 0$.

Dans la Publication ${\mathcal N}^{\mathrm{o}}\, 1$, nous avons considéré que les dimensions supplémentaires sont compactes en forme de disque\footnote{La forme des dimensions supplémentaires ne joue en réalité aucun rôle vis-à-vis de l'évolution classique des couplages. Néanmoins, les calculs sont simplifiés dans le cas du disque. Nous reviendrons au cas d'un tore de rayons $R_1$ et $R_2$ au paragraphe \ref{model6DNJL}.} de rayon $R \gg \Lambda^{-1}$. Nous imposons des conditions de Dirichlet au bord du disque :
\be
\Phi (r =R) = 0 \quad , \label{Dirichlet}
\ee
avec $r = \sqrt{y_1^2 + y_2^2}$.

\begin{figure}[ht!]
\begin{center}
\includegraphics[scale=0.4]{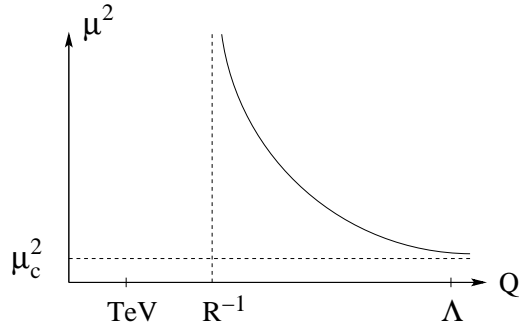}
\caption{\'Evolution du couplage $\mu^2$ avec l'énergie. Lorsque la valeur nue $\mu^2(\Lambda)$ est exactement la valeur (\ref{muc}), un mode non-massif existe à quatre dimensions.}
\label{Running}
\end{center}
\end{figure}

L'évolution (\ref{running}) implique que $\mu^2$ atteint un pôle de Landau en
\be
Q_c = \Lambda \exp \rl -\frac{2 \pi}{\mu^2(\Lambda)}\rr \quad , \label{Qc}
\ee
comme illustré sur la Fig. \ref{Running}.

Si $Q_c$ est suffisamment petit, de telle manière que le couplage $\mu^2$ est perturbatif à l'échelle de compactification $R$, alors le modèle est de type Kaluza-Klein (section \ref{xdim}) et les masses des modes KK de $\Phi$ sont d'ordre $1/R$.

Par contre, si $Q_c \simeq R^{-1}$, alors le mode zéro de $\Phi$ est très léger à quatre dimensions. En $Q_c = R^{-1}$, ce mode est exactement non-massif, tandis que pour $Q_c > R^{-1}$ sa masse devient tachyonique et engendre un mécanisme de type Higgs.

La valeur nue $\mu_c^2$ du couplage correspondant à la transition de phase $Q_c = R^{-1}$ est donc, d'après (\ref{Qc}),
\be
\mu_c^2 \ = \ \frac{2\pi}{\ln \frac{R}{\epsilon}} \quad , \label{muc}
\ee
où $\epsilon$ est l'épaisseur de la brane.

Pour vérifier concrètement l'existence d'une transition de phase en $\mu^2 = \mu_c^2$, il faut résoudre les équations du mouvement pour le mode zéro de $\Phi$ à partir de l'action (\ref{SGW}) pour les deux régions de l'espace tranverse : l'intérieur de la brane, $r \leqslant \epsilon$ ; et l'extérieur de la brane $\epsilon < r \leqslant R$. En imposant les conditions aux bords (\ref{Dirichlet}) et des conditions de correspondance en $r=\epsilon$, on trouve facilement l'expression de la masse quadri-dimensionnelle pour ce mode. On montre en outre que les modes KK sont beaucoup plus lourds que le mode zéro.

Ce modèle très simple est bien entendu motivé par le secteur électrofaible du Modèle Standard. Concrètement, la transition de phase ``à la Higgs'' peut engendrer la brisure de symétrie électrofaible si on peut expliquer pourquoi $\mu^2(\Lambda) > \mu_c^2$, ce qui correspond à une masse tachyonique à quatre dimensions. Un moyen d'expliquer ceci serait de rendre les dimensions supplémentaires dynamiques, c'est-à-dire que le champ scalaire $\Phi$ pourrait être couplé à un module (voir le chapitre \ref{modules}) dont la valeur moyenne dans le vide engendrerait un paramètre de masse $\mu^2(\Lambda)$ dans la bonne gamme de valeurs.


\chapter{Supersymétrie et supergravité\label{SUSY-SUGRA}}


Il existe de nombreuses façons de motiver l'introduction de la supersymétrie comme extension du Modèle Standard. La raison la plus souvent évoquée est le problème de hiérarchie dans le secteur électrofaible \cite{Martin:1997ns}. Le boson de Higgs $H$ est la seule particule scalaire du Modèle Standard. Il est couplé par des interactions de Yukawa $- \lambda_f \psib_L H \psi_R$ à tous les fermions $\psi$, et sa masse reçoit des corrections à une boucle 
\be
\delta_f m_H^2  = - \frac{\al \lambda_f \ar^2}{8\pi^2} \Lambda^2 + \ldots \quad ,
\label{dmHfermions}
\ee
où $\Lambda$ est le cutoff ultraviolet introduit pour régulariser l'intégrale sur le moment interne des fermions dans la boucle\footnote{L'expression (\ref{dmHfermions}) devrait comporter une sommation sur les indices de couleur pour les quarks. Il s'agit ici d'une simple illustration.}. La contribution dominante vient du quark top pour lequel $\lambda_{top} \sim 1$. Par ailleurs, pour que le Modèle Standard soit une théorie unitaire, la masse physique $m_H = m_0 + \delta m_H$ ne peut excéder l'ordre du TeV. Sa valeur moyenne dans le vide valant $246$ GeV, $m_H$ est plutôt de l'ordre d'une centaine de GeV. Si $\Lambda$, qui représente l'échelle ultraviolette de validité du Modèle Standard, est la masse de Planck, cela implique une compensation de plus de $30$ ordres de grandeur entre la masse ``nue'' $m_0$ du potentiel du Higgs et la correction $\delta_f m_H$. Inversement, si on suppose que $\Lambda$ est de l'ordre du TeV de sorte à réduire la compensation non-naturelle, cela signifie que la nouvelle physique est présente à cette échelle.

Le problème de hiérarchie vient du fait que la masse des particules scalaires est très difficile à protéger par des symétries. Les fermions, de leur côté, ont des masses $m_{\psi}$ naturellement petites car ils sont protégés par la symétrie chirale $\psi \rightarrow e^{i\beta \gamma_5}\psi$ que seuls les couplages de Yukawa brisent spontanément. La limite\footnote{Cela revient à prendre la limite $\lambda_f \rightarrow 0$.} $m_{\psi} \rightarrow 0$ augmente donc la symétrie du lagrangien. Les masses $m_{\psi}$ ne peuvent alors pas recevoir de corrections quantiques trop fortes. Pour le Higgs, en revanche, la limite $m_0 \rightarrow 0$ n'entraîne pas d'augmentation de symétrie, ce qui le soumet aux corrections (\ref{dmHfermions}). En associant les scalaires aux fermions dans un même multiplet, on protège automatiquement la masse du boson de Higgs : c'est la supersymétrie.

Pour mieux comprendre ce que cela signifie, supposons qu'un scalaire $\phi$ très lourd soit présent dans la théorie et qu'il soit couplé au Higgs par un terme $- \lambda_s \al H \ar^2 \al \phi \ar^2$. Cela engendre une correction
\be
\delta_s m_H^2 = \frac{\lambda_s}{16\pi^2}\rl \Lambda^2 -2m_{\phi}^2 \ln \frac{\Lambda}{m_{\phi}} \rr \quad ,
\label{dmHbosons}
\ee
où $m_{\phi}$ est la masse de $\phi$. Il est clair que même si $\phi$ est très lourd et découple de la théorie au niveau classique, sa contribution dans les boucles a un effet dominant sur les corrections à la masse du Higgs, et il en est de même pour les particules les plus lourdes auxquelles $H$ soit couplé.

La deuxième conséquence de l'équation (\ref{dmHbosons}), qui n'a plus rien à voir avec $m_{\phi}$, est que le premier terme compense parfaitement la moitié de la contribution fermionique (\ref{dmHfermions}) si $\lambda_s = \al \lambda_f \ar^2$. Donc si chaque fermion du Modèle Standard est apparié à deux scalaires de cette façon, la masse de $H$ ne reçoit plus de correction quadratique mais seulement des corrections logarithmiques du type (\ref{dmHbosons}), et le problème de naturalité est en partie résolu.


\section{La supersymétrie\label{susy}}



\subsection{Algèbre, superespace et superchamps}


Les bosons et les fermions sont associés dans des multiplets supersymétriques. Cela implique que l'algèbre de supersymétrie doit contenir l'algèbre de Poincaré. Effectivement, les théorèmes de Haag-Lopuszanski-Sohnius \cite{Haag:1974qh} et Coleman-Mandula \cite{Coleman:1967ad} établissent que l'algèbre de supersymétrie est la seule algèbre de Lie graduée\footnote{Une algèbre de Lie graduée contient des commutateurs et des anti-commutateurs.} de symétries possibles de la matrice S.

Concrètement, les hypothèses sont :
\begin{enumerate}
\item que la matrice S repose sur une théorie quantique de champs relativistes à quatre dimensions,

\item qu'il n'y a qu'un nombre fini de particules différentes qui puissent être associées à un état de masse donnée,

\item qu'il existe un saut d'énergie entre le vide et les états à une particule.
\end{enumerate}
Le théorème conclut que l'algèbre la plus générale doit contenir le générateur des translations $P_m$, le générateur des rotations $M_{m n}$ et un certain nombre $\mathcal N$ de générateurs de symétries internes $Q^A$. On montre que ces générateurs sont des spineurs, que nous représenterons par des spineurs de Weyl à deux composantes $Q^A_{\alpha}$ et ${\Qb}^A_{\dot \alpha}$ avec $\alpha, \, \dot \alpha = 1, 2$.
Les relations supplémentaires par rapport à l'algèbre de Poincaré sont\footnote{Nous utilisons les conventions de Wess et Bagger \cite{WessBagger}. En particulier, la métrique plate est $\eta_{mn} = \text{diag}\rl -, +,+,+ \rr$.}
\begin{eqnarray}
&&\left\{ Q_{\alpha}^A, {\Qb}_{\dot \beta\, B} \right\} \ = \ 2 \sigma^{m}_{\alpha \dot \beta} P_{m} \delta^A_B \quad , \nonumber \\
&&\left\{ Q_{\alpha}^A, Q_{\beta}^B \right\} \ = \ \varepsilon_{\alpha \beta} Z^{A B} \quad , \label{algebreSusy} \\
&&\left[ Q_{\alpha}^A,  M_{m n}\right] \ = \ \frac{1}{2}\rl \sigma_{m n}\rr_{\alpha}^{\beta} Q_{\beta}^A \quad , \nonumber \\
&&\left[ P_{m}, Q_{\alpha}^A \right] \ = \ \left[ P_{m}, {\Qb}_{\dot \alpha}^A \right] \ = \ 0 \quad , \nonumber
\end{eqnarray}
où $\varepsilon$ est le tenseur antisymétrique à deux indices spinoriels, analogue de la métrique plate $\eta$, et $\sigma_{m n} = \frac{1}{4}\rl \sigma_m {\sigmab}_n - \sigma_n {\sigmab}_m \rr$. Les quantités $Z^{A B}$ sont les charges centrales, elles sont antisymétriques $Z^{A B} = -Z^{B A}$. Dans la suite, nous considèrerons uniquement le cas $\mathcal N = 1$ pour lequel $Z = 0$ par antisymétrie.


\subsubsection*{Transformations de supersymétrie}


La première égalité de l'algèbre traduit non seulement le fait que les générateurs $Q$ et $\Qb$ sont des spineurs de dimension canonique $1/2$, mais surtout que deux transformations de supersymétrie successives forment une translation. La transformation de supersymétrie de paramètres $\xi_{\alpha}$ et ${\xib}_{\dot \alpha}$ pour un champ $\phi$ quelconque s'écrit
\be
\delta_{\xi} \phi \ = \ \rl \xi^{\alpha} Q_{\alpha} + \xib_{\dot \alpha} {\Qb}^{\dot \alpha} \rr \phi \quad ,
\label{transfoSusy}
\ee
ce qui implique que $\xi$ et $\xib$ ont une dimension $-1/2$. Grâce à ces paramètres, l'algèbre de supersymétrie (\ref{algebreSusy}) se réécrit uniquement en termes de commutateurs
\begin{eqnarray}
&&\left[\xi Q, \xib \Qb\right] \ = \ 2 \xi \sigma^m \xib P_m \quad , \nonumber \\
&&\left[\xi Q, \xi Q\right] \ = \ \left[\xib \Qb, \xib \Qb\right] \ = \ 0 \quad , \label{algebreSusy2} \\
&&\left[P^m, \xi Q\right] \ = \ \left[P^m, \xib \Qb\right] \ = \ 0 \quad . \nonumber
\end{eqnarray}
On en déduit que le commutateur de deux transformations de supersymétrie est
\be
\left[\delta_{\xi}, \delta_{\eta}\right]\phi = \rl \delta_{\xi}\delta_{\eta} - \delta_{\eta}\delta_{\xi}\rr \phi = -2i \rl \eta \sigma^m \xib - \xi \sigma^m \bar \eta \rr \dd_m \phi \quad .
\label{commuSusy}
\ee

L'action de $Q$ sur un champ de dimension $d$ donne un champ de dimension $d+ \frac{1}{2}$ : les bosons et les fermions sont appariés par supersymétrie. Pour un champ scalaire $z$, nous définissons son partenaire fermionique $\psi$ par
\be
\delta_{\xi} z \ = \ \sqrt{2} \xi \psi \quad . \label{deltaz}
\ee
Le champ $\psi$ doit lui-même se transformer en un champ de dimension $2$. Le choix le plus général  se ramène à
\be
\delta_{\xi} \psi \ = \ i\sqrt{2}\sigma^m \xib \dd_m z + \sqrt{2}\xi F \quad . \label{deltapsi}
\ee
Si nous voulons que l'algèbre (\ref{commuSusy}) soit close, la transformation de $F$ doit être
\be
\delta_{\xi} F \ = \ i\sqrt{2} \xib {\sigmab}^m \dd_m \psi \quad . \label{deltaf}
\ee
Les champs $\rl z, \psi, F \rr$ forment donc un supermultiplet puisque leurs transformations ferment l'algèbre de supersymétrie. Il contient le même nombre de degrés de liberté fermioniques et bosoniques : $\psi$ est complexe donc il contient 4 degrés de liberté, $z$ en contient 2 et $F$ aussi. Ce multiplet est appelé multiplet chiral ou scalaire.


\subsubsection*{Le superespace}


Ces transformations peuvent être vues d'une autre façon en introduisant le superespace. Il contient, en plus des quatre dimensions usuelles, deux types de coordonnées grassmanniennes $\theta$ et $\thetab$ vérifiant les mêmes relations d'anticommutation que les paramètres de transformation :
\be
\left\{ \theta^{\alpha} , \theta^{\beta} \right\} \ = \ \left\{ \thetab^{\dot \alpha} , \thetab^{\dot \beta} \right\} \ = \ \left\{ \theta^{\alpha} , \thetab^{\dot \beta} \right\} \ = \ 0 \quad .
\label{commuthetas}
\ee
Un point du superespace a des coordonnées $X = \rl x^m, \theta^{\alpha}, \thetab^{\dot \alpha} \rr$.

L'algèbre de supersymétrie (\ref{algebreSusy}) étant une algèbre de Lie, on définit un élément du groupe par
\be
G(x^m, \theta, \thetab) = \exp\left[i\rl -x^m P_m + \theta Q + \thetab \Qb  \rr \right] \quad , \nonumber
\ee
de sorte que la multiplication de deux éléments donne
\be
G(x_1^m, \theta_1, \thetab_1)\, G(x_2^m, \theta_2, \thetab_2) = G(x_1^m + x_2^m + i \theta_2 \sigma^m \thetab_1 - i \theta_1\sigma^m \thetab_2, \theta_1 + \theta_2, \thetab_1 + \thetab_2 ) \quad . \nonumber
\ee

En particulier, une transformation de supersymétrie du type (\ref{transfoSusy}) se déduit de la multiplication
\be
G(0, \xi, \xib) \, G(x^m, \theta, \thetab) = G(x^m + i \theta \sigma^m \xib - i \xi \sigma^m \thetab, \theta + \xi, \thetab + \xib ) \quad . \nonumber
\ee
Cette opération induit une ``super''-translation dans le superespace
\be
\rl x^m , \theta , \thetab \rr \rightarrow \rl x^m + i \theta \sigma^m \xib - i \xi \sigma^m \thetab, \theta + \xi, \thetab + \xib \rr \quad , \label{translationspspace}
\ee
dont une représentation est donnée par les opérateurs différentiels
\begin{eqnarray}
&&Q_{\alpha} \ = \ \frac{\dd}{\dd \theta^{\alpha}} - i \sigma^m_{\alpha \dot \alpha} \thetab^{\dot \alpha} \dd_m \quad , \nonumber \\
&&{\Qb}_{\dot \alpha} \ = \ \frac{\dd}{\dd \thetab^{\dot \alpha}} - i \theta^{\alpha} \sigma^m_{\alpha \dot \alpha}  \dd_m \quad . \label{diffoperators}
\end{eqnarray}
Le superespace nous permet donc de représenter l'algèbre de supersymétrie (\ref{algebreSusy2}) en termes de translations généralisées.

On peut également définir des dérivées covariantes
\begin{eqnarray}
&&D_{\alpha} \ = \ \frac{\dd}{\dd \theta^{\alpha}} + i \sigma^m_{\alpha \dot \alpha} \thetab^{\dot \alpha} \dd_m \quad , \nonumber \\
&&{\Db}_{\dot \alpha} \ = \ - \frac{\dd}{\dd \thetab^{\dot \alpha}} - i \theta^{\alpha} \sigma^m_{\alpha \dot \alpha}  \dd_m \quad . \label{deriveescov}
\end{eqnarray}
qui anticommutent avec les charges (\ref{diffoperators}) :
\be
\left\{D, Q \right\}=\left\{\Db, Q \right\}=\left\{D, \Qb \right\}=\left\{\Db, \Qb \right\}=0 \quad .
\label{DQ+QD}
\ee

Nous définissons un superchamp comme une fonction du superespace. Ce n'est pas un champ physique, et il doit être considéré uniquement à travers son développement en puissance des coordonnées grassmanniennes
\begin{eqnarray}
f (x, \theta, \thetab) &=& z(x) + \theta \phi(x) + \thetab {\chib}(x) + \theta \theta m(x) + \thetab \thetab n(x) \nonumber \\
&+& \theta \sigma^m \thetab A_m(x) + \theta \theta \thetab {\lambdab}(x) + \thetab \thetab \theta \psi(x) + \theta \theta \thetab \thetab d(x) \quad .
\label{superchp}
\end{eqnarray}
Toute puissance supérieure en $\theta$ et $\thetab$ est automatiquement nulle d'après  les relations (\ref{commuthetas}). Les champs $z, \, m, \, n, \, A_m, \, d$ sont bosoniques, ils représentent $16$ degrés de liberté, c'est-à-dire autant que les champs fermioniques $\phi, \, \chi, \, \lambda, \, \psi$. Les transformations $\delta_{\xi} f = \rl \xi Q + \xib \Qb \rr f$ d'un superchamp $f$ se déduisent de celles de ses composantes.

On peut réduire le nombre de composantes des superchamps en imposant des conditions sur $f$. Une condition est $\Db_{\dot \alpha} f = 0$ (ou $D_{\alpha} f^{\dagger}$=0) ; elle définit le superchamp chiral, ou scalaire, dont nous avons déjà étudié les transformations (\ref{deltaz}), (\ref{deltapsi}) et (\ref{deltaf}). La condition $f = f^{\dagger}$ définit le superchamp vectoriel. Dans la suite, nous étudions ces deux types de superchamps en détail. Tout lagrangien supersymétrique renormalisable peut être construit en termes de ces deux seuls types de superchamps.


\subsubsection*{Le superchamp chiral}


En imposant la condition $\Db_{\dot \alpha} \Phi$ à un superchamp (\ref{superchp}), et en remarquant que $\Db_{\dot \alpha} y^m = 0, \, \Db_{\dot \alpha} \theta = 0$ pour $y^m = x^m + i \theta \sigma^m \thetab$, on réduit le nombre de composantes à
\be
\Phi (y, \theta) = z(y) + \sqrt{2} \theta \psi(y) + \theta \theta F(y)  \quad , \nonumber
\ee
c'est-à-dire
\begin{eqnarray}
\Phi (x, \theta, \thetab) &=& z(x) + i \theta \sigma^m \thetab \dd_m z(x) + \frac{1}{4} \theta \theta \thetab \thetab \square z(x)  \nonumber \\
&+&\sqrt{2}\theta \psi(x) -\frac{i}{\sqrt{2}}\theta \theta \dd_m \psi(x) \sigma^m \thetab + \theta \theta F(x) \quad . \label{spchpchiral}
\end{eqnarray}
Le complexe conjugué $\Phi^{\dagger}$ vérifie $D_{\alpha} \Phi^{\dagger}=0$
\be
\Phi^{\dagger} (y^{\dagger}, \thetab) = \zb(y^{\dagger}) + \sqrt{2} \thetab \psib(y^{\dagger}) + \thetab \thetab \Fb(y^{\dagger})  \quad . \nonumber
\ee
La composante $\theta^2 \thetab^2$ du produit $\Phi^{\dagger} \Phi$ est
\be
\left. \Phi^{\dagger} \Phi \ar_{\theta^2 \thetab^2} = F \Fb + \zb \square z - \frac{i}{2}\rl \psi \sigma^m \dd_m \psib - \dd_m \psi \sigma^m \psib \rr + d() \quad , \label{Phi*Phi}
\ee
où $d()$ est une dérivée totale. Ce terme contient les termes cinétiques des champs physiques $z$ et $\psi$. La composante (\ref{Phi*Phi}) se transforme comme une dérivée totale sous la supersymétrie (voir (\ref{deltaz}), (\ref{deltapsi}), (\ref{deltaf})) : elle mène donc à un lagrangien supersymétrique. Il en résulte que $F$ est un champ auxiliaire : il n'a pas de terme cinétique et peut être éliminé par les équations du mouvement.

Par ailleurs, le produit de plusieurs superchamps chiraux entre eux est un superchamp chiral. Leurs composantes $\theta^2$ nous seront utiles dans la suite :
\begin{eqnarray}
&&\left. \Phi_i \Phi_j \ar_{\theta^2} = z_i F_j + z_j F_i - \psi_i \psi_j \quad ,  \label{prodspchpchiraux} \\
&&\left. \Phi_i \Phi_j \Phi_k \ar_{\theta^2} = z_i z_j F_k + z_j z_k F_i + z_k z_i F_j - \psi_i \psi_j z_k - \psi_j \psi_k z_i - \psi_k \psi_i z_j \quad . \nonumber
\end{eqnarray}
D'après  (\ref{deltaf}), la composante $F$ se transforme également en dérivée totale et nous pouvons utiliser un des produits (\ref{prodspchpchiraux}) comme lagrangien supersymétrique.

Nous arrivons au résultat suivant : le lagrangien renormalisable le plus général que l'on puisse construire uniquement à partir de superchamps chiraux $\Phi_i$ est
\be
{\mathscr L} =  \left. \Phi_i^{\dagger} \Phi^i \ar_{\theta^2 \thetab^2} + \left[\left. \rl \lambda_i \Phi^i + \frac{1}{2}m_{ij}\Phi^i \Phi^j + \frac{1}{3}g_{ijk} \Phi^i \Phi^j \Phi^k  \rr \ar_{\theta^2} + {\rm h.c.} \right] \ .
\label{Lchiral}
\ee
Lorsqu'il n'y a qu'une seule ``saveur'' $i$, ce lagrangien décrit le modèle de Wess-Zumino \cite{Wess:1973kz}, qui est le plus simple que l'on puisse construire.

Le lagrangien (\ref{Lchiral}) peut se réécrire comme une intégrale sur le superespace\footnote{Les variables $\theta$ et $\thetab$ étant grassmanniennes, l'intégrale équivaut à une dérivée de sorte que $\int d\theta \rl a\theta +b \rr = a$. On note $d^2 \theta = -\frac{1}{4}\varepsilon_{\alpha \beta}d\theta^{\alpha} d\theta^{\beta}$ et $d^4\theta = d^2\theta d^2 \thetab$. De cette façon, $\int d^2\theta$ sélectionne la composante $\theta^2$ de l'intégrand, et $\int d^4 \theta$ sélectionne la composante $\theta^2\thetab^2$.}
\be
{\mathscr L} \ = \ \int d^4\theta K(\Phi^i, \Phi_j^{\dagger}) \ + \ \left[\int d^2 \theta \, W(\Phi^i) + {\rm h.c.}\right] \quad ,
\label{Zumino}
\ee
où $K$ et $W$ sont le potentiel de K\"ahler et le superpotentiel. Nous reviendrons sur ce point au paragraphe \ref{modeleschiraux}. Dans notre cas,
\begin{eqnarray}
&&K(\Phi^i, \Phi_j^{\dagger}) \ = \ \Phi_i^{\dagger} \Phi^i \quad , \nonumber \\
&&W(\Phi^i) \ = \ \lambda_i \Phi^i + \frac{1}{2}m_{ij}\Phi^i \Phi^j + \frac{1}{3}g_{ijk} \Phi^i \Phi^j \Phi^k \quad .
\label{ZuminoWZ}
\end{eqnarray}
De manière générale, le superpotentiel est une fonction analytique des champs, tandis que le potentiel de K\"ahler est une fonction réelle.

En termes de composantes, le lagrangien (\ref{Lchiral}) s'écrit
\begin{eqnarray}
\mathscr L &=& i \dd_m {\psib}_i {\sigmab}^m \psi^i + \Fb_i F^i + \zb_i \square z^i \label{Lchiralcomp} \\
&+& \left[ \lambda_i F^i + m_{ij} \rl z^i F^j -\frac{1}{2}\psi^i \psi^j \rr  + g_{ijk} \rl z^i z^j F^k - \psi^i \psi^j z^k  \rr  + {\rm h.c.} \right] \quad . \nonumber
\end{eqnarray}

Les champs auxiliaires $F_i$ sont éliminés par les équations du mouvement
\be
\frac{\dd \mathscr L}{\dd \Fb_i} = 0 \quad , \label{eqmouvementF}
\ee
qui se réécrivent, à l'aide du lagrangien (\ref{Lchiralcomp}),
\be
F^i \ = \ - \frac{\dd \overline W}{\dd \zb_i} \quad , \nonumber
\ee
avec $W$ exprimé en fonction des champs scalaires. Après élimination des champs auxiliaires, nous obtenons le lagrangien suivant
\be
\mathscr L = i \dd_m {\psib}_i {\sigmab}^m \psi^i + \zb_i \square z^i +\frac{1}{2}\rl \frac{\dd^2 W}{\dd z^i \dd z^j}\psi^i \psi^j + {\rm h.c.}\rr - \al \frac{\dd W}{\dd z^i}\ar^2 \quad . \label{LchiralW}
\ee
Le dernier terme représente le potentiel scalaire $\mathscr V = \sum \al F^i \ar^2$.


\subsubsection*{Le superchamp vectoriel}


Les vecteurs de jauge sont généralisés en superchamps sur lesquels on impose la contrainte $V = V^{\dagger}$. Les composantes (\ref{superchp}) sont alors réduites à
\begin{eqnarray}
V(x, \theta, \thetab) &=& C(x) + i \theta \chi(x) -i \thetab \chib(x) - \theta \sigma^m \thetab v_m(x) \nonumber \\
&+& \frac{i}{2}\theta^2 \left[ M(x) +iN(x) \right] - \frac{i}{2}\thetab^2 \left[ M(x) -iN(x) \right]  \nonumber \\
&+& i\theta^2 \thetab \left[{\lambdab}(x) +\frac{i}{2}\sigmab^m \dd_m \chi(x) \right] - i \thetab^2 \theta \left[ \lambda(x) +\frac{i}{2} \sigma^m \dd_m \chib(x) \right]  \nonumber \\
&+& \frac{1}{2}\theta^2 \thetab^2 \left[ D(x) + \frac{1}{2}\square C(x)\right] \quad , \nonumber
\end{eqnarray}
avec $M, \, N, \, C, \, D, \, v_m$ réels.

On peut encore diminuer le nombre de composantes en imposant un choix de jauge particulier : celle de Wess-Zumino. En effet, la généralisation des transformations de jauge $U(1)$ habituelles est
\be
V \rightarrow V + \Phi + \Phi^{\dagger}
\label{transfojaugeV}
\ee
où $\Phi$ est un superchamp chiral. En utilisant les composantes $z, \, \psi, \, F$ de ce multiplet, on peut forcer $C, \, \chi, \, M, \, N$ à zéro. On trouve, de plus, que $\lambda$ et $D$ sont invariants de jauge. Le vecteur se transforme comme d'habitude
\be
v_m \rightarrow v_m -i \dd_m \rl z - \zb \rr \quad . \nonumber
\ee
Dans la jauge de Wess-Zumino, le superchamp vectoriel est simplement
\be
V(x, \theta, \thetab) = - \theta \sigma^m \thetab v_m(x) + i\theta \theta \thetab {\lambdab}(x) - i \thetab \thetab \theta \lambda(x) + \frac{1}{2}\theta \theta \thetab \thetab D(x) \quad ,
\label{spchpvect}
\ee
et on voit aisément que $V^n = 0$ pour $n\geqslant 3$.

Par supersymétrie, l'ensemble du multiplet vectoriel est dans la représentation adjointe du groupe de jauge. Par conséquent, les fermions $\lambda$, appelés jauginos, ne sont pas chiraux.

Nous voulons trouver un lagrangien à la fois supersymétrique et invariant de jauge, qui inclue le terme cinétique $v^{mn}v_{mn}$ avec $v_{mn}=\dd_m v_n - \dd_n v_m$. Nous ne pouvons prendre simplement $D^{\alpha}V D_{\alpha}V$ car ce terme n'est pas invariant de jauge. Il faut en réalité introduire des superchamps chiraux
\begin{eqnarray}
&&W_{\alpha} \ = \ - \frac{1}{4} \Db \Db D_{\alpha} V \nonumber \\
&&\overline W_{\dot \alpha} \ = \ - \frac{1}{4} D D \Db_{\dot \alpha} V \label{Walpha}
\end{eqnarray}
qui sont spinoriels et dont on montre facilement qu'ils sont invariants sous (\ref{transfojaugeV}). En composantes,
\be
W_{\alpha} = - i \lambda_{\alpha}(y) + \theta_{\alpha}D(y) -\frac{i}{2} \rl \sigma^m \sigmab^n \theta \rr_{\alpha} v_{mn}(y) \quad .
\label{devWalpha}
\ee

Puisque $W_{\alpha}$ est chiral, la composante $\theta^2$ du produit
\be
\left. W^{\alpha}W_{\alpha}\ar_{\theta^2} = -2i\lambda \sigma^m \dd_m \lambdab -\frac{1}{2}v^{mn} v_{mn}+D^2 +\frac{i}{4}\varepsilon^{mnpq}v_{mn}v_{pq} \nonumber
\ee
se transforme comme une dérivée. Le lagrangien
\begin{eqnarray}
\mathscr L \ &=& \  \frac{1}{4} \left[\int d^2 \theta \, W^{\alpha}W_{\alpha} +  \int d^2 \thetab \, \overline W_{\dot \alpha}\overline W^{\dot \alpha} \right] \nonumber \\
&=&  \frac{1}{2}D^2 - i \lambda \sigma^m \dd_m \lambdab -\frac{1}{4} v^{mn} v_{mn}
\label{Lvect}
\end{eqnarray}
est donc supersymétrique et invariant de jauge. En outre, la composante $D$ du superchamp vectoriel est un champ auxiliaire et peut disparaître de la théorie en résolvant les équations du mouvement.

Remarquons un point important pour la suite : la plus haute composante d'un superchamp se transforme toujours comme une dérivée de manière à clore l'algèbre de supersymétrie. En particulier, 
\be
\delta_{\xi} D = \xib \sigmab^m \dd_m \lambda - \xi \sigma^m \dd_m \lambdab = \dd_m \rl  \xib \sigmab^m \lambda - \xi \sigma^m \lambdab \rr \quad . \nonumber
\ee
Donc, puisque $D$ est supersymétrique et invariant de jauge, rien ne nous empêche de rajouter un terme
\be
\mathscr L_{FI} \ = \ \int d^4 \theta \zeta V \quad ,
\label{LFI}
\ee
avec $\zeta$ constante, dans la théorie. Nous verrons au paragraphe \ref{BrisureSponSusy} qu'un tel terme, appelé terme de Fayet-Iliopoulos, peut briser spontanément la supersymétrie.


\subsection{Théories de jauge supersymétriques\label{theoriejaugeSUSY}}


Notre but dans ce paragraphe est de décrire les interactions entre un secteur de jauge, représenté par le superchamp vectoriel $V$, et la matière, représentée par des superchamps chiraux $\Phi^i$ chargés sous le groupe de jauge $G$.

Les générateurs $T^a$ du groupe, $a=1,\ldots, \text{dim}(G)$, vérifient
\be
\tr (T^a T^b) \ = \ C(r) \delta^{ab} \quad \quad \text{et} \quad \quad \left[T^a, T^b \right] \ = \ i f^{abc} T^c \quad .
\ee

Les superchamps $\Phi_i$ se transforment sous l'action du groupe
\be
\Phi^i \rightarrow \rl e^{-i\Lambda}\rr^i_j \Phi^j \quad \quad , \quad \quad \Phi^{\dagger}_i \rightarrow \Phi^{\dagger}_j \rl e^{i\Lambda^{\dagger}} \rr^j_i \quad ,
\label{transjauge}
\ee
où $\Lambda^i_j = \rl T^a \rr^i_j \Lambda_a$ est une matrice de superchamps chiraux. Il est clair que le terme cinétique $\Phi_i^{\dagger}\Phi^i$ n'est pas invariant de jauge.

Si nous généralisons la transformation de jauge (\ref{transfojaugeV}) de la façon suivante
\be
e^V \rightarrow e^{-i\Lambda^{\dagger}} e^V e^{i\Lambda} \quad ,
\label{transfojaugeexpV}
\ee
avec $V = V^a T^a$, alors le terme
\be
\left. \Phi_i^{\dagger} \rl e^V \rr^i_j \Phi^j \ar_{\theta^2 \thetab^2} \nonumber
\ee
est supersymétrique et invariant de jauge.

En ce qui concerne le superchamp vectoriel, le lagrangien (\ref{Lvect}) est invariant de jauge pourvu qu'on étende la définition de $W_{\alpha}$ en
\begin{eqnarray}
W_{\alpha} &=& -\frac{1}{4} \Db \Db e^{-V} D_{\alpha} e^V = W^a_{\alpha} T^a \quad , \nonumber \\
{\overline W}_{\dot \alpha} &=& -\frac{1}{4}  D D e^{-V} \Db_{\dot \alpha} e^V = \Wb^a_{\dot \alpha} T^a \quad .
\label{WalphaNonAbelien}
\end{eqnarray}
Ces superchamps, matriciels, ne sont pas invariants de jauge $W_{\alpha} \rightarrow e^{-i\Lambda} W_{\alpha} e^{i\Lambda}$. Le lagrangien (\ref{Lvect}) l'est à condition de prendre la trace sur les indices de jauge\footnote{En réalité, dans le cas général, on peut introduire une fonction analytique des superchamps chiraux telle que $f_{ab}(\Phi)W^a W^b$ est invariante de jauge. Dans l'ensemble de la thèse, sauf mention contraire, nous nous restreindrons à $f_{ab} = \delta_{ab}$, ce qui revient à prendre la trace $W^a W^a$.} dans le produit $W^{\alpha}W_{\alpha}$.

Enfin, les termes d'interaction (\ref{Lchiral}) décrits par le superpotentiel (\ref{ZuminoWZ}), sont invariants. La présence des termes linéaires $a_i \Phi^i$ suppose que certains des $\Phi^i$ peuvent être singlets de jauge. Les termes de masse $m_{ij}\Phi^i\Phi^j$ et les couplages cubiques $g_{ijk}\Phi^i\Phi^j\Phi^k$ doivent être invariants de jauge.

Le lagrangien renormalisable et supersymétrique le plus général que l'on puisse construire pour décrire les interactions de jauge non-abéliennes est donc
\begin{eqnarray}
\mathscr L &=& \frac{1}{4C(r)} \left[ \int d^2 \theta \tr \, W^{\alpha}W_{\alpha} + {\rm h.c.} \right] + \int d^4 \theta \, \Phi^{\dagger} e^V \Phi \nonumber \\
&+& \left[\int d^2 \theta \rl a_i \Phi^i + \frac{1}{2}m_{ij}\Phi^i \Phi^j + \frac{1}{3}g_{ijk}\Phi^i \Phi^j \Phi^k \rr + {\rm h.c.}\right] \quad .
\label{Lintjauge}
\end{eqnarray}

Le développement en composantes donne
\begin{eqnarray}
\mathscr L &=& -\frac{1}{4}v_{mn}^a v^{a\, mn}  - i \lambda^a \sigma^m {\mathscr D}_m \lambdab^a -  i \psi^i \sigma^m {\mathscr D}_m {\psib}_i   -\al {\mathscr D}_m z^i \ar^2 \nonumber \\
&+&\frac{1}{2}\rl \frac{\dd^2 W}{\dd z^i \dd z^j}\psi^i \psi^j + {\rm h.c.}\rr + i\sqrt{2}g T^{a\, i}_{\ \ j} \rl \zb_i \psi^j \lambda^a - z^j {\psib}_i {\lambdab}^a  \rr \label{Lgauge} \\
&+& gD^a \zb_i T^{a\, i}_{\ \ j}  z^j + \frac{1}{2}D^aD^a - \frac{\dd W}{\dd z^i}F^i - \frac{\dd \overline W}{\dd \zb_i} \Fb_i + \al F^i\ar^2 \quad , \nonumber
\end{eqnarray}
où l'on a fait explicitement apparaître le couplage de jauge $g$ en redéfinissant le superchamp vectoriel $V \rightarrow 2g V$.

La première ligne du lagrangien contient tous les termes cinétiques. Ils font intervenir les dérivées covariantes et le tenseur champ de jauge
\begin{eqnarray}
&&{\mathscr D}_m z^i \ = \ \dd_m z^i + i g v_m^a T^{a\, i}_{\ \ j} z^j  \quad , \nonumber \\
&&{\mathscr D}_m \psi^i \ = \ \dd_m \psi^i +  i g v_m^a T^{a\, i}_{\ \ j} \psi^j \quad , \nonumber \\
&&{\mathscr D}_m \lambda^a \ = \ \dd_m \lambda^a - g f^{abc} v_m^b \lambda^c \quad , \nonumber \\
&&v_{mn}^a \ = \ \dd_m v_n^a - \dd_n v_m^a -g f^{abc} v_m^b v_m^c \quad . \label{derCov}
\end{eqnarray}

Les équations du mouvement pour les champs auxiliaires donnent
\begin{eqnarray}
&&\Fb_i \ = \ - \frac{\dd W}{\dd z^i} \quad , \nonumber \\
&&D^a \ = \ - g \zb_i T^{a\, i}_{\ \ j} z^j \quad . \label{auxGaugeTheory}
\end{eqnarray}

Après élimination de ces champs non-physiques, le lagrangien (\ref{Lgauge}) prend finalement la forme
\begin{eqnarray}
\mathscr L &=& -\frac{1}{4}v_{mn}^a v^{a \, mn} - i \lambda^a \sigma^m {\mathscr D}_m \lambdab^a -  i \psi^i \sigma^m {\mathscr D}_m {\psib}_i  -\al {\mathscr D}_m z^i \ar^2 \nonumber \\
&+&\frac{1}{2}\rl \frac{\dd^2 W}{\dd z^i \dd z^j}\psi^i \psi^j + {\rm h.c.}\rr + i\sqrt{2}g T^{a\, i}_{\ \ j} \rl \zb_i \psi^j \lambda^a - z^j {\lambdab}^a {\psib}_i  \rr \label{LgaugeSansAux} \\
&-& {\mathscr V}(z^i, \zb_j) \quad , \nonumber
\end{eqnarray}
où $\mathscr V$ est le potentiel scalaire
\begin{eqnarray}
\mathscr V \ &=& \ \sum_i \al F^i \ar^2 + \frac{1}{2}D^a D^a \nonumber \\
&=& \ \sum_i \al \frac{\dd W}{\dd z^i} \ar^2 + \frac{1}{2}g^2 \rl \zb_i T^{a\, i}_{\ \ j} z^j \rr^2 \quad .
\label{potentielSusy}
\end{eqnarray}

Remarquons qu'ici aussi nous pourrions ajouter des termes de Fayet-Iliopoulos (\ref{LFI}). Mais dans une théorie de jauge non-abélienne, les termes $D^a$ donnés par (\ref{auxGaugeTheory}) ne sont pas invariants de jauge, sauf pour les facteurs abéliens que le groupe $G$ contient éventuellement. Pour tout facteur $U(1)$ du groupe de jauge, nous pouvons donc ajouter le terme $\zeta^a D^a$. La principale modification que cela entraîne est que les équations du mouvement de ces composantes deviennent
\be
D^a \ = \ - g \zb_i T^{a\, i}_{\ \ j} z^j -g \zeta^a \quad .
\label{DFI}
\ee


\subsection{Brisure spontanée de la supersymétrie\label{BrisureSponSusy}}


La supersymétrie n'est pas réalisée dans la nature puisqu'aucun superpartenaire n'a jamais été observé pour toutes les particules que nous connaissons. Il faut donc trouver des mécanismes pouvant expliquer sa brisure. Dans ce paragraphe, nous nous penchons sur les conditions sous lesquelles la supersymétrie est brisée spontanément. Pour autant, nous voulons préserver l'invariance de Lorentz contenue dans l'algèbre de supersymétrie. Le mécanisme de brisure doit s'opérer au moyen des champs scalaires contenus dans la théorie puisque ce sont les seuls objets invariants de Lorentz. Nous allons donc étudier le potentiel scalaire (\ref{potentielSusy}) d'une théorie de jauge supersymétrique.

Les vides $\vert \Omega \rangle$ qui préservent la supersymétrie dans une théorie quelconque sont toujours d'énergie nulle. Pour le montrer, reprenons l'algèbre (\ref{algebreSusy}). Elle implique que le hamiltonien d'un système supersymétrique s'écrit en fonction des supercharges
\be
H \ = \ \frac{1}{4} \rl Q_{\alpha} \Qb_{\alpha} +  \Qb_{\alpha} Q_{\alpha} \rr \quad . \nonumber
\ee
L'énergie du vide est
\be
E_0 \ = \ \langle \Omega \vert H \vert \Omega \rangle = \frac{1}{4} \rl \al Q_{\alpha} \vert \Omega \rangle \ar^2 + \al \Qb_{\alpha} \vert \Omega \rangle \ar^2 \rr \quad . \label{E0}
\ee
Cet état est supersymétrique si $Q_{\alpha} \vert \Omega \rangle =0$, ce qui implique que son énergie est $E_0=0$. Inversement, un état $\vert \Omega' \rangle$ d'énergie nulle devra vérifier (\ref{E0}) avec $E_0 = 0$ et donc $Q_{\alpha} \vert \Omega' \rangle =0$.

Nous retrouvons ce résultat à travers le potentiel scalaire (\ref{potentielSusy}) qui est une somme de termes positifs. Tout état supersymétrique a donc une énergie positive ou nulle. La supersymétrie est spontanément brisée si le minimum global a une énergie strictement positive.

Les minima $\langle z^i \rangle$ du potentiel sont donnés par
\be
\left. \frac{\dd \mathscr V}{\dd z^i} \ar_{z^i = \langle z^i \rangle} \ = \ - F^j \frac{\dd^2 W}{dz^i dz^j} -g D^a \zb_j T^{a \, j}_{\ \, i} \ = \ 0 \quad ,
\label{minpotentielSusy}
\ee
et ils ont une énergie
\be
\langle \mathscr V \rangle \ = \ \sum_i \al F^i (\, \langle \zb_j \rangle\,)\ar^2 + \frac{1}{2}\left[ D^a (\, \langle z^i \rangle\, ,\, \langle \zb_j \rangle\,) \right]^2 \quad ,
\label{energieminSusy}
\ee
avec $F^i$ et $D^a$ satisfaisant les équations (\ref{auxGaugeTheory}). La condition pour que la supersymétrie soit brisée spontanément est donc
\be
F^i (\, \langle \zb_i \rangle\,) \ \neq \ 0 \quad \quad \text{ou} \quad \quad D^a (\, \langle z^i \rangle\, ,\, \langle \zb_j \rangle\,) \ \neq \ 0 \quad .
\label{conditionSusybris}
\ee

L'équation du minimum (\ref{minpotentielSusy}) peut être réécrite de la façon suivante
\be
\bigg \langle  \begin{pmatrix} F^i & D^a \end{pmatrix}  \begin{pmatrix} \frac{\dd^2 W}{dz^i dz^j} & g \zb_j T^{a \, j}_{\ \ i} \cr \cr g \zb_i T^{a \, i}_{\ \ j} & 0 \end{pmatrix} \bigg \rangle \ = \ 0 \quad , \nonumber
\ee
où l'on a utilisé l'invariance de jauge du superpotentiel. Cette formulation est particulièrement intéressante car on remarque que la matrice intervenant ici n'est autre que la matrice de masse qui mélange les fermions $\psi^i$ et les jauginos $\lambda^a$ dans le lagrangien (\ref{LgaugeSansAux}) :
\be
\frac{1}{2} \begin{pmatrix} \psi^i & \sqrt{2}i\lambda^a \end{pmatrix}  \begin{pmatrix} \frac{\dd^2 W}{dz^i dz^j} & g \zb_j T^{a \, j}_{\ \ i} \cr \cr g \zb_i T^{a \, i}_{\ \ j} & 0 \end{pmatrix} \begin{pmatrix} \psi^j \cr \cr \sqrt{2}i\lambda^a \end{pmatrix} \quad . \nonumber
\ee

D'après cette constatation et en utilisant (\ref{conditionSusybris}), la brisure spontanée de la supersymétrie entraîne la présence d'un fermion de Golstone, le Goldstino, donné par
\be
\lambda_G \ = \ \sum_i \langle F^i \rangle \psi^i \ + \ \frac{i}{\sqrt{2}}g \langle D^a \rangle \lambda^a \quad .
\label{goldstino}
\ee
Le Goldstino a une masse nulle et ne pourra disparaître du spectre, à l'instar des théories de jauge, que si nous rendons la supersymétrie locale, c'est-à-dire dans une théorie de supergravité. Une autre approche consisterait à penser que la supersymétrie n'est pas brisée spontanément mais explicitement. Mais dans ce cas aussi, il nous faut trouver un mécanisme qui pourrait engendrer des termes non-supersymétriques dans le lagrangien. Nous étudierons cela dans le cadre de la supergravité, section \ref{potentielsugra}.

Si la supersymétrie est brisée spontanément par un $\langle F^i\rangle \neq 0$, alors le Goldstino sera simplement le fermion $\psi^i$ associé. Si par contre, c'est un $D^a$ qui a une valeur moyenne non-nulle, le Goldstino sera le jaugino associé. Dans la suite, nous étudions ces deux types de brisure. Le premier cas se produit dans les modèles de O'Raifeartaigh \cite{O'Raifeartaigh:1975pr}, tandis que le second mécanisme fut proposé par Fayet et Iliopoulos \cite{Fayet:1974jb} et est lié aux termes (\ref{LFI}).


\subsubsection*{Le modèle de Fayet-Iliopoulos}


Considérons le cas le plus simple : celui de l'électrodynamique supersymétrique
\begin{eqnarray}
\mathscr L &=& \frac{1}{4} \left[ \int d^2 \theta \, W^{\alpha}W_{\alpha} + {\rm h.c.} \right] + \int d^4 \theta \rl \Phi_1^{\dagger} e^{+eV} \Phi_1 + \Phi_2^{\dagger} e^{-eV} \Phi_2 \rr \nonumber \\
&+& \left[\int d^2 \theta \,  m\Phi_1 \Phi_2  + {\rm h.c.}\right] + \int d^4 \theta \zeta V \quad ,
\label{LQED}
\end{eqnarray}
où $e$ représente la charge électrique. Les équations du mouvement
\begin{eqnarray}
&&D + \frac{1}{2}\zeta + \frac{e}{2} \rl \al z_1 \ar^2 - \al z_2 \ar^2 \rr =0 \nonumber \\
&&F_1 + m \zb_2 = 0 \nonumber \\
&&F_2 + m \zb_1 = 0 \nonumber
\end{eqnarray}
n'ont pas de solution $D = F_1 = F_2 = 0$ : la supersymétrie est brisée spontanément.
Le potentiel scalaire est
\be
\mathscr V = \frac{1}{8} \zeta^2 + \rl m^2 + \frac{1}{4}e\zeta \rr \al z_1 \ar^2 + \rl m^2 - \frac{1}{4}e\zeta \rr \al z_2 \ar^2 + \frac{e^2}{8} \rl  \al z_1 \ar^2 - \al z_2 \ar^2 \rr^2 \ .
\label{VFI}
\ee

\begin{figure}[ht!]
\begin{center}
\includegraphics[scale=1.2]{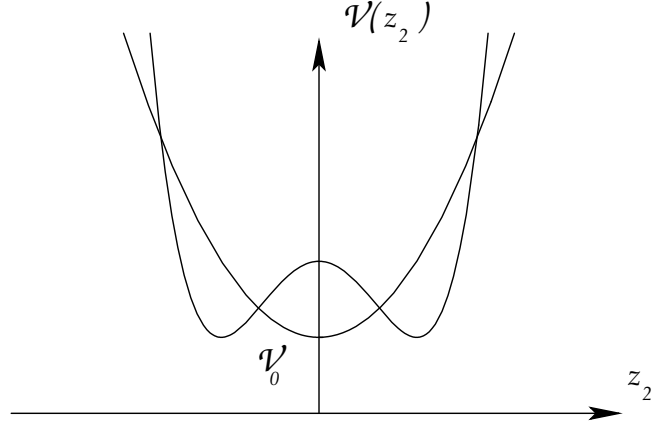}
\caption{Brisure spontanée de la supersymétrie par une composante $D$ : l'énergie du minimum est non-nulle. Dans un cas seule la supersymétrie est brisée, dans l'autre on brise également la symétrie de jauge.}
\label{Dbreaking}
\end{center}
\end{figure}

Deux cas se présentent : $m^2 > \frac{1}{4}e\zeta$ ou $m^2 < \frac{1}{4}e\zeta$.

Si $m^2 > \frac{1}{4}e\zeta$, le minimum est en $z_1=z_2=0$ ; ces deux champs scalaires complexes ont des masses $m_1^2 = m^2+ \frac{1}{4}e\zeta$ et $m_2^2 = m^2- \frac{1}{4}e\zeta$. Le reste du spectre est inchangé, et le jaugino $\lambda$ joue le rôle du Goldstino. La séparation de masse entre les scalaires et les fermions d'un multiplet signe la brisure de supersymétrie.

Plaçons-nous maintenant dans le second cas : $m^2 < \frac{1}{4}e\zeta$. Le potentiel est minimisé en $z_1=0$ et $z_2 = \langle z_2 \rangle \neq 0$ : la symétrie de jauge est spontanément brisée par $z_2$. Ces deux cas sont illustrés sur la Fig. \ref{Dbreaking}.

Les modèles de Fayet-Iliopoulos sont très intéressants et ont conduit à beaucoup de recherche pour briser la supersymétrie. En effet, dans les théories de supergravité issues de théories des cordes, un certain nombre de facteurs abéliens sont présents dans le groupe de jauge. Nous reviendrons sur ces mécanismes dans le paragraphe \ref{Dtermuplifting}.


\subsubsection*{Les modèles de O'Raifeartaigh}


Il nous faut construire un superpotentiel de telle sorte que l'un au moins des $F^i \neq 0$ au minimum, tout en gardant $D^a = 0$. Nous aurons besoin d'un terme linéaire pour nous assurer qu'un des champs scalaires $z^i$ a une valeur moyenne non-nulle. Cela signifie qu'au moins un champ singlet de jauge est présent. Le modèle le plus simple implique au moins trois superchamps chiraux $X, Y, Z$. Le superpotentiel est
\be
W \ = \ Y \rl M^2 - X^2 \rr + \mu Z X + w(X, \Phi^i) \quad ,
\label{WO'rai}
\ee
où $Y$ est le singlet de jauge qui a un terme linéaire $M^2 Y$, et $\Phi^i$ représente les autres superchamps chiraux présents dans la théorie.

Les équations du mouvement
\be
-\Fb_Y \ = \ \frac{\dd W}{\dd Y} \ = \ M^2 - X^2 \quad \quad , \quad \quad -\Fb_Z \ = \ \frac{\dd W}{\dd Z} \ = \ \mu X \quad , \nonumber
\ee
ne peuvent pas s'annuler simultanément et la supersymétrie est brisée spontanément.

Le potentiel scalaire
\be
\mathscr V = \al M^2 - X^2 \ar^2 + \mu^2 \al X \ar^2 + \al \mu Z - 2XY + \frac{\dd w}{\dd X} \ar^2 + \al \frac{\dd w}{\dd z^i} \ar^2 + \frac{1}{2}D^aD^a 
\label{VO'rai}
\ee
est minimisé pour
\be
\langle X^2 \rangle \ = \ M^2 - \frac{\mu^2}{2} \quad \quad , \quad \quad \langle \mu Z - 2XY + \frac{\dd w}{\dd X} \rangle = 0 \quad , \nonumber
\ee
si l'on suppose que $\langle \frac{\dd w}{\dd z^i}\rangle =0$ et $\langle D^a \rangle =0$. L'énergie de ce vide est
\be
\mathscr V_0 \ = \ \mu^2 \rl \frac{\mu^2}{4} + \al M^2 - \frac{\mu^2}{2} \ar \rr \quad . \nonumber
\ee

On montre que seul le spectre du superchamp $X$ est non-supersymétrique. De nombreuses extensions du modèle de O'Raifeartaigh sont possibles, mais elles reposent toujours sur le principe d'un singlet de jauge $Y$ qui force le scalaire d'un autre superchamp $X$ à acquérir une valeur moyenne non-nulle.


\subsection{Les $R$-symétries\label{Rsym}}


Il s'agit de symétries chirales $U(1)$ continues du lagrangien sous lesquelles les coordonnées du superespace se transforment :
\be
\rl x^m , \theta, \thetab \rr \longrightarrow \rl x^m , e^{iR_{\theta}\eta} \theta, e^{-iR_{\theta}\eta} \thetab \rr \quad .
\nonumber
\ee

À partir de ces transformations, nous pouvons déduire les conditions pour lesquelles le lagrangien supersymétrique (\ref{Lintjauge}) est invariant. Soient $R_i$ les charges des superchamps de matière $\Phi^i$, les superchamps vectoriels sont invariants,
\begin{eqnarray}
V^a(x, \theta, \thetab) \ &\longrightarrow &\  V^a(x, e^{iR_{\theta}\eta}\theta, e^{-iR_{\theta}\eta}\thetab) \quad , \nonumber \\
\Phi^i(x, \theta, \thetab) \ &\longrightarrow &\  e^{iR_i\eta}\Phi^i(x, e^{iR_{\theta}\eta}\theta, e^{-iR_{\theta}\eta}\thetab) \quad , \label{Rtransfo} \\
\Phi^{\dagger}_i(x, \theta, \thetab) \ &\longrightarrow &\  e^{-iR_i\eta}\Phi^{\dagger}_i(x, e^{iR_{\theta}\eta}\theta, e^{-iR_{\theta}\eta}\thetab) \quad .  \nonumber
\end{eqnarray}
Le choix des charges $R_i$ est arbitraire et caractérise la $R$-symétrie.

La conséquence première des $R$-symétries est que les composantes d'un multiplet ne se transforment pas de la même façon
\be
\begin{array}{ccc}
z^i \ \rightarrow \  e^{iR_i\eta}z^i \quad &,& \quad \zb_i \ \rightarrow \  e^{-iR_i\eta}\zb_i \quad , \\
\psi^i \ \rightarrow \  e^{i\rl R_i-R_{\theta}\rr\eta}\psi^i \quad &,& \quad \psib^i \ \rightarrow \  e^{-i\rl R_i-R_{\theta}\rr\eta}\psib^i \quad ,  \\
F^i \ \rightarrow \  e^{i\rl R_i-2R_{\theta}\rr \eta}F^i \quad &,& \quad \Fb_i \ \rightarrow \  e^{-i\rl R_i-2R_{\theta}\rr \eta}\Fb_i \quad ,   \\
\lambda^a \ \rightarrow \ e^{-iR_{\theta}\eta} \lambda^a \quad &,& \quad \lambdab^a \ \rightarrow \ e^{iR_{\theta}\eta} \lambdab^a \quad . 
\end{array}
\label{Rtransfocomp}
\ee

Les vecteurs $v^a_m$ et les champs auxiliaires $D^a$ sont invariants. Le superchamp chiral de jauge $W^{a}_{\alpha}$ a la même transformation que sa plus basse composante $\lambda^a$ d'après (\ref{devWalpha}).

L'intégrale grassmannienne $\int d^2\theta \, \theta^2 = 1$ doit être invariante, donc 
\be
d^2 \theta \rightarrow e^{-2iR_{\theta}\eta} d^2 \theta \quad . \nonumber
\ee

En écrivant le lagrangien le plus général possible
\be
\mathscr L \ = \ \int d^4 \theta K(\Phi^{\dagger}_i e^V \Phi^i) \ + \ \left[\int d^2 \theta \rl W(\Phi^i) + \tr \, W^{\alpha}W_{\alpha} \rr  + {\rm h.c.} \right] \quad , \nonumber
\ee
nous voyons que le potentiel de K\"ahler est toujours $R$-invariant. Le superchamp chiral de jauge $W_{\alpha}$ a la charge $-R_{\theta}$. Le superpotentiel W, comme le produit $W^{\alpha}W_{\alpha}$, a une charge $R_W = 2R_{\theta}$ indépendante des charges $R_i$. Par conséquent, le choix des charges des superchamps de matière permet d'empêcher certains couplages dans le superpotentiel.

Dans le Modèle Standard Supersymétrique Minimal (MSSM), qui est l'extension minimale du Modèle Standard, un choix de $R$-symétrie est $R_i = R_{\theta}=1$ pour les superchamps décrivant les quarks et les leptons, et $R_1 = R_2 = 0$ pour les superchamps $H_1$ et $H_2$ décrivant les deux doublets de Higgs\footnote{Le superpotentiel étant une fonction analytique des superchamps, les couplages de Yukawa $y_u \bar u Q H$, $y_d \bar d Q H^{*}$ et $y_e \bar e L H^*$ du Modèle Standard doivent être généralisés. Il faut donc introduire deux doublets de Higgs, de charges opposées. Dans le superpotentiel, le premier doublet couple aux quarks de type up $y_u \bar u Q H_1$, et le second couple aux quarks de type down $y_d \bar d Q H_2$, ainsi qu'aux leptons $y_e \bar e Q H_2$.}. Ce choix permet d'empêcher tous les couplages trilinéaires violant les nombres leptonique et baryonique et pouvant donner lieu à une désintégration du proton. Néanmoins ce choix empêche également un terme de masse $\mu H_1 H_2$, qui l'équivalent du terme de masse $\mu^2 H^2$ à l'origine de la brisure électrofaible.

De manière générale, la présence d'une $R$-symétrie continue en supersymétrie globale pose les problèmes suivants \cite{Nelson:1993nf} :
\begin{itemize}
 \item[$\bullet$] soit la $R$-symétrie n'est pas brisée, et alors les jauginos ne peuvent pas avoir de masse car ils ne sont pas chiraux,

 \item[$\bullet$] soit la $R$-symétrie est brisée spontanément, auquel cas le spectre contient un boson de Goldstone de masse nulle.
\end{itemize}
En particulier, on ne peut pas briser spontanément la supersymétrie dans le MSSM par des mécanismes de Fayet-Iliopoulos ou O'Raifeartaigh car ceux-ci contiennent des $R$-symétries\footnote{Celles-ci sont respectivement $R_1 = R_2 = 1$ pour le modèle de Fayet-Iliopoulos, et $R_X = 0$ et $R_Y = R_Z = 2$ pour le modèle de O'Raifeartaigh, si on a choisi $R_{\theta} = 1$.}.

On considère plutôt la présence d'une parité $R = (-1)^{3(B-L)+2S}$, avec $B$ et $L$ les nombres baryonique et leptonique, et $S$ le spin. Cette symétrie $\mathbb Z_2$ assigne une charge $1$ aux particules du Modèle Standard et aux doublets de Higgs, et une charge $-1$ à leur superpartenaire. On la considère souvent comme le résidu d'une $R$-symétrie continue brisée à haute énergie. La $R$-parité a plusieurs avantages très importants. D'abord, elle empêche les termes conduisant à la désintégration du proton, et autorise le terme de masse $\mu H_1 H_2$. Ensuite, elle implique la production des sparticules par paires. Par conséquent, la plus légère d'entre elles (appelée LSP pour ``lightest super-particule'') est forcément stable si la $R$-parité est conservée. La LSP est alors un candidat naturel pour décrire la matière noire de l'Univers.


\subsection{Modèles chiraux et géométrie de K\"ahler\label{modeleschiraux}}


Dans cette partie, nous développons un certain nombre de propriétés des théories chirales supersymétriques \cite{Zumino:1979et}. Cela nous sera particulièrement utile lorsque nous étudierons la supersymétrie locale, section \ref{sugra}, mais ces relations sont plus aisées à dériver dans le cas global, et restent valides en supergravité.

Comme nous l'avons rapidement évoqué, le lagrangien le plus général que l'on puisse écrire à partir de superchamps chiraux $\Phi^i$ a la forme
\be
{\mathscr L} \ = \ \int d^4\theta K(\Phi^i, \Phi_j^{\dagger}) \ + \ \left[\int d^2 \theta \, W(\Phi^i) + {\rm h.c.}\right] \quad ,
\label{Lchiralgeneral}
\ee
avec, en toute généralité,
\begin{eqnarray}
&&K = \sum a_{i_1,\ldots,i_N}^{\quad \quad \ j_1,\ldots,j_M} \Phi^{i_1} \ldots \Phi^{i_N} \Phi^{\dagger}_{j_1} \ldots \Phi^{\dagger}_{j_M} \equiv \sum a_{i_1,\ldots,i_N}^{\quad \quad \ j_1,\ldots,j_M} K_M^N \quad , \nonumber \\
&&W = \sum b_{i_1,\ldots,i_P}\Phi^{i_1} \ldots \Phi^{i_P} \quad , \nonumber
\end{eqnarray}
où $K_M^N$ est la notation abrégée pour $K^{i_1,\ldots,i_N}_{\quad \quad \ j_1,\ldots,j_M}$.

En utilisant les développements en composantes des superchamps chiraux, on peut calculer les composantes $\theta^2$ de $W$ et $\theta^2 \thetab^2$ de $K$ pour trouver l'expression du lagrangien en composantes. À partir de (\ref{spchpchiral}) et (\ref{prodspchpchiraux}), nous voyons que le superpotentiel s'écrit
\be
W(\Phi^i) = W(z) + \sqrt{2}\theta \psi^i \frac{\dd W(z)}{\dd z^i} + \theta \theta \left[F^i\frac{\dd W(z)}{\dd z^i} - \frac{1}{2}\psi^i \psi^j\frac{\dd^2 W}{\dd z^i \dd z^j} \right] \quad ,
\label{Wgeneral}
\ee
où $W(z) = \sum b_{i_1,\ldots,i_P}z^{i_1} \ldots z^{i_P}$ est le superpotentiel exprimé en termes des champs scalaires.
Pour écrire le développement du potentiel de K\"ahler, il est plus simple de commencer par le développement de chaque monôme $K_M^N$. Il faut interpréter le produit $\Phi^{\dagger}\Phi$ donné par l'expression (\ref{Phi*Phi}) pour trouver
\begin{eqnarray}
\left. K_M^N \ar_{\theta^2 \thetab^2} &=&  -\frac{1}{2}\frac{\dd^3 K_M^N(z,\zb)}{\dd z^i \dd \zb^j \dd \zb^k} F^i \psib^j \psib^k  -\frac{1}{2}\frac{\dd^3 K_M^N(z,\zb)}{\dd \zb^i \dd z^j \dd z^k} \Fb^i \psi^j \psi^k  \nonumber \\
&+& \frac{1}{4}\frac{\dd^4 K_M^N(z,\zb)}{\dd z^i \dd z^j \dd \zb^k\dd \zb^l} \psi^i \psi^j \psib^k \psib^l -\frac{\dd^2 K_M^N(z,\zb)}{\dd z^i \dd \zb^j } \dd_m z^i \dd^m \zb^j \nonumber \\
&-& i \frac{\dd^2 K_M^N(z,\zb)}{\dd z^i \dd \zb^j } \psib^j \sigmab^m \dd_m \psi^i - i \frac{\dd^3 K_M^N(z,\zb)}{\dd z^i \dd z^j \dd \zb^k} \psib^k \sigmab^m \psi^i \dd_m z^j \nonumber \\
&+& \frac{\dd^2 K_M^N(z,\zb)}{\dd z^i \dd \zb^j}F^i \Fb^j\quad . \nonumber
\end{eqnarray}
L'expression pour le potentiel de K\"ahler complet a la même forme en remplaçant $K_M^N(z,\zb)$ par $K(z,\zb)$.

Il se trouve que ces expressions sont grandement simplifiées par l'introduction de la quantité
\be
K_{i \bar j} = \frac{\dd^2 K(z^i, \zb^j)}{\dd z^i \dd \zb^j} \quad ,
\label{Kahlermetric}
\ee
Cette grandeur est la métrique de K\"ahler. En effet, les champs scalaires $z^i$ et $\zb^j$ définissent une variété de K\"ahler, qui est un type de variété de Riemann analytique. La métrique de K\"ahler doit être définie positive et inversible. Elle permet alors de monter ou baisser les indices des tenseurs
\begin{eqnarray}
&&V^i = K^{i \bar j} V_{\bar j} \quad \quad , \quad \quad V^{\bar j} = K^{i \bar j} V_i \quad , \nonumber \\
&&V_i = K_{i \bar j} V^{\bar j} \quad \quad , \quad \quad V_{\bar j} = K_{i \bar j} V^i \quad . \nonumber
\end{eqnarray}
On peut définir une dérivée covariante $\nabla_i, \, \nablab_{\bar j}$ qui doit respecter la structure analytique de la variété. En outre, elle doit être compatible avec la métrique
\be
\nabla_k K_{i \bar j} = 0 = \nablab_{\bar k} K_{i \bar j} \quad . \nonumber
\ee
Ces restrictions impliquent que la connection $\Gamma$ de la variété a pour seules composantes non-nulles $\Gamma^i_{jk}$ et $\Gamma^{\bar i}_{\bar j \bar k}$
\be
\Gamma^k_{ij} \ = \ K^{k \bar l} \frac{\dd K_{j \bar l}}{\dd z^i} \ = \ \Gamma^k_{ji} \quad .
\label{connectionKahler}
\ee
On peut également définir un tenseur de courbure
\be
\left[\nabla_i, \nabla_j\right] V_k = R^l_{ijk} V_l \quad , \quad \left[\nabla_i, \overline{\nabla}_{\bar j}\right] V_k = R^l_{i\bar j k} V_l \quad , \nonumber
\ee
dont on montre que les seules composantes non-nulles sont
\be
R_{i\bar j k \bar l} \ = \ K_{m\bar l} \ \frac{\dd \Gamma^m_{ik}}{\dd \zb^j} \ = \ \frac{\dd^2 K_{k \bar l}}{\dd z^i \dd \zb^j} \ - \ K^{m \bar n}\ \frac{\dd K_{m\bar l}}{\dd \zb^j} \ \frac{\dd K_{k \bar n}}{\dd z^i} \quad , \label{courbureKahler}
\ee
et son complexe conjugué. Le tenseur de courbure est antisymétrique sous l'échange de deux indices de types différents. La métrique, la connection et la courbure sont invariantes sous certaines transformations du potentiel de K\"ahler
\be
K(z, \zb) \longrightarrow K(z,\zb) + F(z) + \overline F(\zb) \quad ,
\label{transfoKahler}
\ee
appelées transformations de K\"ahler, où la fonction $F$ est analytique.

À l'aide de ces objets, le lagrangien (\ref{Lchiralgeneral}) s'écrit en termes des composantes
\begin{eqnarray}
\mathscr L &=& K_{i \bar j} F^i \Fb^j + \frac{1}{4} \frac{\dd^2 K_{i \bar j}}{\dd z^k \dd \zb^l} \psi^i \psi^k \psib^j \psib^l - F^i \left[\frac{1}{2}K_{i \bar m}\Gamma^{\bar m}_{\bar j \bar k}\psib^j \psib^k - \frac{\dd W}{\dd z^i}\right] \nonumber \\
&-& \Fb^i \left[\frac{1}{2}K_{m \bar i}\Gamma^{m}_{j k} \psi^j  \psi^k - \frac{\dd \overline W}{\dd \zb^i}\right] - K_{i \bar j}\dd_m z^i \dd^m \zb^j - i K_{i \bar j} \psib^j \sigmab^m D_m \psi^i \nonumber \\
&-& \frac{1}{2} \frac{\dd^2 W}{\dd z^i \dd z^j} \psi^i \psi^j - \frac{1}{2} \frac{\dd^2 \overline W}{\dd \zb^i \dd \zb^j} \psib^i \psib^j \quad , \nonumber
\end{eqnarray}
où $D_m \psi^i = \dd_m \psi^i + \Gamma^i_{jk}\dd_m z^j \psi^k$ est une dérivée d'espace-temps covariante sur la variété de K\"ahler pour $\psi^i$. Les champs auxiliaires vérifient
\be
K_{i \bar j}F^i - \frac{1}{2}K_{k \bar j}\Gamma^{k}_{m l} \psi^m  \psi^l - \frac{\dd \overline W}{\dd \zb^j} \ = \ 0 \quad . \nonumber
\ee

L'élimination de ces derniers nous donne le résultat final
\begin{eqnarray}
\mathscr L &=&  - K_{i \bar j}\dd_m z^i \dd^m \zb^j - i K_{i \bar j} \psib^j \sigmab^m D_m \psi^i + \frac{1}{4}R_{i\bar j k \bar l}\, \psi^i \psi^k \psib^j \psib^l \nonumber \\
&-& \frac{1}{2} D_i W_j \psi^i \psi^j - \frac{1}{2} D_{\bar i} \Wb_{\bar j} \psib^i \psib^j - K^{i \bar j} W_i \Wb_{\bar j} \quad , \label{LZuminocomp}
\end{eqnarray}
où
\be
W_i \ = \ \frac{\dd W}{\dd z^i} \quad \quad \text{et} \quad \quad D_i W_j = \dd_i W_j \ - \ \Gamma^k_{i j}W_k \quad .
\label{DiW}
\ee

Le troisième terme du lagrangien (\ref{LZuminocomp}) est un terme non-renormalisable d'interactions à quatre fermions. Dans le modèle de Wess-Zumino (\ref{Lchiral}), le potentiel de K\"ahler est canonique $K = \Phi^{\dagger}_i \Phi^i$. La métrique de K\"ahler est plate dans ce cas, $K_{i \bar j} = \delta_{i \bar j}$, et le tenseur de courbure est identiquement nul. Ces termes ne sont donc pas présents. Dans les modèles de supersymétrie locale que nous considèrerons au paragraphe \ref{intjaugesugra}, nous ne ferons aucune hypothèse sur le potentiel de K\"ahler, et le lagrangien contiendra ces interactions à quatre fermions.

Le résultat (\ref{LZuminocomp}) montre que la donnée de la métrique de K\"ahler et du superpotentiel spécifient entièrement le lagrangien d'une théorie supersymétrique. En supergravité, nous tirerons avantage du fait que le potentiel de K\"ahler n'est pas défini de manière unique (éq. (\ref{transfoKahler})).


\section{La supergravité\label{sugra}}



\subsection{L'inclusion naturelle de la gravitation\label{gravitation}}


Pour briser spontanément la supersymétrie, nous avons vu qu'il est plus intéressant de la jauger, c'est-à-dire de rendre les paramètres $\xi$ et $\xib$ des transformations (\ref{transfoSusy}) dépendant de $x$. À partir de l'algèbre (\ref{commuSusy}), il est clair que le produit de deux transformations de supersymétrie induira alors une translation $\xi(x) \sigma^m \bar \eta(x) \dd_m$ dépendant du point $x$. Ceci est une transformation générale des coordonnées et nous nous attendons à ce que la gravitation intervienne dans les théories de supersymétrie locale : c'est la raison pour laquelle elle est appelée supergravité.

Pour jauger la supersymétrie, nous allons nous inspirer de ce qui se produit habituellement quand on rend une symétrie locale. Considérons le lagrangien d'un fermion de Dirac (4 composantes) libre
\be
\mathscr L \ = \ i\chib \gamma^m \dd_m \chi \quad . \nonumber
\ee
Il est invariant sous $\chi \rightarrow e^{-i\alpha} \chi$ qui est une symétrie $U(1)$ globale. Dès lors qu'on jauge cette symétrie, $\dd_m \alpha(x) \neq 0$, le lagrangien n'est plus invariant. La procédure de Noether nous apprend que le courant associé est $J^m = \chib \gamma^m \chi$ et que la variation du lagrangien est
\be
\delta \mathscr L \ = \ \rl \dd_m \alpha \rr J^m \quad . \nonumber
\ee
Pour compenser ce terme, on doit introduire le vecteur $A^m$ de spin $1$, qui se transforme selon $\delta A_m = \dd_m \alpha(x)$, et ajouter le lagrangien $-A_mJ^m$. On intègre ce dernier à la théorie en promouvant la dérivée $\dd_m$ à une dérivée covariante $\mathscr D_m = \dd_m +iA_m$. Le lagrangien
\be
\mathscr L \ = \ i\chib \gamma^m \mathscr D_m \chi \nonumber
\ee
est invariant sous la symétrie de jauge $U(1)$. Bien entendu, on doit également rajouter à la théorie le terme cinétique $-\frac{1}{4}F^{mn}F_{mn}$ du champ de jauge.

Suivons la même procédure dans le cas de la supersymétrie \cite{Nilles:1983ge}. Le lagrangien supersymétrique d'un multiplet chiral $\rl z, \chi \rr$ libre est
\be
\mathscr L \ = \ -\al \dd_m z \ar^2 - i \chi \sigma^m \dd_m \chib \quad , \nonumber
\ee
où nous avons éliminé le champ auxiliaire par l'équation du mouvement triviale $F=0$. En reprenant les transformations (\ref{deltaz}) et (\ref{deltapsi}) avec $\xi \rightarrow \xi(x)$, nous trouvons que la variation du lagrangien sous la supersymétrie locale est
\be
\delta_{\xi} \mathscr L \ = \ \text{dérivée totale} \ + \ \sqrt{2} \rl \chi \sigma^m \sigmab^n \dd_n \zb \dd_m \xi + {\rm h.c.} \rr \quad ,
\label{deltaLsusylocale}
\ee
ce qui montre que le lagrangien n'est pas invariant. Il faut donc introduire un nouveau champ. Puisque les générateurs de supersymétrie sont spinoriels, il est naturel que cette nouvelle particule soit un fermion qui, comme dans le cas $U(1)$ vu précédemment, se transforme selon
\be
\delta_{\xi} \psi^{\alpha}_m \ = \ \frac{1}{\kappa}\dd_m \xi^{\alpha}(x)
\label{transfogravitino}
\ee
Il est appelé gravitino, et a un spin $3/2$. La constante $\kappa$ a une dimension $-1$ ; elle est introduite pour que le lagrangien $\psi_m J^m$ compensatoire
\be
\mathscr L_N = -\sqrt{2}\kappa \rl \chib \sigmab^m \sigma^n \psib_n \dd_m z  + {\rm h.c.} \rr \nonumber
\ee
ait les bonnes dimensions.

Toutefois, nous n'avons pas encore rendu la supersymétrie locale. En effet, par supersymétrie, le gravitino doit avoir un superpartenaire qui n'apparaît pas ici. Ceci est confirmé par le calcul de la variation du champ $\chi$ dans le lagrangien de Noether :
\be
\delta_{\xi} \mathscr L_N = 2i \kappa \, \psib_m \sigmab^n \xi \, T^m_n + {\rm h.c.} \quad , \label{deltaLNoether}
\ee
où
\be
T_{mn} \ = \ \frac{1}{2}\rl \dd_m \zb \dd_n z + \dd_n \zb \dd_m z \rr - \frac{1}{2} \eta_{mn} \dd_p \zb \dd^p z \nonumber
\ee
est le tenseur énergie-impulsion du champ scalaire.

De nouveau, la variation (\ref{deltaLNoether}) peut être compensée en introduisant un tenseur $g_{mn}$ de transformation $\delta g_{mn} \sim i \kappa \xi \sigma_n \psib_m + {\rm h.c.}$ et en ajoutant le lagrangien
\be
\mathscr L_g \ = \ g_{mn}T^{mn} \quad . \nonumber
\ee

Ce nouveau champ est le graviton, de spin $2$, naturellement couplé au tenseur énergie-impulsion. Le graviton $g_{mn}$ et le gravitino $\psi_m^{\alpha}$ forment le multiplet gravitationnel (ou de supergravité) indispensable dans toute théorie de supersymétrie locale. Le graviton est le médiateur des interactions gravitationnelles tandis que son partenaire est le médiateur des interactions de supergravité. La constante $\kappa$ a maintenant une interprétation physique naturelle : il s'agit de la masse de Planck $\kappa^{-1} = M_P$.


\subsection{Le multiplet de supergravité\label{multipletsugra}}


Nous nous intéressons au modèle de supergravité le plus simple : celui du multiplet gravitationnel seul (sans matière ni interaction de jauge). L'action décrivant le graviton est l'action d'Einstein-Hilbert
\be
\mathscr L_{EH} \ = \ -\frac{1}{2\kappa^2} \sqrt{\det g} \, R \quad .
\label{LEH}
\ee

Pour exprimer l'action du gravitino, nous avons besoin d'introduire les vierbein $e_m^a$ et la connection de spin $\omega_m^{ab}$. Les indices $m, n$ sont des indices d'Einstein ``globaux'', tandis que les indices $a, b$ sont des indices de Lorentz traduisant la platitude locale. La connexion de spin est nécessaire dès qu'une théorie de relativité inclut des fermions. Les vierbein sont définis par
\be
g_{mn} (x) \ = \ e_m^a(x) e_n^b(x) \eta_{ab} \quad ,
\label{vierbein}
\ee
où $\eta$ est la métrique plate. On en déduit que $\sqrt{\det g} = \det e$. Les vierbein inverses sont définis par $g^{mn}=e_a^m e_b^n \eta^{ab}$.

Les matrices de Dirac $\gamma$ s'écrivent
\be
\gamma_m(x) = e_m^a(x) \gamma_a \quad . \nonumber
\ee
Alors l'algèbre de Clifford $\left\{\gamma_a, \gamma_b\right\}=-2\eta_{ab}$ devient $\left\{\gamma_m, \gamma_n\right\}=-2g_{mn}$.

En relativité générale, la connection de spin standard $\tilde \omega_m^{ab}$ peut toujours être exprimée en fonction des vierbein car le terme de torsion
\be
T_{mn}^a \ = \ \dd_m e_n^a - \dd_n e_m^a + \tilde \omega_{mn}^a - \tilde \omega_{nm}^a \nonumber
\ee
est nul. On trouve
\begin{eqnarray}
&&\tilde \omega_{mnl}(e) = e_{ma} e_{nb} \tilde \omega_l^{ab} \quad \quad , \quad \quad \text{avec} \nonumber \\
&&\tilde \omega_{mnl} = \frac{1}{2}\left[e_{na}\rl \dd_m e_l^a-\dd_le_m^a\rr - e_{la}\rl \dd_n e_m^a-\dd_m e_n^a\rr -e_{ma}\rl \dd_l e_n^a-\dd_n e_l^a\rr  \right] \ . \nonumber
\end{eqnarray}
En supergravité, la torsion n'est pas nulle, elle est contrainte par les identités de Bianchi, ce qui implique que la connection contient également des termes dépendant du gravitino :
\begin{eqnarray}
\omega_{mnl} &=& \tilde \omega_{mnl} + \frac{i}{4}\left[e_{na}\rl \psi_l \sigma^a\psib_m -\psi_m \sigma^a\psib_l \rr \right.  \label{spinconnection} \\
&-& \left. e_{la}\rl \psi_m \sigma^a\psib_n -\psi_n \sigma^a\psib_m \rr -e_{ma}\rl\psi_n \sigma^a\psib_l -\psi_l \sigma^a\psib_n \rr  \right] \ . \nonumber
\end{eqnarray}

En utilisant les relations (\ref{vierbein}) et (\ref{spinconnection}), on réexprime l'action (\ref{LEH}) uniquement en termes des vierbein :
\be
\mathscr L_{EH} \ = \ -\frac{1}{2\kappa^2} \det e \, \mathscr R \quad ,
\label{LEHvierbein}
\ee
où $\mathscr R$ est le scalaire de Ricci incorporant la connection
\be
\mathscr R \ = \ e_a^m e_b^n \rl \dd_m \omega_n^{ab} - \dd_n \omega_m^{ab} + \omega_n^{ac} \omega_{m\, c}^{b} - \omega_m^{ac} \omega_{n\, c}^{b} \rr \quad . \nonumber
\ee

L'action pour le gravitino est l'action de Rarita-Schwinger
\be
\mathscr L_{RS} \ = \ -\frac{1}{2} \det e \, \varepsilon^{mnpq} \rl \psi_m \sigma_n {\mathcal D}_p \psib_q - \psib_m \sigmab_n {\mathcal D}_p \psi_q \rr \quad .
\label{LRS}
\ee
Cette expression fait intervenir la dérivée covariante ${\mathcal D}_m  = \dd_m - \frac{1}{2} \omega_m^{ab}\sigma_{ab}$, avec $\sigma^{ab} = \frac{1}{4}\rl \sigma^a \sigmab^b - \sigma^b \sigmab^a \rr$.

Enfin, les transformations des champs sous la supersymétrie sont
\begin{eqnarray}
&&\delta e_m^a \ = \ i \kappa \rl \psi_m \sigma^a \xib - \xi \sigma^a \psib_m \rr \quad , \nonumber \\
&&\delta \psi_m^{\alpha} \ = \ -\frac{2}{\kappa} {\mathcal D}_m \xi^{\alpha} +\ldots \quad , \nonumber
\end{eqnarray}
où $\dots$ désigne des termes additionnels faisant intervenir des champs auxiliaires. En effet, le multiplet de supergravité hors couche de masse contient également un champ scalaire complexe $M$ et un champ vectoriel réel $b_a$, sans quoi les degrés de liberté ne sont pas équilibrés entre les bosons et les fermions \cite{Deser:1976eh}.


\subsection{Interactions de jauge en supergravité\label{intjaugesugra}}


Nous présentons maintenant le lagrangien de supergravité \cite{Freedman:1976xh, Cremmer:1982en} incluant des superchamps chiraux $\Phi^i = \rl z^i, \chi^i, F^i\rr$ et des superchamps vectoriels $V^{(a)} = \rl v_m^{(a)}, \lambda^{(a)}, D^{(a)}\rr$ ; l'indice $(a)$ de jauge ne doit pas être confondu avec l'indice $a$ de Lorentz. Nous décomposons le lagrangien comme suit
\be
\mathscr L_{sugra} = \mathscr L_G + \mathscr L_{cin} + \mathscr L_{int} - \det e \, \mathscr V_{sugra} \quad . \label{Lsugra}
\ee

$\mathscr L_G$ est le lagrangien de supergravité\footnote{Dans la suite, nous prenons $\kappa = 1$.} pure
\be
\mathscr L_G = -\frac{1}{2} \det e \, \mathscr R + \det e \, \varepsilon^{mnpq} \psib_m \sigmab_n \DD_p \psi_q \quad .
\label{LG}
\ee

$\mathscr L_{cin}$ contient les termes cinétiques covariants des champs de matière et du secteur de jauge
\begin{eqnarray}
\mathscr L_{cin} &=& - \det e \pl K_i^{\bar j}\DD_m z^i \DD^m \bar z_j + i K_i^{\bar j}\, \chib_j \sigmab^m \DD_m \chi^i \right. \nonumber \\
&&\quad \quad \quad \quad + \left. \frac{1}{4} F_{mn}^{(a)} F^{mn\, (a)} + i \lambdab^{(a)} \sigmab^m \DD_m \lambda^{(a)}   \pr \quad ,
\label{Lcinsugra}
\end{eqnarray}
où $g$ est le couplage de jauge et $K_i^{\bar j}$ est la métrique de K\"ahler introduite dans la section \ref{modeleschiraux}.

$\mathscr L_{int}$ contient tous les termes d'interaction
\begin{eqnarray}
\mathscr L_{int} &=&  \det e \pl  \sqrt{2} K_i^{\bar j} \overline X^{(a)}_{\bar j} \chi^i \lambda^{(a)} + \sqrt{2} K_i^{\bar j} X^{(a)\, i} \chib_j \lambdab^{(a)} \right. \nonumber \\
&-& \frac{1}{2} D^{(a)} \psi_m \sigma^m \lambdab^{(a)} + \frac{1}{2} D^{(a)} \psib_m \sigmab^m \lambda^{(a)} \nonumber \\
&-& \frac{1}{\sqrt{2}} K_i^{\bar j} \DD_n \zb_j \chi^i \sigma^m \sigmab^n \psi_m - \frac{1}{\sqrt{2}} K_i^{\bar j} \DD_n z^i \chib_j \sigmab^m \sigma^n \psib_m \nonumber \\
&+& \frac{i}{4} \left[ \psi_m \sigma^{ab} \sigma^m \lambdab^{(a)} + \psib_m \sigmab^{ab} \sigmab^m \lambda^{(a)} \right] F_{ab}^{(a)} \nonumber \\
&+& \frac{1}{4} K_i^{\bar j} \left[ i \varepsilon^{klmn}\psi_k \sigma_l \psib_m + \psi_m \sigma^n \psib^m \right] \chi^i \sigma_n \chib_j \label{Lintsugra} \\
&-& \frac{1}{8} \left[ K_i^{\bar j}K_k^{\bar l} - 2 R_{ik}^{\bar j \bar l} \right] \chi^i \chi^k \chib_j \chib_l + \frac{1}{8} K_i^{\bar j} \chib_j \sigmab^m \chi^i \lambdab^{(a)}\sigmab_m \lambda^{(a)} \nonumber \\
&-& \frac{3}{16} \lambda^{(a)} \sigma^m \lambdab^{(a)} \lambda^{(b)} \sigma_m \lambdab^{(b)} \nonumber \\
&-& e^{K/2} \left[ \Wb \psi_a \sigma^{ab} \psi_b + W \psib_a \sigmab^{ab} \psib_b  + \frac{i}{\sqrt{2}} D_i W \chi^i \sigma^a \psib_a \right. \nonumber \\
&+& \left.  \left. \frac{i}{\sqrt{2}} \Db^{\bar i} \Wb \chib_i \sigmab^a \psi_a + \frac{1}{2}  D_i  D_j W \chi^i \chi^j + \frac{1}{2} \Db^{\bar i} \Db^{\bar j} \Wb \chib_i \chib_j \right] \pr \nonumber
\end{eqnarray}

Enfin, le potentiel scalaire est
\be
\mathscr V_{sugra} \ = \ e^{K} \pl K_i^{\bar j} \, D^i W  \Db_{\bar j} \, \Wb  \ - \ 3 \Wb W \pr + \frac{1}{2} D^{(a)} D^{(a)} \quad ,
\label{Vsugra}
\ee
et nous avons ``sorti'' le déterminant de la métrique par convention. La dérivée $D_i$ est
\begin{eqnarray}
&&D_i W \ = \ W_i \ + \ K_i W \quad , \label{DiWsugra} \\
&&D_i D_j W = W_{ij} + K_{ij} W + K_i D_j W + K_j D_i W - K_i K_j W - \Gamma^i_{jk} D_k W \quad , \nonumber
\end{eqnarray}
où $W_{ij} = \dd^2 W / \dd z^i \dd z^j$. La connection $\Gamma$ a été définie dans la section \ref{modeleschiraux}, équation (\ref{connectionKahler}).

Les lagrangiens (\ref{LG}), (\ref{Lcinsugra}) et (\ref{Lintsugra}) contiennent les dérivées $\DD_m$ qui sont covariantes par rapport aux transformations de jauge, d'espace-temps et par rapport aux transformations de K\"ahler.

Pour voir ce dernier point, considérons le potentiel $\mathscr V_{sugra}$. Il est invariant sous les transformations de K\"ahler pourvu qu'elles soient généralisées et qu'elles s'accompagnent de transformations du superpotentiel
\begin{eqnarray}
&&K(\Phi, \Phi^{\dagger}) \rightarrow  K(\Phi, \Phi^{\dagger}) + F(\Phi) + F^{\dagger}(\Phi^{\dagger}) \quad , \nonumber \\
&& W(\Phi) \rightarrow e^{-F(\Phi)} W(\Phi) \quad , \label{transfoKahlerSugra}
\end{eqnarray}
qui transforment les dérivées (\ref{DiWsugra}) selon $D_i W \rightarrow e^{-F} D_i W$. Les termes cinétiques (\ref{Lcinsugra}) sont invariants si les fermions ont une transformation de super-Weyl
\be
\chi^i \rightarrow e^{\frac{i}{2} \text{Im} F} \chi^i \quad \quad , \quad \quad \psi_m \rightarrow e^{-\frac{i}{2} \text{Im} F} \psi_m \quad ,
\label{superWeylfermions}
\ee
ce qui induit un terme de dérivée covariante supplémentaire, comme énoncé ci-dessus. On peut montrer que les transformations (\ref{superWeylfermions}) sont automatiquement induites par les transformations de K\"ahler (\ref{transfoKahlerSugra}) accompagnées d'une transformation de Weyl de poids $F$ pour le multiplet de supergravité. Le lagrangien total $\mathscr L_{sugra}$ est invariant sous les transformations de K\"ahler-Weyl.

Dans le lagrangien, nous avons éliminé les champs auxiliaires
\begin{eqnarray}
F^i \ &=& \ - e^{K/2}\,  \rl K^{-1} \rr^i_{\bar j} \,  \Db^{\bar j} \Wb \quad , \nonumber \\
D^{(a)} \ &=& \ -g K_i  \, T^{(a)\, i}_{\ \ \ j} z^j \quad ,
\label{champsauxsugra}
\end{eqnarray}
où $K^{-1}$ est la métrique de K\"ahler inverse. Ces expressions sont similaires à celles trouvées dans la limite globale (\ref{auxGaugeTheory}).

Penchons-nous de manière plus détaillée sur le secteur de jauge. On peut montrer que la variété de K\"ahler possède un groupe d'isométries $G$ de dimension $d$ qui laisse invariante la métrique. Il est généré par les vecteurs de Killing holomorphes
\be
X^{(a)} \ = \ X^{(a)\, i} \frac{\dd}{\dd z^i} \quad \quad , \quad \quad \overline X^{(a)} \ = \ \overline X^{(a)}_j \frac{\dd}{\dd \zb_j} \quad , \nonumber
\ee
avec $(a)=1,\ldots, d$, qui vérifient $\left[X^{(a)}, X^{(b)}\right] = - f^{abc} X^{(c)}$, où $f^{abc}$ sont les constantes de structure de $G$. La dérivée de Lie définie à partir de ces vecteurs doit être nulle sur la métrique de K\"ahler. Par conséquent, il existe $d$ quantités réelles $D^{(a)}$, appelées potentiels de Killing, telles que
\be
K_{i}^{\bar j} \overline X^{(a)}_{\bar j} = i \ \frac{\dd D^{(a)}}{\dd z^i} \quad \quad , \quad \quad K_{i}^{\bar j} X^{i \, (a)} = -i \ \frac{\dd D^{(a)}}{\dd \zb_j} \quad , \nonumber
\ee
et qui, comme on le voit, sont définies à des constantes près
\be
D^{(a)} \ \longrightarrow \ D^{(a)} \ + \ c^{(a)} \quad .
\label{FIsugra}
\ee

Le potentiel de K\"ahler, contrairement à la métrique, n'est pas invariant sous les isométries
\be
\delta K = \rl \epsilon^{(a)} X^{(a)} + \bar \epsilon^{(a)} \overline X^{(a)} \rr K \quad . \nonumber
\ee
Si les paramètres $\epsilon^{(a)}$ sont réels, ceci équivaut à une transformation de K\"ahler (\ref{transfoKahler}). S'ils sont complexes, alors on peut réarranger l'expression précédente de façon à faire apparaître une transformation de K\"ahler
\be
\delta K = \epsilon^{(a)} F^{(a)} + \bar \epsilon^{(a)} \Fb^{(a)}  -  i \rl \epsilon^{(a)} - \bar \epsilon^{(a)}  \rr D^{(a)} \quad . \nonumber
\ee
avec $F^{(a)} = X^{(a)} K + i D^{(a)}$. Le dernier terme ressemble étrangement à une transformation de jauge.

L'idée principale est de tirer parti de cette transformation pour trouver le secteur de jauge d'une théorie de supergravité. On promeut les paramètres $\epsilon^{(a)}$ à des superchamps chiraux $\Lambda^{(a)}$. On montre alors que pour qu'une théorie chirale soit invariante sous les isométries de la variété de K\"ahler, on doit introduire un superchamp vectoriel $V = V^{(a)} T^{(a)}$, et $G$ n'est autre que le groupe de jauge. Ce processus n'est pas aussi simple que cela : il faut complexifier le groupe $G$ et vérifier que les termes cinétiques de supersymétrie $\Phi^{\dagger} e^V \Phi$ sont bien invariants. Pour l'essentiel, c'est ainsi que se construit le secteur de jauge en supergravité. L'indétermination (\ref{FIsugra}) des champs auxiliaires $D$ peut donner lieu à des termes de Fayet-Iliopoulos et donc briser spontanément la supersymétrie.

Dans le paragraphe, nous considérons les brisures spontanées de supersymétrie à la O'Raifeartaigh, c'est-à-dire que seuls les champs auxiliaires $F$ peuvent avoir des valeurs moyennes non-nulles.


\subsection{Le potentiel scalaire et la brisure de supersymétrie\label{potentielsugra}}


Le potentiel scalaire (\ref{Vsugra}) avec $D^{(a)}=0$  se réécrit
\begin{eqnarray}
\mathscr V_{sugra} &=& e^{K} \pl K_i^{\bar j} \, D_i W  \Db^{\bar j} \, \Wb  \ - \ 3 \Wb W \pr  \nonumber \\
&=& K_i^{\bar j} F^i \Fb_j  - 3 e^K \al W \ar^2 \quad .
\label{VsugrasansD}
\end{eqnarray}
Ce n'est plus une somme de termes positifs, donc :
\begin{center}
\textit{En supergravité, l'énergie peut être négative.}
\end{center}

Ce potentiel peut être réécrit en utilisant la fonction 
\be
G = K + \ln \al W \ar^2 \quad ,
\label{GKlnW}
\ee
ce qui donne
\be
\mathscr V_{sugra} \ = \ e^G \pl \ G_i^{\bar j}\  G^i \ \overline{G}_{\bar j}\ - \ 3 \ \pr \quad ,
\label{VsugraenfonctiondeG}
\ee
où $G_i^{\bar j}= \dd^2 G / \dd z^i \dd \zb_{\bar j}$ et $G_i = \dd G / \dd z^i$ sont reliés à $K_i^{\bar j}$, $K_i$, $K_{\bar j}$ et aux dérivées covariantes de $W$ par (\ref{GKlnW}). Dans une théorie chirale sans secteur de jauge, on peut montrer que le lagrangien s'exprime uniquement en terme du potentiel $G$.

La condition de brisure de la supersymétrie est 
\be
F_i \sim D_i W \neq 0 \quad . \nonumber
\ee
Dans ce cas, comme nous le voyons d'après l'équation (\ref{deltapsi}), le fermion $\chi^i$ a une transformation par un terme constant
\be
\delta_{\xi} \chi^i = \ldots - \rl \sqrt{2} e^{K/2} K^{i \, \bar j} \Db_{\bar j} \Wb \rr \xi \quad ,
\ee
ce qui signifie que $\chi^i$ est le Goldstino (à un facteur multiplicatif près). Les interactions (\ref{Lintsugra}) donnent lieu à une absorbtion de $\chi^i$ par le gravitino. À travers ce mécanisme\footnote{Il est aisé de voir à partir du lagrangien d'interaction (\ref{Lintsugra}) - deuxième et deux dernières lignes - et des champs auxiliaires (\ref{champsauxsugra}) que l'expression du Goldstino (\ref{goldstino}) reste valable en supergravité dans le cas général.}, appelé super-Higgs, le gravitino acquiert une masse
\be
m_{3/2}^2 \ = \ e^{K} \al W \ar^2  \ = \ e^G \quad .
\label{m3/2}
\ee
Dans ce cas le potentiel se réécrit simplement
\be
\mathscr V_{sugra} = K_i^{\bar j} F^i \Fb_j  - 3 m_{3/2}^2 M_P^2 \quad , \nonumber
\ee
où nous avons réintroduit la constante $\kappa$. Ceci implique que :
\begin{center}
\textit{En supergravité, il est possible de briser spontanément la supersymétrie tout en ayant une constante cosmologique nulle.}
\end{center}

En forçant la constante cosmologique $\langle \mathscr V \rangle = 0$, et en définissant l'échelle de brisure de la supersymétrie par
\be
M_{SUSY}^4 \ = \ \langle K_i^{\bar j} F^i \Fb_j \rangle \quad , \nonumber
\ee
on voit que la masse du gravitino paramétrise la brisure de supersymétrie
\be
m_{3/2}^2 \ = \ \frac{M_{SUSY}^4}{3 M_P^2} \quad . \nonumber
\ee


\subsubsection*{Les termes de brisure douce}


À travers les différents couplages du lagrangien d'interaction (\ref{Lintsugra}), on voit que la brisure spontanée de supersymétrie dans le cadre d'une théorie de supergravité donne lieu à des termes non-supersymétriques dans le lagrangien. On peut montrer par des théorèmes de non-renormalisation que ces termes n'introduisent pas de nouvelles divergences quadratiques. En ce sens, ils sont appelés ``termes de brisure douce'' \cite{Girardello:1981wz, Kaplunovsky:1993rd, Dudas:2005vv}. Ces termes sont :
\begin{itemize}

\item[$\bullet$] des masses pour les champs scalaires $M^2_{i \bar j} z^i \zb^{\bar j}$ et $\rl M^2_{ij} z^i z^j + {\rm h.c.}\rr$,

\item[$\bullet$] des masses pour les jauginos $\rl M_a \lambda^{(a)} \lambda^{(a)} + {\rm h.c.}\rr$,

\item[$\bullet$] des couplages trilinéaires $A_{ijk} z^i z^j z^k$.\\
\end{itemize}

La matrice de masse des scalaires se déduit du potentiel (\ref{VsugraenfonctiondeG})
\be
\begin{pmatrix} M^2_{i \bar j} & M^2_{i j} \cr M^2_{\bar i \bar j} & M^2_{\bar i j} \end{pmatrix} \quad . \nonumber
\ee

En imposant une constante cosmologique nulle, on montre que les termes de masse pour les scalaires sont\footnote{On considère toujours que la supersymétrie est brisée par des composantes $F$, donc $D^{(a)}=0$.}
\begin{eqnarray}
&&M^2_{i \bar j} \ = \ \langle \ e^G \rl G_{i \bar j}  - R_{i \bar j \alpha \bar \beta} G^{\alpha} G^{\bar \beta} \rr \  \rangle \quad , \nonumber \\
&&M^2_{i j} \ = \ \langle \ e^G \rl 2 \nabla_i G_j + G^{\alpha} \nabla_i \nabla_j G_{\alpha} \rr \  \rangle \quad .
\label{scalarsoftmass}
\end{eqnarray}
Dans ces expressions, nous avons indicés par des lettres latines $i, j$ les champs ne contribuant pas à la brisure de supersymétrie ($F^i=0$), et par des lettres grecques $\alpha, \beta$ ceux y contribuant ($F^{\alpha}\neq 0$).

Les couplages trilinéaires sont
\be
A_{ijk} \ = \ \langle \ e^G \rl 3 \nabla_i \nabla_j G_k + G^{\alpha} \nabla_i \nabla_j \nabla_k G_{\alpha} \rr \  \rangle \quad . \nonumber
\ee

Enfin, les masses des jauginos les plus générales sont
\be
M_a \ = \ \frac{1}{2} \rl \text{Re} f_a \rr^{-1} m_{3/2} f_{a\, \alpha} G^{\alpha} \quad ,
\label{gauginosoftmass}
\ee
où $f_a$ est la fonction cinétique de jauge (voir note de bas de page, paragraphe \ref{theoriejaugeSUSY}).

Notons enfin qu'après l'absorbtion du Goldstino par le gravitino, les masses des fermions peuvent être diagonalisées dans le lagrangien (\ref{Lintsugra}). On trouve que la matrice de masse est
\be
m_{i j} \ = \ \langle \ e^G \rl \nabla_i G_j + \frac{1}{3} G_i G_j \rr \ \rangle  \quad . \nonumber
\ee

La séparation de masse entre les fermions et les bosons a bien lieu. Dans les expressions (\ref{scalarsoftmass}), le premier terme $G_{i\, \bar j}$ et $2 \nabla_i G_j$ de chaque expression constitue la masse supersymétrique, présente dans le lagrangien, tandis que le second terme est la masse de brisure douce. Tout l'enjeu est donc de réaliser la brisure spontanée de supersymétrie de sorte que ces termes de brisure douce élèvent les masses des scalaires à des énergies non-atteintes à l'heure actuelle.


\chapter{Brisures dynamiques de symétries\label{BrisureDynaSym}}


\section{Brisure de symétrie chirale à la Nambu-Jona-Lasinio}



\subsection{Le modèle non-supersymétrique}


Un exemple simple de brisure dynamique de symétrie par formation d'états liés est le modèle de Nambu-Jona-Lasinio (NJL, \cite{Nambu:1961tp}). Il s'agit d'une théorie fermionique contenant une interaction de Fermi 
\be
\mathscr L_{NJL} \ = \ i \psib^a \gamma^{\mu} \dd_{\mu} \psi^a + \frac{G}{4} \ccl \rl \psib^a \psi^a \rr^2 - \rl \psib^a \gamma_5 \psi^a \rr^2 \ccr \quad ,
\label{LNJL}
\ee
où $a=1, \ldots, N$ est un indice de saveur interne et $G$ est une constante dimensionnée $\ccl G \ccr = -2$.

Ce modèle possède un groupe de symétries globales $SU(N)\times U(1)_V \times U(1)_A$ avec les transformations suivantes
\be
\psi^a \stackrel{U(1)_A}{\longrightarrow} e^{i\alpha\gamma_5} \psi^a \quad \quad , \quad \quad \psi^a \stackrel{U(1)_V}{\longrightarrow} e^{i\beta} \psi^a \quad ,
\label{transfoNJL}
\ee
qui se réécrivent
\be
\begin{array}{ccc}
\psi^a_L \stackrel{U(1)_A}{\longrightarrow} e^{-i\alpha} \psi^a_L \quad \quad &,& \quad \quad \psi^a_L \stackrel{U(1)_V}{\longrightarrow} e^{i\beta} \psi^a_L \quad , \\
\psi^a_R \stackrel{U(1)_A}{\longrightarrow} e^{i\alpha} \psi^a_R \quad \quad &,& \quad \quad \psi^a_R \stackrel{U(1)_V}{\longrightarrow} e^{i\beta} \psi^a_R \quad ,
\end{array}
\nonumber
\ee
en utilisant la notation $\psi^a_{R, L} = \frac{1 \pm \gamma_5}{2}\psi^a_{R, L}$. Le lagrangien (\ref{LNJL}) en fonction des composantes gauche et droite est
\be
\mathscr L_{NJL} \ = \ i \psib^a_R \sigma^{\mu} \dd_{\mu} \psi^a_R + i \psib^a_L \sigma^{\mu} \dd_{\mu} \psi^a_L + G \rl \psib^a_R \psi^a_L \rr \rl \psib^b_L \psi^b_R \rr \quad .
\label{LNJL2}
\ee

L'interaction à quatre fermions étant non-renormalisable, ce modèle admet une échelle de validité $\Lambda$, qui sera le cutoff des intégrales divergentes dans les boucles. La self-énergie $\Sigma$ des fermions est définie par la résolution de l'équation de Dirac
\be
\begin{array}{ccc}
i\gamma^{\mu} p_{\mu} +  \Sigma(p, m, G, \Lambda) \ = \ 0 \quad ,  \\
\left. \Sigma(p, m, G, \Lambda)\ar_{i\gamma^{\mu} p_{\mu} + m \ = \ 0} \ = \ m \quad ,
\end{array}
\label{selfenergie}
\ee
puisqu'il n'y a pas de terme de masse dans le lagrangien (\ref{LNJL2}).

\begin{figure}[ht!]
\begin{center}
\includegraphics[scale=0.5]{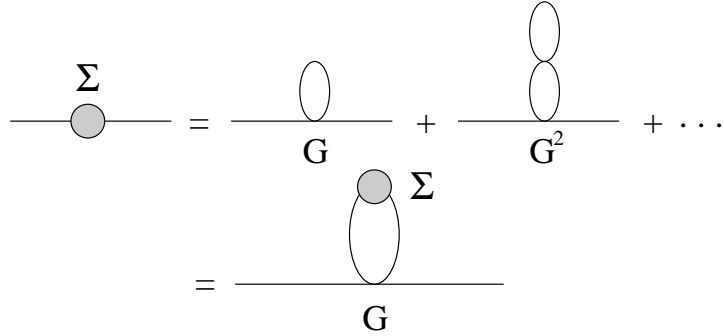}
\caption{Resommation des graphes donnant la self-énergie à l'ordre dominant.}
\label{Selfenergy}
\end{center}
\end{figure}

Les équations (\ref{selfenergie}) définissent la masse $m$, et la détermination de la self-énergie passe par le calcul des diagrammes présentés sur la Fig. \ref{Selfenergy}. À l'ordre dominant, ces diagrammes peuvent être resommés et on obtient $m$ par une équation implicite
\be
m \ = \ \Sigma \ = \ - \frac{2 N G}{\rl 2 \pi \rr^4} \ i \int d^4p \frac{m}{p^2 - m^2} \quad . \label{eqngapNJL}
\ee
Cette équation\footnote{L'équation (\ref{eqngapNJL}) est souvent appelée équation de gap, par analogie avec la supraconductivité.} admet la solution $m=0$, ainsi qu'une seconde solution non-triviale qui s'exprime en fonction du couplage $G$ et du cutoff $\Lambda$ permettant de régulariser l'intégrale. Après calcul explicite, on trouve
\be
\frac{1}{G} \ = \ \frac{N \Lambda^2}{8 \pi^2} \pl 1 - \frac{m^2}{\Lambda^2} \ln \frac{\Lambda^2}{m^2} \pr \quad .
\label{eqnmasseNJL}
\ee
Puisque l'on veut $m^2 < \Lambda^2$, l'équation (\ref{eqnmasseNJL}) n'a de solution que si
\be
G \geqslant G_c \ \equiv \ \frac{8 \pi^2}{N \Lambda^2} \quad ,
\label{gcritiqueNJL}
\ee
où $G_c$ est le couplage critique.

La solution $m = - G \langle \psib^b \psi^b \rangle \neq 0$ donnée par l'équation (\ref{eqnmasseNJL}) brise spontanément la symétrie chirale $U(1)_A$ car un terme de masse $m\psib^a \psi^a$ n'est pas invariant sous (\ref{transfoNJL}). Le spectre contient alors une particule pseudo-scalaire de masse nulle, qui correspond au boson de Goldstone accompagnant la brisure de symétrie chirale, et un scalaire de masse $2m$. Cette dernière correspond à l'apparition, en dessous de l'échelle $\Lambda$, d'un état lié $\langle \psib^a_L \psi^a_R \rangle$, comme le montre le lagrangien (\ref{LNJL2}).

On peut décrire de manière effective le lagrangien NJL en introduisant un champ scalaire $\phi$ sans dynamique de la façon suivante
\be
\mathscr L_{eff} \ = \ i \psib^a_R \sigma^{\mu} \dd_{\mu} \psi^a_R + i \psib^a_L \sigma^{\mu} \dd_{\mu} \psi^a_L + g_0 \rl \phi \psib^a_R \psi^a_L  + {\rm h.c.} \rr  - m_0^2 \al \phi \ar^2 \quad .
\label{LNJLeff}
\ee
L'équation du mouvement pour $\phi$ permet facilement de retrouver le lagrangien (\ref{LNJL2}) avec
\be
\phi \ = \ \frac{g_0}{m_0^2} \psib^a_L \psi^a_R \quad \quad , \quad \ \text{et} \quad \quad G \ = \ \frac{g_0^2}{4 m_0^2} \quad .
\label{solutionsEqPhiNJL}
\ee

Comme nous l'avons évoqué au début du chapitre \ref{SUSY-SUGRA}, il est très difficile de protéger un champ scalaire par des symétries. À travers les couplages de Yukawa du lagrangien (\ref{LNJLeff}), le champ $\phi$ acquiert donc un terme cinétique de normalisation $Z_{\phi}$, une masse $m^2_{\phi}$ et un couplage d'auto-interaction quartique $\lambda_{\phi}$ donnés par
\begin{eqnarray}
&&Z_{\phi}(Q) \ = \ \frac{Ng_0^2}{\rl 4\pi \rr^2} \ln \frac{\Lambda^2}{Q^2} \quad , \nonumber \\
&&m^2_{\phi}(Q) \ = \ m_0^2 - \frac{2Ng_0^2}{\rl 4\pi \rr^2} \rl \Lambda^2-Q^2\rr \quad , \label{ZmgPhiNJL} \\
&&\lambda_{\phi}(Q) \ = \ \frac{2Ng_0^4}{\rl 4\pi \rr^2} \ln \frac{\Lambda^2}{Q^2} \quad , \nonumber 
\end{eqnarray}
à une certaine échelle $Q$. Le lagrangien généré à basse énergie est
\begin{eqnarray}
\mathscr L_{eff} \ &=& \ i \psib^a_R \sigma^{\mu} \dd_{\mu} \psi^a_R + i \psib^a_L \sigma^{\mu} \dd_{\mu} \psi^a_L + g_0 \rl \phi \psib^a_R \psi^a_L  + {\rm h.c.} \rr \nonumber \\
&+& Z_{\phi} \al \dd_{\mu} \phi \ar^2 - m^2_{\phi} \al \phi \ar^2 - \frac{\lambda_{\phi}}{2} \al \phi^{\dagger}\phi \ar^2 \quad . \label{LNJLeffphi}
\end{eqnarray}

D'après (\ref{ZmgPhiNJL}), lorsque $Q \rightarrow \Lambda$, les quantités $Z_{\phi}$ et $\lambda_{\phi}$ tendent vers zéro. Ceci montre que $\phi$ n'est pas un champ fondamental de la théorie mais qu'il s'agit d'un état composite. À haute énergie, c'est-à-dire au dessus de l'échelle $\Lambda$, l'état lié $\psib^a_L \psi^a_R$ disparaît et on doit retrouver le lagrangien (\ref{LNJL2}) en remplaçant les paramètres selon (\ref{solutionsEqPhiNJL}).

On redéfinit le champ $\phi \rightarrow \phi \cdot Z_{\phi}^{-1/2}$ de sorte que le lagrangien (\ref{LNJLeffphi}) est canoniquement normalisé si on redéfinit également les couplages
\be
g = \frac{g_0}{Z_{\phi}^{1/2}} \quad \quad , \quad \quad \lambda^{\prime} = \frac{\lambda_{\phi}}{Z_{\phi}^2} \quad . \label{rescaleglambda}
\ee
On trouve alors que $\lambda^{\prime} \rightarrow \infty$ lorsque $Q \rightarrow \Lambda$.

À la fin des années $1980$, l'idée d'utiliser ces modèles pour expliquer la brisure électrofaible et de considérer le champ de Higgs comme un état composite top-antitop connut un certain enthousiasme \cite{Nambu:1988mr}. En jaugeant la symétrie chirale étudiée ci-dessus, et en écrivant une interaction à quatre fermions uniquement pour le quark top,
\be
\mathscr L \ = \ \mathscr L_{\text{Modèle Standard}} \ + \ G\rl \psib^{i\,a}_L t_R^a \rr \rl \psi^{i\,b}_L {\bar t}_R^b \rr \quad , \nonumber
\ee
on engendre dynamiquement la brisure $SU(2)\times U(1) \rightarrow U(1)$ via la condensation $\langle \bar t t \rangle$. De la même manière que dans le Modèle Standard, les champs de Goldstone liés à cette brisure sont absorbés par les bosons de jauge et ceux-ci acquièrent une masse. L'interaction à quatre fermions est alors comprise comme la reminiscence à basse énergie de l'échange d'un boson massif $W^{\pm}$ ou $Z$ qui a découplé de la théorie.

L'intérêt de ce modèle est évident : il permet d'éviter l'introduction d'une nouvelle particule dans le Modèle Standard et explique naturellement la brisure de la symétrie électrofaible. Malheureusement, ces modèles sont maintenant exclus. En effet, la masse du quark top dépend de l'échelle de structure $\Lambda$. Les calculs à une boucle (voir Bardeen, Hill, Lindner \cite{Nambu:1988mr}) montrent que pour des valeurs de $\Lambda$ allant de $10^4$ à $10^{19}$ GeV, la masse du quark top ne descend pas en dessous de $220$ GeV, alors que les dernières mesures donnent $m_t = 170$ GeV. De plus, les paramètres (\ref{ZmgPhiNJL}) ne sont pas renormalisables. De ce fait, on ne résout pas le problème de hiérarchie dans le Modèle Standard. Pour palier ce dernier point, l'idée la plus naturelle est de rendre le modèle de Nambu-Jona-Lasinio supersymétrique.


\subsection{Supersymétrisation des modèles NJL}


On inclut les fermions $\psi^a$ dans des superchamps chiraux $\Phi^a$ (voir section \ref{susy}). L'extension minimale du modèle NJL est \cite{Buchmuller:1982ty}
\be
\mathscr L_{SNJL} \ = \ \int d^4 \theta \rl \Phib^a \Phi^a \ + \ G \Phi^a \Phi^a \Phib^b \Phib^b \rr \quad .
\label{LSNJL}
\ee
Les transformations des superchamps sous la symétrie chirale $U(1)_A$ sont
\be
\Phi^a \stackrel{U(1)_A}{\longrightarrow} e^{i\alpha} \Phi^a \quad \quad , \quad \quad \Phib^a \stackrel{U(1)_A}{\longrightarrow} e^{-i\alpha} \Phib^a \quad . \nonumber
\ee
De plus, le lagrangien ci-dessus contient une $R$-symétrie continue puisque, comme nous l'avons dans la section \ref{Rsym}, le potentiel de K\"ahler est toujours invariant.

Si l'on suit la même procédure que dans le cas non-supersymétrique, on définit la masse du multiplet $\Phi^a$ et on calcule les corrections de la fonction à deux points. L'équation de gap trouvée dans le cas de deux champs \cite{Buchmuller:1982ty} est
\be
1 \ = \ G^2  \int \frac{d^4p}{\rl 2 \pi \rr^4}\frac{d^4q}{\rl 2 \pi \rr^4} \frac{\al m \ar^2}{\rl p^2 + \al m \ar^2 \rr \rl q^2 + \al m \ar^2 \rr \rl \rl m-p-q\rr^2 + \al m \ar^2 \rr} \quad . \nonumber
\ee
En calculant explicitement l'intégrale avec le cutoff $\Lambda$, on s'aperçoit immédiatement que l'équation ci-dessus n'admet qu'une seule solution $m=0$ car l'intégrale est négative. La supersymétrie protège donc la symétrie chirale et empêche la formation d'un condensat.

On lève cette difficulté en introduisant un terme de brisure douce \cite{Buchmuller:1984yr}. Le lagrangien (\ref{LSNJL}) est modifié en
\be
\mathscr L_{SNJL} \ = \ \int d^4 \theta \ccl \Phib^a \Phi^a \rl 1 - \Sigma^2 \theta^2 \thetab^2 \rr \ + \ G \Phi^a \Phi^a \Phib^b \Phib^b \ccr \quad .
\label{LSNJLsoft}
\ee
Avec cette modification, la symétrie chirale est dynamiquement brisée par le condensat $\langle \psi^a \psi^a \rangle$, et on retrouve les mêmes propriétés que dans le modèle NJL originel. En particulier, l'équation de gap conduit à la condition $G \geqslant G_c$  avec le couplage critique
\be
\frac{1}{G_c} \ = \ \frac{N\Sigma^2}{4 \pi^2} \, \ln \frac{\Lambda^2}{\Sigma^2} \quad .
\label{gcritiqueSNJL}
\ee
Cette expression traduit le fait que la divergence quadratique à l'origine du couplage critique (\ref{gcritiqueNJL}) a été réduite en divergence logarithmique par supersymétrie.

En outre, on peut montrer que la version supersymétrique du lagrangien à basse énergie (\ref{LNJLeff}) met en jeu deux superchamps chiraux $H_1$ et $H_2$. Le lagrangien correspondant est
\begin{eqnarray}
\mathscr L_{eff} \ &=& \ \int d^4 \theta \ccl \Phib^a \Phi^a \rl 1 - \Sigma^2 \theta^2 \thetab^2 \rr \ + \ H^{\dagger}_1 H_1 \ccr \nonumber \\
&+& \pl \ \int d^2 \theta H_2 \rl m H_1 - g \Phi^a \Phi^a \rr + {\rm h.c.}\pr \quad .
\label{LSNJLeff}
\end{eqnarray}

L'équation du mouvement pour $H_2$ implique que $H_1 = \frac{g}{m} \Phi^a \Phi^a$, qui est la généralisation de (\ref{solutionsEqPhiNJL}). En injectant ce résultat dans (\ref{LSNJLeff}), on retrouve le lagrangien (\ref{LSNJLsoft}) avec $G = \frac{g^2}{m^2}$. C'est donc $H_1$ qui joue le rôle du champ composite, tandis que $H_2$ est juste un multiplicateur de Lagrange. L'équation du mouvement pour $H_1$ implique $H_2 = \frac{1}{4m}\Db^2 H_1^{\dagger} = \frac{g}{4m^2} \Db^2 \rl \Phib^a \Phib^a \rr$.

L'inconvénient majeur du modèle NJL supersymétrique avec brisure douce est que pour une échelle $\Sigma \ll \Lambda$, le couplage minimum requis pour former un condensat est extrêmement fort : $G_c \gg 1/\Lambda^2$, ce qui constitue un problème de naturalité assez sévère. C'est la raison pour laquelle ces idées furent abandonnées pour expliquer l'origine de la brisure électrofaible dans le MSSM.


\subsection{Dualité avec des modèles à dimensions supplémentaires\label{model6DNJL}}


Nous allons montrer que le modèle de Nambu-Jona-Lasinio (\ref{LNJL2}) et sa version supersymétrique (\ref{LSNJLsoft}) sont duaux de théories avec dimensions supplémentaires (Publication ${\mathcal N}^{\mathrm{o}}\, 2$).

Pour ce faire, nous reprenons le modèle à six dimensions développé dans la section \ref{6Dmodel} et dans la Publication ${\mathcal N}^{\mathrm{o}}\, 1$
\begin{eqnarray}
&& \mathcal S \ = \ \int d^4 x d^2 y \ \ccl \frac{1}{2} (\partial_M \phi)^2 \ - \  V_\delta (\phi) \ccr \quad , \nonumber \\
&& V_\delta (\phi) \ = \ \rl -\frac{\mu^2}{2}\phi^2 + \frac{\lambda}{4} \phi^4 \rr \cdot \delta^2 (y)  \quad ,
\label{model6d2}
\end{eqnarray}
que nous compactifions sur l'orbifold $T^2/\mathbb Z_2$ (voir paragraphe \ref{Orbifolds}). Le choix de conditions aux bords de type Scherk-Schwarz
\be
\phi \rl y_1 + 2\pi R_1, y_2 \rr \ = \ - \phi \rl y_1, y_2 \rr \quad , \nonumber
\ee
est analogue aux conditions de Dirichlet (\ref{Dirichlet}) prises sur le disque\footnote{En effet, en prenant $y_1 = - \pi R_1$, on trouve simplement que $\phi (\pi R_1, y_2) = 0$.}.
Par ailleurs, nous choisissons la parité $+1$ sous l'action de l'orbifold pour assurer la présence d'un mode zéro. Le champ se décompose comme suit
\be
\phi (x^{\mu}, \overrightarrow{y}) = \sum_{k_1, k_2 \in \mathbb Z} \frac{1}{\sqrt{4 \pi^2 R_1 R_2}} \cos \ccl \frac{\rl k_1+ 1/2\rr y_1}{R_1} + \frac{k_2 y_2}{R_2} \ccr \phi_{k_1, k_2}(x^{\mu}) \ . \label{decompphiOrbifold}
\ee
En injectant (\ref{decompphiOrbifold}) dans l'action (\ref{model6d2}) et en diagonalisant la matrice de masse $\mathcal M$ sur les modes propres $\Phi_m$, on trouve une équation aux valeurs propres $\mathcal M \Phi_m = m \Phi_m$ avec
\be
\frac{4\pi^2 R_1 R_2}{\mu^2} \ = \ \sum_{k_1, k_2 \in \mathbb Z} \ccl \rl \frac{k_1 + 1/2}{R_1}\rr^2 + \rl \frac{k_2}{R_2}\rr^2-m^2 \ccr^{-1} \quad . \nonumber
\ee
Le couplage critique correspond à un mode de masse nulle
\be
\frac{4\pi^2 R_1 R_2}{\mu_c^2} \ = \ \sum_{k_1, k_2 \in \mathbb Z} \ccl \rl \frac{k_1 + 1/2}{R_1}\rr^2 + \rl \frac{k_2}{R_2}\rr^2 \ccr^{-1} \quad . \nonumber
\ee
Au delà de cette valeur critique, $\mu^2 > \mu_c^2$, la masse $m^2$ devient négative et on observe la transition de phase développée dans la Publication ${\mathcal N}^{\mathrm{o}}\, 1$.\\

Le modèle (\ref{model6d2}) peut être généré en couplant le champ scalaire à un ensemble de $N$ fermions $\psi^a$ localisés à l'origine des dimensions supplémentaires par une interaction $\rl g \psi^a \psi^a \phi(\overrightarrow{y}=\overrightarrow{0}) + {\rm h.c.} \rr$. Comme dans le modèle effectif (\ref{LNJLeff}), le champ $\phi$ acquiert un terme cinétique, une masse et un couplage quartique par correction à une boucle de sa fonction à deux points. On trouve un paramètre de masse
\be
\mu^2 \ = \ \frac{N g}{4 \pi^2} \Lambda^2 \quad \quad , \label{muparcouplagefermions}
\ee
où nous avons supposé que $R_1 = R_2 \equiv R$. Pour $\mu^2 > \mu_c^2 = 4 \pi / \ln \rl \Lambda^2 R^2 \rr$, le champ développe une valeur moyenne $\langle \phi \rangle \neq 0$ qui brise spontanément la symétrie chirale des fermions $\psi^a$.

La version duale de ce modèle se trouve en découplant le champ scalaire $\phi$ de la théorie au niveau des arbres. Ceci produit un lagrangien localisé à quatre dimensions de la forme (\ref{LNJL2}) avec
\be
G \ = \ \frac{g^2}{2 \pi^2 R_1 R_2} \sum_{k_1, k_2 \in \mathbb Z} \ccl \rl \frac{k_1 + 1/2}{R_1}\rr^2 + \rl \frac{k_2}{R_2}\rr^2 \ccr^{-1} \simeq \frac{g^2}{4 \pi} \ln \rl \Lambda^2 R^2 \rr \quad . \nonumber
\ee
La brisure dynamique de symétrie chirale à la NJL a lieu pour $G > G_c$ avec $G_c^{-1} =  N \Lambda^2 / 4 \pi^2 $. On trouve donc la même condition que dans la description à six dimensions $\mu^2 > \mu_c^2$ en remplaçant $\mu^2$ et $\mu_c^2$ par leur valeur.

Si l'on reprend l'évolution classique des couplages comme nous l'avions fait dans la Publication ${\mathcal N}^{\mathrm{o}}\, 1$, on montre clairement que le champ $\phi$ ou plutôt son mode zéro est le champ composite introduit dans le lagrangien (\ref{LNJLeffphi}). Les deux modèles sont donc duaux : ils décrivent la même théorie. Rappelons que notre échelle $\Lambda$ est directement reliée à la régularisation de la singularité en $\overrightarrow{y} = \overrightarrow{0}$, et non pas à l'échelle de structure à laquelle le scalaire $\phi$ cesse d'être composite : dans ce modèle, $Z_{\phi}(Q)\rightarrow 0$ lorsque $Q \rightarrow R^{-1}$ et non pas $\Lambda$.

Dans la Publication ${\mathcal N}^{\mathrm{o}}\, 2$, nous avons également étudié la version supersymétrique du modèle, et montré que la dualité est présente là aussi. Malheureusement, le problème de naturalité pour $G_c \gg R^{2}$ évoqué à la fin du paragraphe précédent n'est pas résolu.

Le modèle (\ref{model6d2}) avec $\mu$ donné par (\ref{muparcouplagefermions}) donne lieu à la brisure de symétrie chirale à l'ordre des arbres. Le modèle NJL, quant à lui, reproduit cette brisure en faisant appel à des phénomènes non-perturbatifs. On peut donc tirer avantage de la dualité puisque dans la version dimensions supplémentaires de la théorie, les calculs sont grandement simplifiés. Nous pensons que ce type de dualité entre des phénomènes non-perturbatifs à quatre dimensions et des modèles avec dimensions supplémentaires est générique.


\section{QCD Supersymétrique\label{SUSYQCD}}


Dans cette section, nous donnons une seconde illustration de la manière dont des phénomènes nonperturbatifs, ici la condensation de jauginos, peuvent engendrer un potentiel effectif à basse énergie. Nous verrons notamment comment cela peut induire la brisure dynamique d'une symétrie.

Considérons une théorie de Yang-Mills supersymétrique dont le groupe de jauge est $SU(N_c)$, où $N_c$ représente le nombre de couleurs. Le multiplet chiral de jauge
\be
W^{a} = \left( \lambda^{a}, A_{\mu}^{a}\right) \quad , \nonumber
\ee
avec $a=1,\ldots,N_c^2-1$, décrit les gluons $A_{\mu}^{a}$ et leur partenaire, les gluinos $\lambda^{a}$, que nous appellerons aussi jauginos de manière plus générique.

La matière est représentée par $N_f$ saveurs de quarks $\psi$ et d'antiquarks $\widetilde \psi$ dans les représentations $N_c$ et $\overline{N_c}$ de $SU(N_c)$, associés à leur partenaire scalaire $z,\ \widetilde z$ dans les multiplets chiraux
\begin{eqnarray}
 Q^i_{\alpha} = (z^i_{\alpha},\psi^i_{\alpha}) \quad , \nonumber \\
{\widetilde Q}^{\alpha}_i = ({\widetilde z}^{\alpha}_i,{\widetilde \psi}^{\alpha}_i) \quad , \nonumber 
\end{eqnarray}
 avec $i=1,\ldots,N_f$ et $\alpha=1,\ldots,N_c$.

L'évolution de la constante de couplage $g$ avec l'énergie est donnée par sa fonction bêta à une boucle
\be
\beta(g) \ = \ \mu \frac{d g}{d \mu} \ = \ -\frac{g^3}{16 \pi^2} (3N_c-N_f) \quad .
\label{betaelectrik}
\ee
Cette théorie est donc asymptotiquement libre si $N_f < 3N_c$, et libre dans l'infra-rouge pour $N_f > 3N_c$. Nous nous plaçons dans le premier cas ; de telles théories sont appelées SUSY-QCD.

Comme dans le cas de la QCD du Modèle Standard, il existe une échelle dynamique, $\Lambda$, qui est un invariant du groupe de renormalisation, donnée par
\be
\Lambda \ = \ \mu \exp \rl -\frac{8 \pi^2}{(3N_c - N_f) g^2(\mu)} \rr \quad .
\label{Lambda}
\ee
avec, donc, $d \Lambda / d\mu = 0$. Cette échelle représente l'énergie à partir de laquelle la théorie des perturbations n'est plus une bonne approximation, c'est-à-dire $g(\Lambda) \sim 1$.


\subsection{L'indice de Witten\label{indice}}


En 1982, E. Witten \cite{Witten:1982df} montra qu'une telle théorie possède $N_c$ vides supersymétriques non-dégénérés. Pour ce faire, il calcula la quantité
\be
\tr (-1)^F \ = \ \text{nombre de bosons} \ - \ \text{nombre de fermions} \ = \ n_B - n_F
\label{indiceWitten}
\ee
pour une théorie de SUSY-QCD de groupe de jauge $SU(N_c)$. Comme nous le verrons, ce résultat est indépendant de la matière introduite, et donc de $N_f$.

La donnée de $\tr (-1)^F$, appelé ``indice de Witten'', permet d'appréhender si la supersymétrie est brisée ou non dans une théorie donnée. Un point clé de cette détermination est de se placer à volume fini. Dans ce cas, le spectre est discret et les états propres du système peuvent être comptés. Un état bosonique $\vert b \rangle$ vérifie $(-1)^F \vert b \rangle = + \vert b \rangle$ tandis qu'un état fermionique $\vert f \rangle$ vérifie $(-1)^F \vert f \rangle = - \vert f \rangle$. à partir de l'ensemble du spectre d'une théorie, nous retrouvons l'expression (\ref{indiceWitten}).

Seuls les états d'énergie nulle contribuent à l'indice de Witten. En effet, une supercharge $Q$ et son conjugué $\Qb$ vérifient l'algèbre
\begin{eqnarray}
&&Q^2 \ = \ \Qb^2 \ = \ 0 \quad , \nonumber \\
&& Q\Qb \ + \ \Qb Q \ = \ H \quad ,
\label{QQH}
\end{eqnarray}
où $H$ est l'hamiltonien du système. L'indice de Witten anticommute avec les supercharges : $(-1)^F Q = - Q (-1)^F$.

Considérons un état propre bosonique de l'hamiltonien $H \vert  b \rangle = E  \vert  b \rangle$. L'action de l'opérateur $Q$ sur cet état est
\be
Q \vert  b \rangle = \sqrt{E} \vert  b' \rangle \quad \text{et} \quad \Qb \vert  b \rangle = 0 \quad .
\label{Qib}
\ee
En appliquant $Q$ sur la deuxième équation, nous obtenons
\be
0 = Q \Qb \vert  b \rangle = ( H -  \Qb Q ) \vert  b \rangle = E  \vert  b \rangle  - \sqrt{E} \Qb \vert  b' \rangle \ .
\nonumber
\ee
Donc l'action de $Q$ et $\Qb$ sur l'état $\vert  b' \rangle$ est
\be
Q \vert  b' \rangle = 0 \quad \text{et} \quad \Qb \vert  b' \rangle = \sqrt{E} \vert  b \rangle  \quad , \nonumber
\ee
si et seulement si $E \neq 0$. Il s'ensuit que $\vert  b' \rangle$ est d'énergie $E$ et que les deux états sont dégénérés. Si on applique $(-1)^F$ à la première égalité de (\ref{Qib}), on trouve que $\vert  b' \rangle$ est un état fermionique.

Nous venons de montrer que les états d'énergie non-nulle sont associés en paires fermion-boson par la supersymétrie, et ils ne contribuent donc pas à l'indice de Witten. Nous appellerons $n_B$ et $n_F$ le nombre de bosons et de fermions dans l'état d'énergie nulle.

Clairement, si $\tr (-1)^F \neq 0$, alors le niveau d'énergie nulle est occupé et la supersymétrie n'est pas brisée. En revanche, si $\tr (-1)^F = 0$, deux cas de figure se présentent : si $n_B = n_F \neq 0$, nous sommes dans un cas particulier de la configuration précédente, et la supersymétrie n'est pas brisée ; si $n_B = n_F = 0$, la supersymétrie est brisée puisque l'état d'énergie nulle n'est pas occupé. Rien ne permet de différencier ces deux cas, et il est donc plus intéressant de contraindre les mécanismes de brisure de la supersymétrie en étudiant les cas où $\tr (-1)^F \neq 0$. En effet, si l'état d'énergie nulle n'est pas occupé, cela signe une brisure \textit{spontanée} de la supersymétrie. En outre, la donnée de $\tr (-1)^F = 0$ ne nous apprend rien sur une brisure dynamique de la supersymétrie.

Quelle est la limite de volume infini une fois que l'on connaît l'indice de Witten d'une théorie ? Si l'on a trouvé que $\tr (-1)^F = 0$, la supersymétrie est peut-être brisée à volume fini, mais il se peut tout à fait qu'elle soit restaurée dans la limite $V \rightarrow \infty$. En revanche, si $\tr (-1)^F \neq 0$ pour tout volume fini, alors la supersymétrie n'est pas brisée dans la limite de volume infini.

Une question que l'on peut se poser est : comment évolue l'indice de Witten lorsque l'on varie continûment les paramètres de la théorie ? Principalement deux cas peuvent se produire :
\begin{enumerate}
\item une particule d'énergie non-nulle est déplacée vers l'état fondamental. Dans ce cas, $n_B$ et $n_F$ sont augmentés d'une unité puisque l'état originel étant d'énergie non-nulle, par supersymétrie, le superpartenaire de la particule descend aussi vers l'état fondamental.

\item une particule de l'état fondamental acquiert une énergie non-nulle. De même, par supersymétrie, un superpartenaire doit lui être apparié. Donc $n_B$ et $n_F$ perdent chacun une unité.
\end{enumerate}

Dans les deux cas, l'indice de Witten est invariant. En revanche, il se peut que de nouveaux états soient créés lorsqu'on varie les paramètres. Le modèle le plus simple dans lequel ceci se produise est celui d'un champ scalaire $\phi$ dont le potentiel
\be
V(\phi) \ = \ \rl m\phi - g^2 \phi^2 \rr^2 \nonumber
\ee
admet deux minima d'énergie nulle en $\phi = 0$ et $\phi = m/g$ si $g \neq 0$, tandis qu'il n'en admet qu'un seul ($\phi = 0$) pour $g = 0$. L'indice de Witten est donc discontinu. En réalité, lorsque $g$ varie de zéro à une valeur non-nulle, le comportement asymptotique du potentiel passe d'une forme $\sim \phi^2$ à une forme $\sim \phi^4$, et le second minimum $m/g$ est ramené continûment de l'infini. La condition pour laquelle $\tr (-1)^F$ est invariant sous un changement des paramètres est que l'on n'introduise pas dans la théorie de nouveaux termes susceptibles de modifier le comportement asymptotique de celle-ci. Autrement dit, on ne doit pas rajouter de termes plus grands que les termes déjà présents dans la théorie.

Considérons enfin le problème des particules de masse nulle. Si les particules sont non-massives par ``accident'' car aucune symétrie n'en est responsable, alors nous pouvons calculer l'indice de Witten en supposant $m \neq 0$ puis prendre la limite $m \rightarrow 0$ à la fin du calcul. Si, par contre, les particules sont protégées par une symétrie qui les empêche d'acquérir une masse, alors le problème est plus délicat à traiter. C'est cette dernière possibilité que nous allons considérer pour SUSY-QCD puisque c'est ce qui se produit dans les théories de jauge.

La difficulté ici est que nous travaillons à volume fini. Il faut donc imposer des conditions aux bords pour les champs. L'algèbre de supersymétrie incluant l'invariance par translation, nous devons imposer des conditions périodiques, et qui plus est, elles doivent être imposées aux bosons et aux fermions de la même façon, y compris pour le champ de jauge. D'autre part, le champ de jauge reste non-massif tant que la symétrie de jauge n'est pas brisée. Sa composante de moment nul contribue donc à $\tr (-1)^F$. Dans la limite de volume infini, on fixe en général la jauge pour éliminer la composante $A_0$. En volume fini, la difficulté réside dans le fait de respecter les conditions de périodicité. On peut montrer que pour une symétrie de jauge abélienne, il est encore possible de fixer la jauge de sorte à éliminer $A_0$. Dans ce cas, seuls les jauginos participent à l'indice de Witten.

En présence d'un groupe de jauge non-abélien, il y a de nombreuses façons de satisfaire les conditions aux bords tout en fixant la jauge pour éliminer la composante de moment nul. Néanmoins il reste toujours une composante de spin zéro dans l'état fondamental, donc $\tr (-1)^F = 1$ pour chaque choix de condition aux bords. Techniquement, le champ de jauge d'énergie nulle peut toujours être écrit $A_{\mu} = -i (\dd_{\mu} U) U^{-1}$ pour une certaine matrice $U$. Les conditions de périodicité sur $A_{\mu}$ se transcrivent en conditions sur $U$ de manière non-bijective. On montre que, pour une théorie $SU(N_c)$, il y a $N_c$ choix différents de conditions sur $U$ (qui reviennent essentiellement à un choix de phase $\exp\rl\frac{2i\pi k}{N_c}\rr$) menant de manière équivalente aux bonnes conditions aux bords pour $A_{\mu}$. Finalement, $\tr (-1)^F = N_c$.

Enfin, si l'on ajoute un certain nombre $N_f$ de multiplets chargés sous $SU(N_c)$, aucune symétrie n'empêche des termes de masses (qui sont invariants de jauge) d'être présents pour ces champs de matière. On calcule donc l'indice de Witten en supposant ces champs massifs, et on prend la limite de masse nulle à la fin du calcul. Clairement, ces champs ne contribuent pas à $\tr (-1)^F$ puisqu'ils ont au moins l'énergie de leur masse. Dans ce qui suit, nous allons retrouver la prédiction de Witten et expliciter les $N_c$ vides de la théorie.


\subsection{\'Etude à haute énergie et détail des symétries}


La théorie ultraviolette \cite{Affleck:1983mk} est décrite par le lagrangien
\be
\mathscr{L} = \frac{1}{4g^2}\left(\int d^2 \theta \tr \, W^{a}W^{a} + {\rm h.c.} \right) + \frac{1}{g^2} \int d^4 \theta \rl Q^{\dagger} e^{V} Q + {\widetilde Q}^{\dagger} e^{-V} {\widetilde Q}\rr
\label{LSQCD}
\ee
où $V$ est le supermultiplet vectoriel.

Au niveau classique, le lagrangien (\ref{LSQCD}) est invariant sous le groupe de symétries globales $U(N_f)_L\times U(N_f)_R \times U(1)_R$, que l'on peut décomposer en $SU(N_f)_L\times SU(N_f)_R\times U(1)_B\times U(1)_A\times U(1)_R$. Le facteur $U(1)_B$ représente le nombre baryonique, et $U(1)_A$ est chiral.

La symétrie chirale $U(1)_A$ et la $R$-symétrie $U(1)_R$ diffèrent de par la transformation des coordonnées grassmanniennes, comme nous l'avons vu au paragraphe \ref{Rsym}. Soit $\omega$ le paramètre de la transformation chirale et $\eta$ celui de la $R$-symétrie. Les transformations des superchamps sous ces symétries sont :
\begin{itemize}
 \item[$\bullet$] $U(1)_A$
\be
\begin{array}{ccc}
Q(x, \theta) \ &\longrightarrow &\ e^{i\omega} Q(x, \theta) \quad ,  \\
{\widetilde Q}(x, \theta) \ &\longrightarrow &\ e^{i\omega} Q(x, \theta) \quad ,
\end{array}
\label{U1A}
\ee
\item[$\bullet$] $U(1)_R$
\be
\begin{array}{ccc}
Q(x, \theta) \ &\longrightarrow &\  Q(x, e^{-i\eta}\theta) \quad ,  \\
{\widetilde Q}(x, \theta) \ &\longrightarrow &\ {\widetilde Q}(x, e^{-i\eta}\theta) \quad ,  \\
W^{a}(x, \theta) \ &\longrightarrow &\ e^{i\eta} W^{a}(x, e^{-i\eta}\theta) \quad .
\end{array}
\label{U1R}
\ee
\end{itemize}

Ici, nous avons choisi de paramètrer la $R$-symétrie de sorte que le superpotentiel a une charge $2$. Cette définition n'est pas unique, et il est toujours possible de recombiner $U(1)_A\times U(1)_R$ de façon à modifier ce choix.

Au niveau quantique, les deux symétries chirales possèdent des anomalies. La variation du lagrangien sous celles-ci est
\be
\delta_{U(1)_A} \mathscr{L} \ = \ \omega \ \tr \rl Q_A  T_{a}^2\rr \, \rl \frac{g^2}{32 \pi^2}F_{a} {\widetilde F}_{a}\rr \quad , \nonumber
\ee
où $Q_A$ représente ici la charge $U(1)_A$ des fermions se propageant dans la boucle, et $T_a$ sont les générateurs du groupe de jauge. L'expression correspondante pour $U(1)_R$ est obtenue en remplaçant $\omega$ par $\eta$ et $Q_A$ par $Q_R$. Pour chaque symétrie, les anomalies sont donc
\begin{eqnarray}
&& \delta_{U(1)_A} \mathscr{L} = 2 N_f \omega \rl \frac{g^2}{32 \pi^2} F_{a} {\widetilde F}_{a} \rr \quad , \nonumber \\
&& \delta_{U(1)_R} \mathscr{L} = \rl 2N_c-2 N_f \rr \eta \rl \frac{g^2}{32 \pi^2} F_{a} {\widetilde F}_{a}\rr \quad .
\label{anomalies}
\end{eqnarray}

Nous voyons clairement d'après  (\ref{anomalies}) qu'il existe une combinaison de ces deux symétries qui n'engendre pas d'anomalie au niveau quantique. En sommant les deux variations, on obtient

\be
\delta \mathscr{L} \ = \ \pl 2 N_f \omega + \rl 2N_c-2N_f \rr \eta \pr \, \rl \frac{g^2}{32 \pi^2} F_{a} {\widetilde F}_{a}\rr \quad ,
\label{R+A}
\ee
de sorte que $\delta \mathscr{L} = 0$ si
\be
\omega \ = \ \frac{N_f - N_c}{N_f}\, \eta \quad . \nonumber
\ee

Sous cette combinaison, que nous appellerons $U(1)'_R$, les superchamps se transforment de la façon suivante :
\be
\begin{array}{ccc}
Q(x, \theta) \ &\longrightarrow &\ e^{i\frac{N_f - N_c}{N_f} \eta} \ Q(x, e^{-i\eta}\theta) \quad ,  \\
{\widetilde Q}(x, \theta) \ &\longrightarrow &\ e^{i\frac{N_f - N_c}{N_f} \eta} \ {\widetilde Q}(x, e^{-i\eta}\theta) \quad ,  \\
W^{a}(x, \theta) \ &\longrightarrow &\ e^{i\eta} \ W^{a}(x, e^{-i\eta}\theta) \quad .
\end{array}
\label{U1sansanomalie}
\ee

Après avoir détaillé les symétries chirales, nous avons donc montré que les théories SUSY-QCD avec $N_c$ couleurs et $N_f$ saveurs sont invariantes sous le groupe de symétries globales
\be
G = SU(N_f)_L\times SU(N_f)_R\times U(1)_B\times U(1)'_R 
\label{Gglobal}
\ee
au niveau quantique. Les quarks gauches sont dans la représentation $N_f$ de $SU(N_f)_L$ et ont un nombre baryonique $1$. Les anti-quarks, ou quarks droits, sont dans la représentation $\overline{N_f}$ de $SU(N_f)_R$ et ont un nombre baryonique $-1$.

Le lagrangien (\ref{LSQCD}) ne contient pas de composante $F$ à l'ordre des arbres. D'une part, nous n'avons pas ajouté de terme de masse aux quarks, et d'autre part, les jauginos sont protégés par la $R$-symétrie (\ref{U1R}). Le potentiel scalaire est donc
\be
V \ = \ \frac{1}{2} D^{a} D^{a} \quad ,
\label{Vsansmasse}
\ee
avec
\be
D^{a} \ = g^2 \left[\ z^i_{\alpha}  \rl T^{a} \rr^{\alpha}_{\beta}  \rl  z^{\dagger} \rr^{\beta}_i \ - \ \rl {\widetilde z}^{\dagger}\rr^i_{\alpha} \rl T^{a} \rr^{\alpha}_{\beta}  {\widetilde z}^{\beta}_i \ \right] \quad .
\label{Dsansmasse}
\ee

Les solutions supersymétriques $\langle D^a \rangle = 0$ peuvent correspondre à des valeurs moyennes dans le vide non-nulles pour les scalaires $z$ et $\widetilde z$. Selon que $N_f < N_c$ ou $N_f \geqslant N_c$, certaines des symétries globales et le groupe de jauge sont totalement ou partiellement brisés. Il nous faut distinguer les deux différents cas :

\vskip0.3cm

\begin{enumerate}

\item  $N_f < N_c$ :

La solution la plus générale, à laquelle on peut toujours se ramener par des opérations qui préservent les symétries de jauge et globales, est 
\be
z^i_{\alpha} = \rl {\widetilde z}^{\dagger}\rr^i_{\alpha} = \left\{ \begin{array}{c} v_{(i)} \delta^i_{\alpha} \quad \text{pour} \ 1 \leqslant \alpha \leqslant N_f \\ 0 \quad \text{pour} \ \alpha > N_f  \end{array} \right. \quad .
\label{vevNf<Nc}
\ee

Ces minima brisent spontanément le groupe de jauge $SU(N_c)$ vers $SU(N_c-N_f)$. En particulier, si $N_f = N_c -1$, le groupe de jauge est totalement brisé.

Certaines symétries globales sont également brisées spontanément par ces vides. Le paramètre d'ordre invariant de jauge $\langle z^i_{\alpha} {\widetilde z}^{\alpha}_j \rangle = v_{(i)}^2 \delta^i_j$ permet de trouver le groupe de symétrie restant. Si, par exemple, tous les $v_{(i)}$ sont égaux, le groupe $G$ est brisé\footnote{La $R$-symétrie $U(1)'_R$ est spontanément brisée puisque le produit $z {\widetilde z}$ se transforme sous (\ref{U1sansanomalie}).} vers $SU(N_f)\times U(1)_B$.

\vskip0.3cm

\item $N_f \geqslant N_c$ :

La solution la plus générale est de la forme
\begin{eqnarray}
z^i_{\alpha}  &=& \left\{ \begin{array}{ccc} v_{(i)} \delta^i_{\alpha} \quad \text{pour} \ 1 \leqslant i \leqslant N_c \\ 0 \quad \text{pour} \ i > N_c  \end{array} \right. \quad \quad , \nonumber \\ \label{vevNf>Nc} \\
\rl {\widetilde z}^{\dagger}\rr^i_{\alpha} &=&  \left\{ \begin{array}{ccc} \sqrt{\al v_{(i)} \ar^2 - c^2} \ \delta^i_{\alpha} \quad \text{pour} \ 1 \leqslant i \leqslant N_c \\ 0 \quad \quad \quad \quad \quad \ \text{pour} \ i > N_c  \end{array} \right. \quad . \nonumber
\end{eqnarray}

Ce cas est plus délicat car les symétries de jauge et globales se mélangent pendant la brisure. Pour commencer, $U(1)'_R$ n'est pas totalement brisé. La $R$-symétrie
\begin{eqnarray}
Q^i_{\alpha}(x, \theta) \ &\longrightarrow &\  \left\{ \begin{array}{c} Q^i_{\alpha}(x, e^{i\epsilon}\theta) \quad \quad \text{ pour} \ 1 \leqslant i \leqslant N_c \\ e^{-i\epsilon} Q^i_{\alpha}(x, e^{i\epsilon}\theta) \quad \text{pour} \ N_c < i \leqslant N_f \end{array} \right.  \ , \nonumber \\ \nonumber \\
{\widetilde Q}_i^{\alpha}(x, \theta) \ &\longrightarrow &\  \left\{ \begin{array}{c} {\widetilde Q}_i^{\alpha}(x, e^{i\epsilon}\theta) \quad \quad \text{pour} \ 1 \leqslant i \leqslant N_c \\ e^{-i\epsilon} {\widetilde Q}_i^{\alpha}(x, e^{i\epsilon}\theta) \quad \text{pour} \ N_c < i \leqslant N_f \end{array} \right. \ , \label{U1chapeau} \\ \nonumber \\
W^{a}(x, \theta) \ &\longrightarrow &\ e^{-i\epsilon} W^{a}(x, e^{i\epsilon}\theta) \quad , \nonumber
\end{eqnarray}
que nous appellerons ${\hat U(1)}_R$, est préservée et n'engendre pas d'anomalie au niveau quantique. Pour $c=0$ et tous les $v_{(i)}$ égaux, le groupe total est brisé vers
\be
SU(N_c)\times SU(N_f - N_c)\times SU(N_f-N_c)\times U(1)'_B\times {\hat U(1)}_R \nonumber \quad .
\ee
Nous ne discuterons pas plus avant de ce cas ; nous expliquerons dans la section \ref{TheorieEff} pourquoi les méthodes employées pour le traitement du cas $N_f < N_c$ ne peuvent pas être appliquées ici.

\end{enumerate}

\vskip0.3cm

En dernier lieu, on peut rendre les quarks massifs en ajoutant à la théorie un terme de la forme
\be
\mathscr{L}_{mass} \ = \ - \left( \int d^2 \theta m^i_j {\widetilde Q}^{\alpha}_i Q^j_{\alpha}  + {\rm h.c.} \right) \quad ,
\label{Lmass}
\ee
qui brise explicitement le groupe de symétries globales en $SU(N_f)\times U(1)_B$ si la matrice de masse est proportionnelle à l'identité, ou le brise complètement si les $m^i_j$ sont génériques. Comme nous l'avons vu, même en l'absence de ces termes, dans la limite $m^i_j \rightarrow 0$, les symétries axiales sont brisées au niveau quantique car elles engendrent des anomalies. 
En plus de ces symétries continues, le terme de masse (\ref{Lmass}) préserve une $R$-symétrie discrète ${\mathbb Z}_{2N_c}$
\be
\begin{array}{ccc}
Q(x, \theta) \ &\longrightarrow &\ e^{-\frac{i\pi n}{N_c}} \ Q(x, e^{\frac{i\pi n}{N_c}}\theta) \quad ,  \\
{\widetilde Q}(x, \theta) \ &\longrightarrow &\ e^{-\frac{i\pi n}{N_c}} \ {\widetilde Q}(x, e^{\frac{i\pi n}{N_c}}\theta) \quad ,  \\
W^{a}(x, \theta) \ &\longrightarrow &\ e^{-\frac{i\pi n}{N_c}} \ W^{a}(x, e^{\frac{i\pi n}{N_c}}\theta) \quad ,
\end{array}
\label{Z2Nc}
\ee
avec $n=1,\ldots,2N_c\,$.

Par la suite, nous considèrerons que ce terme est présent dans la théorie. Avec l'introduction de ce superpotentiel, le potentiel scalaire contient la trace des modules au carré des termes $F^{\alpha}_j = -m^i_j {\widetilde z}^{\alpha}_i$ et ${\widetilde F}_{\alpha}^i = m^i_j  z^j_{\alpha}$, en plus des termes $D^a$ donnés par l'expression (\ref{Dsansmasse}). On trouve alors un espace classique de vides supersymétriques : $z = {\widetilde z} = 0$.

Comment évolue cette théorie à basse énergie ? En particulier, nous voulons savoir si les effets non-perturbatifs qui ne manqueront pas d'apparaître peuvent briser des symétries et si oui lesquelles. Enfin, signalons que nous ne intéressons pas ici à une brisure de la supersymétrie ; ce thème sera traité principalement dans la section \ref{ISS}.


\subsection{La théorie effective\label{TheorieEff}}


À basse énergie, la théorie est fortement couplée et il n'est plus possible de faire appel à la théorie des perturbations pour la décrire de manière exacte. Néanmoins, en utilisant la symétrie de jauge, les symétries globales ainsi que les anomalies, il est possible d'en obtenir une description effective \cite{Veneziano:1982ah} lorsque $N_f < N_c$.

Pour être invariant de jauge et sous $SU(N_f)_L\times SU(N_f)_R\times U(1)_B$, le lagrangien effectif ne peut s'exprimer qu'en fonction des deux superchamps
\begin{eqnarray}
U \ &=& \ \frac{g^2}{16 \pi^2}  W^{a}W^{a} \quad , \nonumber \\
M^i_j \ &=& \ Q^i_{\alpha} {\widetilde Q}^{\alpha}_j \quad ,
\label{UM}
\end{eqnarray}
où le facteur définissant $U$ est juste un choix de normalisation. Le champ $U$ est appelé ``glueball'', et les champs $M$ sont les ``mésons'', à l'instar des noms donnés aux champs équivalents en QCD dans le Modèle Standard.
Par ailleurs, les anomalies étant des phénomènes infra-rouges, elles doivent être présentes dans la théorie à basse énergie.

Le lagrangien effectif sera séparé en trois contributions
\be
\mathscr{L}_{eff} \ = \  \mathscr{L}_{cin} +  \left\{ \int d^2 \theta \rl W_{an} +  W_{mass} \rr + {\rm h.c.} \right\}\quad , \nonumber
\ee
qui contiennent respectivement les termes cinétiques, les anomalies, et les termes de masse. Dans la suite, nous nous concentrons sur les deux derniers termes. Nous mentionnerons rapidement la détermination de $ \mathscr{L}_{cin}$ à la fin de ce paragraphe.

Le terme de masse est déduit directement de son expression à haute énergie (\ref{Lmass})
\be
W_{mass} \ = \ - \tr \rl mM \rr \quad .
\label{Wmass}
\ee

Comme nous l'avons vu au paragraphe précédent, le groupe de symétries globales peut être spontanément brisé par les valeurs moyennes dans le vide de certains scalaires. Quoiqu'il en soit, ces brisures ne sont pas explicites et le lagrangien à basse énergie doit être invariant sous le groupe (\ref{Gglobal}). D'après  les idées de 't Hooft, lorsque deux théories de champs décrivent un système physique à haute et basse énergies, même si les degrés de liberté sont différents de l'une à l'autre, les anomalies des symétries globales communes aux deux doivent être les mêmes. Nous allons nous appuyer sur ce fait pour déduire le superpotentiel effectif. Nous aurons l'occasion de vérifier explicitement la concordance d'anomalies à la fin de cette section, ainsi qu'au chapitre suivant, paragraphe \ref{magnetik}.

Pour déterminer $W_{an}$, nous devons réexprimer les variations (\ref{anomalies}) en fonction des champs $U$ et $M$. À des termes nuls dans les équations du mouvement près, en utilisant (\ref{devWalpha}), $ i \rl \int d^2 \theta W^{a}W^{a} - {\rm h.c.}\rr = \frac{1}{2}F^{a} {\widetilde F}^{a}$ de sorte que $\frac{g^2}{32 \pi^2} F^{a} {\widetilde F}^{a} =i\rl U - U^{\dagger}\rr$. Il en résulte que les anomalies (\ref{anomalies}) sont proportionnelles à la composante $F$ du superchamp $U$. C'est la raison pour laquelle le terme portant les anomalies à basse énergie est un superpotentiel. Les expressions (\ref{anomalies}) se réécrivent
\begin{eqnarray}
&& \delta_{U(1)_A} W_{an} \ = \ 2 i N_f \, \omega  \, U \quad , \nonumber \\
&& \delta_{U(1)_R} W_{an} \ = \ 2 i \rl N_c- N_f \rr \eta \, U \quad .
\label{anomalies2}
\end{eqnarray}

Par ailleurs, connaissant les charges (\ref{U1A}) et (\ref{U1R}) des champs fondamentaux sous $U(1)_A$ et $U(1)_R$, nous savons que
\be
\begin{array}{ccc}
\delta_{U(1)_A} U = 0 \quad \quad &,& \quad \quad  \delta_{U(1)_R} U = 2i \eta U  \quad ,  \\
\delta_{U(1)_A} M = 2i \omega M \quad \quad &,& \quad \quad  \delta_{U(1)_R} M = 0  \quad \quad .
\end{array}
\label{deltaU1UM}
\ee

Nous déduisons des équations (\ref{anomalies2}) et (\ref{deltaU1UM}), ainsi que du fait que
\begin{eqnarray}
&&\delta_{U(1)_A} W_{an} \ = \ \tr \rl \frac{\dd W_{an}}{\dd M} \delta_{U(1)_A} M \rr + \frac{\dd W_{an}}{\dd U} \delta_{U(1)_A} U\quad , \nonumber \\
&&\delta_{U(1)_R} W_{an} \ = \ \tr \rl \frac{\dd W_{an}}{\dd M} \delta_{U(1)_R} M \rr + \frac{\dd W_{an}}{\dd U} \delta_{U(1)_R} U  \quad , \nonumber
\end{eqnarray}
les égalités suivantes :
\begin{eqnarray}
&&\tr \rl M \frac{\dd W_{an}}{\dd M} \rr \ = \   N_f  U \quad , \label{dWan1} \\
&&U \frac{\dd W_{an}}{\dd U} \ = \ \rl N_c - N_f \rr U \quad . \label{dWan2}
\end{eqnarray}

En dernier lieu, nous utilisons la combinaison (\ref{U1sansanomalie}) de $U(1)_A$ et $U(1)_R$ qui ne produit pas d'anomalie. Le superpotentiel doit avoir une charge $2$ sous cette symétrie, et les champs $M$ et $U$ ont des charges qui valent respectivement $2(N_f-N_c)/N_f$ et $2$. Nous obtenons donc
\be
(N_f-N_c) \tr \rl M \frac{\dd W_{an}}{\dd M} \rr + N_f U \frac{\dd W_{an}}{\dd U} \ = \ N_f W_{an} \quad .
\label{dWan3}
\ee
Nous ne pouvons directement injecter les équations (\ref{dWan1}) et (\ref{dWan2}) dans l'expression (\ref{dWan3}) : cela impliquerait $W_{an} = 0$. En effet, il y a une redondance dans les trois équations que nous venons de dériver puisque $U(1)'_R$ est déduit de $U(1)_A$ et $U(1)_R$.

En se servant de (\ref{dWan1}) et de (\ref{dWan3}), on obtient un système que l'on peut intégrer. Finalement,
\be
W_{an} \ = \ U \pl \ln \rl \frac{U^{N_c-N_f}  \det M}{\Lambda^{3N_c - N_f}}\rr  - \rl N_c-N_f \rr \pr \quad ,
\label{Wan}
\ee
où $\Lambda$, l'échelle dynamique (\ref{Lambda}), a été introduite afin que l'argument du logarithme soit adimensionné.

Par des arguments de symétrie et grâce au fait que c'est le superpotentiel qui porte les anomalies, nous avons pu déterminer la théorie effective à basse énergie. Il est important de souligner que ce résultat est exact.

L'étude de 
\be
W \ = \ U \pl \ln \rl \frac{U^{N_c-N_f}  \det M}{\Lambda^{3N_c - N_f}}\rr  - \rl N_c-N_f \rr \pr \ - \ \tr \rl mM \rr
\label{Weff}
\ee
permet de déterminer les minima du potentiel. En particulier, la supersymétrie est conservée si $\frac{\dd W}{\dd z^i} = 0$ pour tous les champs scalaires $z^i$ présents.

La plus basse composante du multiplet chiral de jauge étant $\left. W^{a}\ar_{\rm scal} = \lambda^{a}$, il s'ensuit que $U \vert_{\rm scal}$ est le scalaire issu de la composition des jauginos. Le scalaire $\left. M^i_j \ar_{\rm scal}$ est, lui, le paramètre d'ordre $z^i_{\alpha} {\widetilde z}^{\alpha}_j$.

Les vides supersymétriques sont donnés par
\begin{eqnarray}
&&\left. \frac{\dd W}{\dd U}\ar_{\rm scal} = \ln \rl \frac{U^{N_c-N_f}  \det M }{\Lambda^{3N_c - N_f}}\rr  = 0 \quad , \label{dWU} \\
&& \left. \frac{\dd W}{\dd M^j_i}\ar_{\rm scal} = U \rl M^{-1} \rr^i_j - m^i_j = 0 \quad , \label{dWM}
\end{eqnarray}
dont les solutions sont
\begin{eqnarray}
&& \langle M^i_j  \rangle \ = \ \rl m^{-1}\rr^i_j \langle U \rangle  \quad , \nonumber \\
&& \langle U  \rangle \ = \ \Lambda^{\frac{3N_c -N_f}{N_c}} \rl \det m \rr^{1/N_c} \quad .
\label{solsusy}
\end{eqnarray}

Ce résultat montre que les valeurs moyennes dans le vide des mésons sont données de manière unique par celle des glueballs. 
Nous avons vu que le fait de rendre les quarks massifs préservait une $R$-symétrie discrète $\mathbb{Z}_{2N_c}$ donnée par les transformations (\ref{Z2Nc}). Sous cette symétrie du lagrangien effectif, $U$ et $M^i_j$ se transforment par une phase $e^{\frac{2i\pi n}{N_c}}$ avec $n=1,\ldots,N_c$. Nous voyons donc que cette symétrie est spontanément brisée par la valeur moyenne non-nulle de $U$, qui signe la condensation des jauginos. Le résultat (\ref{solsusy}) correspond à $N_c$ vides supersymétriques distincts comme le prédisait le calcul de l'indice de Witten (section \ref{indice}).

Notons que si toutes les masses $m^i_j \rightarrow 0$, les gluinos ne se condensent pas, et les mésons sont envoyés à l'infini : la théorie est singulière en ces points. En particulier, on ne retrouve pas la prédiction de Witten qui doit être valide même lorsque les quarks sont non-massifs.

Pour des énergies inférieures à l'échelle de condensation, le superchamp $U$ découple de la théorie. Par conservation de la supersymétrie, sa valeur classique est
\be
\frac{\dd W}{\dd U} = 0 \quad \Rightarrow \quad \langle U_k \rangle = \rl \frac{\Lambda^{3N_c - N_f}}{\det M} \rr^{\frac{1}{N_c - N_f}} e^{\frac{2ik}{N_c-N_f}} \quad ,
\label{Uclassique}
\ee 
avec $k=1,\ldots,(N_c-N_f)$. Cela permet d'obtenir le superpotentiel effectif des mésons
\be
W_{dyn} \ = \ - \rl N_c - N_f \rr  e^{\frac{2ik}{N_c-N_f}} \ \rl \frac{\Lambda^{3N_c - N_f}}{\det M}\rr^{\frac{1}{N_c-N_f}} - \tr \ mM \quad .
\label{Wdyn}
\ee
Ce superpotentiel est généré dynamiquement par la condensation des jauginos. Les solutions supersymétriques sont, comme dans (\ref{solsusy}),
\be
\langle M^i_j \rangle \ = \  \Lambda^{\frac{3N_c -N_f}{N_c}} \rl \det m \rr^{1/N_c} \rl m^{-1}\rr^i_j  \quad . \label{Mij}
\ee
La présence des $N_c$ vides supersymétriques réside dans la puissance $1/N_c$ du déterminant.


\subsubsection*{Le lagrangien $\mathscr L_{cin}$}


Les termes cinétiques de $M$ et de $U$ sont beaucoup moins contraints par les symétries. Puisque le lagrangien est supersymétrique et invariant sous $G$, il s'exprime comme une composante $D$ d'une fonction de $M^{\dagger} M$ et de $U^{\dagger}U$. Les glueballs ont une dimension canonique $3$ donc le terme cinétique de $U$ ne peut être que $\int d^4 \theta \rl U^{\dagger}U \rr^{1/3}$. Le terme cinétique pour les mésons est moins bien contrôlé. Le lagrangien le plus simple est
\begin{eqnarray}
&&\mathscr{L}_{cin} \ = \ \int d^4 \theta K_{cin} \quad \quad \text{avec} \nonumber \\
&&K_{cin} \ = \ \frac{9}{\alpha} \rl U^{\dagger}U \rr^{1/3} + \frac{1}{\gamma} \rl M^{\dagger} M \rr \rl U^{\dagger}U \rr^{-1/3} \quad ,
\label{Lcin}
\end{eqnarray}
et $\alpha = \mathcal{O}(N_c^{-4/3})$, $\gamma = \mathcal{O}(N_c^{1/3})$. En minimisant ces termes avec le superpotentiel effectif (\ref{Weff}), on s'attend à $\langle U \rangle = \mathcal{O}(N_c)$, et $\langle M \rangle = \mathcal{O}(N_c)$.

En terme des champs scalaires,
\begin{eqnarray}
\mathscr{L}_{cin} &=& \frac{1}{\alpha} \rl U^* U \rr^{-2/3} \al \dd U \ar^2 + \frac{1}{\gamma}  \rl U^* U \rr^{-1/3} \al \dd M \ar^2 \nonumber \\
&+& \pl \frac{1}{\gamma} \rl M^* M \rr \rl U^* U \rr^{-1/3} \frac{\dd M}{M} \frac{\dd U^*}{U^*} + {\rm h.c.} \pr \quad . \nonumber
\end{eqnarray}
Les termes croisés et le terme cinétique des mésons scalaires $M$ sont d'ordre $1/N_c$, tandis que le terme cinétique de $U$ est $\mathcal{O}(1)$. Lorsque l'on fait tendre $N_c \rightarrow \infty$, on retrouve bien une théorie de Yang-Mills pure. En effet, dans cette limite, avec $N_f$ et $N_c g^2$ fixés, le secteur de jauge devient dominant dans les boucles, et la théorie doit se comporter comme une théorie de jauge.

Néanmoins, le terme (\ref{Lcin}) n'est pas le plus général  possible. Cette indétermination relative n'empêche certes pas de calculer les minima du potentiel, mais le spectre de la théorie dépend à priori de la forme des termes cinétiques. Le manque de contrôle sur le potentiel de K\"ahler constitue une des limitations des méthodes de lagrangien effectif.


\subsubsection*{Le cas $N_f \geqslant N_c$}


D'après  l'évolution de la constante de couplage (\ref{betaelectrik}), la théorie $N_f \geqslant N_c$ sera moins fortement couplée dans l'infra-rouge que la théorie $N_f < N_c$. Autrement dit, l'échelle dynamique qui signe la condensation des gluinos sera plus faible. Considérons néanmoins qu'il existe une hiérarchie $m^i_j \ll \Lambda$ suffisante pour que l'on se concentre exclusivement sur les mésons.

Le superpotentiel étant analytique, il ne peut pas dépendre de l'opérateur invariant $M^{\dagger} M$. Le seul objet invariant de jauge et analytique possible est $\det M$. De plus, si on ajoute la condition d'invariance sous $U(1)'_R$, équations (\ref{U1sansanomalie}), le superpotentiel ne peut contenir que $(\det M)^{\frac{1}{N_f-N_c}}$ car il doit avoir une charge $2$. Enfin, pour que le superpotentiel ait les bonnes dimensions, nous introduisons l'échelle dynamique de la théorie.

Par de simples arguments de symétrie et d'analyticité, nous venons de montrer d'une autre façon que le seul superpotentiel permis à très basse énergie pour une théorie SUSY-QCD est
\be
W \ = \ a \ \Lambda^{\frac{3N_c - N_f}{N_c-N_f}} \rl \det M \rr^{\frac{1}{N_f-N_c}} \quad ,
\label{Wargsym}
\ee
où $a$ est une constante\footnote{Il a été montré que $a = \rl N_f - N_c \rr$.}.

Réécrivons de façon plus détaillée le déterminant de $M$ en termes des quarks :
\be
\det M \ = \ \sum_{\alpha_1 = 1}^{N_c} \ldots \sum_{\alpha_{N_f} = 1}^{N_c} \epsilon_{i_1\ldots i_{N_f}} \ \epsilon^{j_1\ldots j_{N_f}} \ Q^{i_1}_{\alpha_1}{\widetilde Q}^{\alpha_1}_{j_1}\ \ldots \ Q^{i_{N_f}}_{\alpha_{N_f}}{\widetilde Q}^{\alpha_{N_f}}_{j_{N_f}} \quad .
\label{detM}
\ee
Dans cette expression, $\epsilon$ est le tenseur totalement antisymétrique.
Chaque terme de chaque somme (y compris sur $i_k$ et $j_k$) s'écrit
\be
Q^{i_1}_{\alpha_1}{\widetilde Q}^{\alpha_1}_{j_1}\ \ldots \ Q^{i_{N_c}}_{\alpha_{N_c}}{\widetilde Q}^{\alpha_{N_c}}_{j_{N_c}} \ Q^{i_{N_c+1}}_{\sigma_1}{\widetilde Q}^{\sigma_1}_{j_{N_c+1}} \ \ldots \ Q^{i_{N_f}}_{\sigma_{N_f-N_c}}{\widetilde Q}^{\sigma_{N_f-N_c}}_{j_{N_f}} \quad ,
\label{redondance}
\ee
où les $\sigma_n$ sont $(N_f - N_c)$ éléments de l'ensemble $\left\{1,\ldots,N_c\right\}$ et où les $\alpha_k$ ne sont pas sommés.

Il y a donc au moins une redondance dans les indices de couleurs pour tout $N_f > N_c$. Supposons que $\sigma_1 = \alpha_1$. Le terme (\ref{redondance}) est (entre autres) invariant sous l'échange $i_1 \leftrightarrow i_{N_c+1}$, tandis que le tenseur $\epsilon_{i_1\ldots i_{N_f}}$ est, lui, antisymétrique.

Tous les termes du développement de $\det M$ s'annulent donc deux à deux. Il en résulte que le superpotentiel (\ref{Wargsym}) est identiquement nul.

Toutefois, comme nous l'avons vu dans le paragraphe \ref{indice}, la théorie contient $N_c$ vides supersymétriques quelque soit $N_f$. Un superpotentiel doit être dynamiquement engendré à basse énergie \cite{Seiberg:1994bz}.

Jusqu'ici, nous avons omis la présence de baryons
\begin{eqnarray}
B_{i_{N_c+1},\ldots,i_{N_f}} \ = \ \epsilon_{i_1\ldots i_{N_f}}Q^{i_1}\cdots Q^{i_{N_c}} \quad , \nonumber \\
{\widetilde B}^{j_{N_c+1},\ldots,j_{N_c}} \ = \ \epsilon^{j_1\ldots j_{N_f}} {\widetilde Q}_{j_1}\cdots {\widetilde Q}_{j_{N_c}} \quad .
\label{baryonsNf>Nc}
\end{eqnarray}
Bien entendu, sans terme de masse, les solutions des équations du mouvement sont $B = {\widetilde B} = 0$. Si la théorie contient un terme de masse pour les baryons, alors aux énergies inférieures à ces masses, ils découplent de la théorie et on obtient un potentiel effectif pour les mésons qui dépend des paramètres de masse des baryons.

Sous le nombre baryonique et la $R$-symétrie (\ref{U1sansanomalie}), les mésons $M$ et les baryons $B$ et ${\widetilde B}$ ont des charges respectives $\rl0,\ 2(N_f - N_c)\rr$, $\rl N_f,\ (N_f - N_c)\rr$ et $\rl - N_f,\ (N_f - N_c)\rr$.

Nous nous proposons de dériver le lagrangien effectif dans le cas particulier $N_f = N_c$. Dans ce cas, $\det M$ n'est pas nul à priori mais la constante $a = N_f - N_c$ du superpotentiel (\ref{Wargsym}) l'est. La démonstration qu'un potentiel effectif est engendré à basse énergie peut également être faite pour $N_f = N_c + 1$.

Lorsque $N_f = N_c$, les baryons (\ref{baryonsNf>Nc}) se réduisent simplement à
\begin{eqnarray}
B \ = \ \epsilon_{i_1\ldots i_{N_f}}Q^{i_1}\cdots Q^{i_{N_f}} \quad , \nonumber \\
{\widetilde B} \ = \ \epsilon^{j_1\ldots j_{N_f}} {\widetilde Q}_{j_1}\cdots {\widetilde Q}_{j_{N_f}} \quad .
\label{baryonsNf=Nc}
\end{eqnarray}

Leur $R$-charge, ainsi que celle des quarks et des mésons, est nulle.

On voit clairement, en utilisant (\ref{detM}), que le système au niveau classique est soumis à la contrainte
\be
\det M \ - \ B {\widetilde B} \ = \ 0 \quad .
\label{detMBBtildeClass}
\ee

Au niveau quantique, cette contrainte devient
\be
\det M \ - \ B {\widetilde B} \ = \ \Lambda^{2N_f}
\label{detMBBtildeQuant}
\ee
en raison de phénomènes non-perturbatifs de type instanton.

Certaines solutions de l'équation (\ref{detMBBtildeQuant}) brisent spontanément le groupe $G$. Pour chacune d'entre elles, nous devons vérifier que les symétries préservées produisent les mêmes anomalies qu'à haute énergie.
\begin{enumerate}

\item $B={\widetilde B} = 0$ et $M^i_j = \Lambda^2 \delta^i_j$

La symétrie $SU(N_f)_L\times SU(N_f)_R$ est brisée en $SU(N_f)$ diagonal. Les groupes abéliens $U(1)_B$ et $U(1)'_R$ ne sont pas brisés. Il faut donc vérifier les anomalies du groupe préservé $SU(N_f)\times U(1)_B\times U(1)'_R$.

À haute énergie, les fermions se propageant dans les boucles sont les $2N_f^2$ quarks et anti-quarks, et les $N_f^2-1$ jauginos :
\be
\begin{array}{ccc}
\psi \quad && \quad \rl N_f, \ 1 , \ -1 \rr  \\
\widetilde \psi \quad &&\quad \rl \overline{N_f}, \ -1 , \ -1 \rr  \\
\lambda^a \quad &&\quad \rl 1, \ 0 , \ 1 \rr
\end{array}
\nonumber
\ee

À basse énergie, ce sont les fermions des multiplets $M, \ B, \ {\widetilde B}$
\be
\begin{array}{ccc}
&&\Psi_M \quad \quad \rl N_f^2-1, \ 0 , \ -1 \rr \\
&&\chi_B \quad \quad \rl 1, \ -N_f , \ -1 \rr \\
&&\chi_{\widetilde B} \quad \quad \rl 1, \ N_f , \ -1 \rr
\end{array}
\nonumber
\ee

Les anomalies non-triviales sont
\begin{eqnarray}
SU(N_f)^2 U(1)'_R &\quad&  -N_f \left[ d^{(2)}(N_f) + d^{(2)}(\overline{N_f}) \right] = -d^{(2)}(N_f^2 - 1) \nonumber \\
U(1)^{\prime \, 3}_R &\quad&  2N_f^2 (-1)^3 + (N_f^2 - 1) = (N_f^2 - 1) (-1)^3 -2 \nonumber \\
U(1)_B^2 U(1)'_R &\quad&  -2 N_f^2 = - (-N_f)^2 - N_f^2 \nonumber \\
U(1)'_R &\quad&  -2N_f^2 + (N_f^2 - 1) = - (N_f^2 - 1) - 2\nonumber
\end{eqnarray}
avec $d^{(2)}(R)$ l'opérateur de Casimir quadratique dans la représentation $R$ du groupe $SU(N_f)$. On trouve que les anomalies concordent parfaitement.

\item $B = -{\widetilde B} = \Lambda^{N_f}$ et $M^i_j = 0$

La symétrie baryonique $U(1)_B$ est spontanément brisée. Les symétries chirales $SU(N_f)\times SU(N_f)\times U(1)'_R$ sont préservées. Les nombres quantiques des fermions à haute énergie sont connus, tandis que les fermions à basse énergie sont
\begin{eqnarray}
\Psi_M \quad && \quad \rl N_f, \ \overline{N_f} , \ -1 \rr \nonumber \\
\frac{\chi_B - \chi_{\widetilde B}}{\sqrt{2}} \quad && \quad \rl 1, \ 1 , \ -1 \rr \nonumber
\end{eqnarray}

Les anomalies sont 
\begin{eqnarray}
SU(N_f)^3 &\quad&  N_f d^{(3)}(N_f) \nonumber \\
SU(N_f)^2 U(1)'_R &\quad&  - N_f d^{(2)}(N_f) \nonumber \\
U(1)^{\prime \, 3}_R &\quad&  -N_f^2 -1 \nonumber \\
U(1)'_R &\quad&  -N_f^2 -1 \nonumber
\end{eqnarray}
avec $d^{(3)}(N_f)$ l'opérateur de Casimir cubique dans la représentation $N_f$. Là encore, on trouve que les anomalies sont les mêmes.
\end{enumerate}

Si l'on n'avait pas tenu compte de l'effet non-perturbatif, et qu'on avait considéré que l'équation (\ref{detMBBtildeClass}) était encore valide au niveau quantique, alors le point $B = {\widetilde B} = \Lambda^{N_f},\  M^i_j = \Lambda^2 \delta^i_j$ aurait été solution. Dans ce cas, la $R$-symétrie aurait été préservée, ainsi que $SU(N_f)$ diagonal. On aurait trouvé que les anomalies ne concordaient pas entre les degrés de liberté à haute et à basse énergies.

Nous introduisons un multiplicateur de Lagrange $X$ dans le superpotentiel effectif
\be
W \ = \ X \rl \det M - B {\widetilde B} - \Lambda^{2N_f} \rr \quad ,
\label{WXNf=Nc}
\ee
de sorte que l'équation du mouvement pour $X$ impose la condition (\ref{detMBBtildeQuant}).

Le lagrangien au niveau classique contient les termes de masse autorisés par les symétries. Il existe trois possibilités : les baryons sont massifs mais pas les mésons ; les mésons sont massifs mais pas les baryons ; les mésons et les baryons sont massifs.

Considérons le second cas : au potentiel (\ref{WXNf=Nc}), nous ajoutons le terme $\tr \rl m M \rr$. Les équations du mouvement donnent $B = \widetilde B = 0$. Les deux autres équations, en revanche, impliquent
\begin{eqnarray}
&&\det M \ = \ \Lambda^{2N_f} \quad , \nonumber \\
&&X M^{-1} \det M + m \ = \ 0 \quad , \nonumber
\end{eqnarray}
c'est-à-dire
\begin{eqnarray}
&&X_n \ = \ e^{\frac{2 i \pi n}{N_f}} \rl \det m \rr^{\frac{1}{N_f}} \Lambda^{-2(N_f-1)} \quad , \nonumber \\
&&\langle M^i_j \rangle_n \ = \ e^{\frac{2 i \pi n}{N_f}} \Lambda^{2} \rl \det m \rr^{1/N_f} \rl m^{-1}\rr^i_j \quad , \nonumber
\end{eqnarray}
avec $n = 1,\ldots,N_f$. Ce résultat est conforme aux solutions (\ref{Mij}) lorsque $N_f = N_c$, et on vérifie de nouveau la valeur de l'indice de Witten.

Supposons pour plus de simplicité que la matrice de masse est diagonale\footnote{Il est en fait toujours possible de redéfinir les superchamps $Q$ et $\widetilde Q$ pour que la matrice $m$ soit diagonale.} $m^i_j = m_i \delta^i_j$. Dans le cas où $p$ des $N_f$ masses sont nulles, on peut découpler les quarks massifs et le multiplicateur $X$ de la théorie. À basse énergie, on trouve un superpotentiel effectif
\be
W_{eff} \ = \ \rl N_f - p \rr \rl \frac{\Lambda_L^{3N_f - p}}{\det' M} \rr^{\frac{1}{N_f - p}}  \quad , \nonumber
\ee
où $\det' M$ porte uniquement sur les champs de masse nulle, et où l'échelle effective est donnée par
\be
\Lambda_L = \Lambda^{\frac{2N_f}{3N_f - p}} \rl \prod_i m_i \rr^{\frac{1}{3N_f - p}} \quad . \nonumber
\ee
Le superpotentiel trouvé a exactement la même forme que (\ref{Wdyn}). Le même type de raisonnement peut être suivi dans le cas $N_f = N_c + 1$. Lorsque $N_f \geqslant N_c + 2$, par contre, il faut faire appel à d'autres outils.


\chapter[Dualités de Seiberg et brisure de supersymétrie]{Dualités de Seiberg et brisure dynamique de supersymétrie\label{DualiteSeiberg}}


\section{La théorie magnétique\label{magnetik}}


Dans le chapitre \ref{BrisureDynaSym}, section \ref{SUSYQCD}, nous avons étudié les théories SUSY-QCD à basse énergie en utilisant une méthode de lagrangien effectif. Comme nous l'avons vu, le contrôle sur les termes cinétiques (potentiel de K\"ahler) est difficile. Dans ce paragraphe, nous développons la dualité de ces théories entre elles pour des choix judicieux du nombre de couleurs et de saveurs \cite{Seiberg:1994pq}. En particulier, nous nous plaçons dans le cas $N_f \geqslant N_c + 2$, pour lequel nous avons vu que le superpotentiel (\ref{Wargsym}) est identiquement nul.

Les charges des superchamps sous les symétries globales sont
\be
\begin{matrix} &  SU(N_f)_L & SU(N_f)_R & U(1)_B & U(1)_R \cr
          & \cr
         Q &  N_f  & 1 &  1 & \displaystyle{\frac{N_f-N_c}{N_f}}\cr
         & \cr
         \widetilde Q & 1 & \overline{N_f} & -1 & \displaystyle{\frac{N_f-N_c}{N_f}}\cr
\end{matrix}
\label{nbquantelec}
\ee
où $U(1)_R$ désigne la $R$-symétrie (\ref{U1sansanomalie}) qui ne produit pas d'anomalie.

Les opérateurs invariants de jauge qui nous intéressent à basse énergie sont les mésons
\be
M^i_j \ = \ Q^i_{\alpha} {\widetilde Q}^{\alpha}_j \quad ,
\label{mesonselec}
\ee
ainsi que les baryons
\begin{eqnarray}
B^{i_1,\ldots,i_{N_c}} \ = \ Q^{i_1}\cdots Q^{i_{N_c}} \quad , \nonumber \\
{\widetilde B}_{j_1,\ldots,j_{N_c}} \ = \ {\widetilde Q}_{j_1}\cdots {\widetilde Q}_{j_{N_c}} \quad ,
\label{baryonselec}
\end{eqnarray}
où les indices sont antisymétrisés.

À très basse énergie, la théorie est très fortement couplée. Pour mieux comprendre comment elle s'y comporte, nous devons inclure le terme à deux boucles dans la fonction bêta (\ref{betaelectrik})
\be
\beta (g) = -\frac{g^3}{16 \pi^2} \frac{3N_c - N_f + N_f \gamma (g^2)}{1 - N_c \frac{g^2}{8 \pi^2}} \quad , \ \text{où} \quad \gamma (g^2) = -\frac{g^2}{8 \pi^2} \frac{N_c^2 - 1}{N_c}
\label{betaexacte}
\ee
est la dimension anormale.

La théorie admet un point fixe $N_c g_*^2 \simeq \frac{8 \pi^2}{3} \frac{3 N_c - N_f}{N_c}$ pour $3N_c/2 \leqslant N_f \leqslant 3N_c\,$.

Le fait que l'algèbre superconforme contienne la $R$-symétrie (\ref{U1sansanomalie}) implique que la dimension d'un opérateur chiral doit être égale à $3R/2$, avec $R$ la charge de l'opérateur\footnote{Pour s'en convaincre, considérer une théorie $K(\Phi, \Phi^{\dagger}) = \Phi^{\dagger}\Phi$ et $W(\Phi)=\lambda \Phi^3$. Elle est invariante d'échelle $\rl x, \theta, \thetab \rr \rightarrow \rl e^{-\epsilon} x, e^{-\epsilon/2}\theta, e^{-\epsilon/2}\thetab\rr$ et possède également une $R$-symétrie de charge $R_{\Phi}=2R_{\theta}/3$ (voir paragraphe \ref{Rsym}). Alors les anomalies engendrées par ces deux symétries s'annulent si et seulement si la charge de $\theta$ est $R_{\theta}=1$.}. Par conséquent, la dimension $D$ des mésons est
\be
D (Q{\widetilde Q}) \ = \ \frac{3}{2} R (Q{\widetilde Q}) \ = \ \frac{3}{2} \rl R(Q) + R({\widetilde Q})\rr \ = \ \frac{3\rl N_f - N_c\rr}{N_f} \quad . \label{dimM}
\ee
Une autre façon de retrouver ce résultat est de reprendre la dimension anormale autour du point fixe $\gamma(g_*^2) = \frac{N_f - 3N_c}{N_f}$ ; les mésons ont donc une dimension
\be
D(Q{\widetilde Q}) = D(Q)+ D(\widetilde Q) + \gamma = 2 + \gamma  \quad . \nonumber
\ee

Si on suppose que la théorie est superconforme à basse énergie, alors elle n'est pas unitaire ($D < 1$) si $N_f < 3N_c /2$. Il s'ensuit que la théorie n'est pas, en réalité, conforme dans ce domaine de saveurs. Elle doit être décrite par un autre groupe de jauge à basse énergie, et les mésons doivent intégrer cette version ``duale'' en tant que champs fondamentaux, de dimension $D=1$.

Nous nous plaçons donc dans le cas $N_c+2 \leqslant N_f \leqslant 3N_c/2$. La théorie à basse énergie avec $N_f$ saveurs dans ce domaine est décrite dans sa version magnétique par le groupe de jauge $SU(N_f - N_c)$. À défaut de pouvoir en donner une démonstration rigoureuse, nous allons prouver que le groupe de symétries globales étant le même, les anomalies entre les deux théories sont les mêmes. Nous en concluerons que la théorie $SU(N_c)$, appelée ``électrique'', et la théorie $SU(N_f - N_c)$, appelée ``magnétique'', sont duales l'une de l'autre.

La fonction bêta à une boucle du couplage de jauge $g'$ de la théorie magnétique avec $N_f$ familles de quarks et d'anti-quarks est

\be
\beta(g') \ = \ -\frac{g^{\prime \, 3}}{16 \pi^2} (3N_c-2N_f) \quad .
\label{betamagnetik}
\ee
Le régime perturbatif est atteint à basse énergie pour $N_f \leqslant 3N_c /2$ et tout l'intérêt de la dualité passe par l'utilisation de la théorie des perturbations dans la théorie magnétique pour calculer les observables intéressantes dans le langage de la théorie électrique.

Lorsque $N_f$ décroît, à $N_c$ fixé, la théorie électrique devient de plus en plus fortement couplée, tandis que la théorie magnétique devient libre dans l'infra-rouge. En ce sens, les deux théories se comportent comme les théories électrique et magnétique habituelles.

En résumé,
\begin{enumerate}

\item La théorie est (super)conforme dans le domaine $3N_c/2 < N_f < 3N_c\,$. Là, les deux versions de la théorie, électrique et magnétique, sont asymptotiquement libres.

\item Lorsque $N_f < 3N_c/2$, la théorie n'est plus conforme. Elle est décrite à basse énergie par sa version magnétique de groupe $SU(N_f-N_c)$, qui est libre dans l'infra-rouge pour ce domaine de saveurs.
\end{enumerate}

Les quarks (\ref{nbquantelec}), associés au groupe de jauge $SU(N_c)$ ne sont pas les quarks fondamentaux de la théorie duale. Nous définissons ces supermultiplets par $q = \rl \ffi, \chi \rr$ et $\widetilde q = \rl \ffitil , \widetilde \chi \rr$ qui sont dans les représentations $N_f - N_c$ et $\overline{N_f - N_c}$ du groupe de jauge magnétique.

Les baryons (\ref{baryonselec}) doivent être déduits de ces degrés de liberté de manière polynomiale, $B \sim q^{N_f - N_c}$ et $\widetilde B \sim \widetilde q^{N_f - N_c}$, ce qui permet de fixer les nombres quantiques sous le groupe de symétries globales. Ayant fixé la charge baryonique et la $R$-charge de sorte à pouvoir construire les ``anciens'' baryons à partir de ces quarks, il n'est plus possible d'exprimer les quarks électriques (\ref{nbquantelec}) en fonction des quarks magnétiques par un développement polynômial. Il n'est donc pas possible non plus de construire les mésons, ce qui renforce ce que nous avions pressenti ci-dessus en calculant la dimension des mésons. Nous les incluons dans la théorie magnétique en tant que champs fondamentaux, singlets de jauge du groupe $SU(N_f - N_c)$, et nous leur assignons les mêmes nombres quantiques globaux que dans la théorie électrique.

Le contenu en champs de matière de la théorie duale est donc :

\be
\begin{matrix} &  SU(N_f)_L & SU(N_f)_R & U(1)_B & U(1)_R \cr
          & \cr
         q & 1 & \overline{N_f} &  \displaystyle{\frac{N_c}{N_f - N_c}} & \displaystyle{\frac{N_c}{N_f}}\cr
         & \cr
         \widetilde q &  N_f  & 1 & -\displaystyle{\frac{N_c}{N_f - N_c}} & \displaystyle{\frac{N_c}{N_f}}\cr
	& \cr
         M &  N_f  & \overline{N_f} & 0 & \displaystyle{\frac{2\rl N_f-N_c \rr }{N_f}}\cr
\end{matrix}
\quad , \label{nbquantmagn}
\ee
auquel il faut rajouter le secteur de jauge de $SU(N_f-N_c)$.

Si le choix de ces nombres quantiques paraît raisonnable pour retrouver les degrés de liberté électriques, rien ne nous garantit la dualité annoncée jusqu'ici. Il faut pour cela vérifier que les anomalies produites par le groupe global $SU(N_f)\times SU(N_f)\times U(1)_B\times U(1)_R$ commun aux deux théories soient les mêmes. Dans la théorie originelle, les $2N_f N_c$ fermions $\psi$ et $\widetilde \psi$ ont des charges $R = \frac{N_f - N_c}{N_f} - 1 = -\frac{N_c}{N_f}$ et les $N_c^2-1$ jauginos ont une charge $1$. Dans la théorie duale, les $2 N_f \rl N_f-N_c\rr$ fermions $\chi$ et $\widetilde \chi$ ont une charge $R = \frac{N_c - N_f}{N_f}$, les fermions $\Psi_M$ associés aux mésons ont une charge $\frac{N_f - 2N_c}{N_f}$, et les jauginos ont une charge $1$. On trouve
\begin{eqnarray}
U(1)^{3}_R &\quad&  -2 \frac{N_c^4}{N_f^2}  + N_c^2 -1 \nonumber \\
U(1)_R &\quad& - N_c^2 - 1 = -2 N_f (N_f - N_c ) \nonumber \\
U(1)_B^2 U(1)_R &\quad& -2 N_c^2 \nonumber \\
SU(N_f)^3 &\quad& N_c d^{(3)}(N_f) \nonumber \\
SU(N_f)^2 U(1)_B &\quad& N_c d^{(2)}(N_f) \nonumber \\
SU(N_f)^2 U(1)_R &\quad& -\frac{N_c^2}{N_f} d^{(2)}(N_f) \nonumber
\end{eqnarray}

La concordance de ces anomalies représente plus de contraintes que nous n'avions de liberté à fixer les nombres quantiques (\ref{nbquantmagn}).

Jusqu'à présent, nous avons montré que l'on pouvait retrouver les degrés de liberté de la théorie $SU(N_c)$ à haute énergie en fonction de ceux de $SU(N_f-N_c)$ à basse énergie, et que les anomalies étaient les mêmes dans les deux théories. Montrons que les deux sont duales dans le sens où en appliquant deux fois la dualité, on retrouve la théorie électrique.

On part d'une théorie $SU(N_c)$ avec $N_f$ saveurs. En appliquant la dualité une première fois, nous obtenons une théorie $SU(N_f-N_c)$ avec $N_f$ saveurs de quarks et des mésons $M^i_j$ singlets de jauge, décrits par le superpotentiel invariant de jauge $\sim \tr \rl \widetilde q M q \rr$. En appliquant la dualité de nouveau, nous retrouvons une théorie $SU(N_c)$ avec $N_f$ saveurs, mais nous avons deux types de mésons singlets de jauge $M$ et $M'$, et le tout est décrit par le superpotentiel
\be
W \ = \ \tr \left[\rl M - Q {\widetilde Q} \rr M' \right] \quad , \label{Wdoubledual}
\ee
qui a bien une charge $R=2$.

Clairement, les champs $M$ et $M'$ sont massifs. Le champ $M'$ découple de la théorie, et l'on retrouve alors la définition $M^i_j  = Q^i {\widetilde Q}_j$.

Nous avons donc montré que pour $N_c +2 \leqslant N_f \leqslant 3N_c/2$, la théorie magnétique décrite par le groupe de jauge $SU(N_f - N_c)$ et les degrés de liberté (\ref{nbquantmagn}) est duale de la théorie de SUSY-QCD développée au paragraphe \ref{SUSYQCD}.


\section{Expression de la dualité\label{dualite}}


Puisque la théorie duale est libre dans l'infra-rouge, elle possède une échelle
\be
\Lambda_m \ = \ \mu \exp \rl -\frac{8 \pi^2}{(3N - N_f) g^{\prime \, 2}(\mu)} \rr \quad ,
\label{Lambdam}
\ee
qui doit pouvoir être reliée au pôle de Landau $\Lambda$ de la théorie électrique. Au-delà de l'énergie $\Lambda_m$, non seulement la théorie n'est plus perturbative mais surtout la dualité ne s'applique plus.

Le superpotentiel invariant de jauge que l'on peut construire à partir des degrés de liberté magnétiques (\ref{nbquantmagn}) est
\be
W \ = \ \frac{1}{\widehat \Lambda}\tr \rl {\widetilde q} M q \rr \ + m \tr M  \quad ,
\label{Wmagn}
\ee
où $m$ a été prise proportionnelle à l'identité pour plus de simplicité.

La théorie étant perturbative et continue autour de l'origine, le potentiel de K\"ahler peut être développé de manière canonique
\be
K \ = \ \frac{1}{\beta} \tr \rl q^{\dagger} q + {\widetilde q}^{\dagger} {\widetilde q} \rr + \frac{1}{\alpha \al \Lambda \ar^2} \tr M^{\dagger}M \quad .
\label{Kmagn}
\ee

Dans les expressions (\ref{Wmagn}) et (\ref{Kmagn}), nous avons introduit deux échelles d'énergie. En effet, si nous voulons que les mésons soient directement reliés aux superchamps $Q \widetilde Q$ de la théorie électrique, cela veut dire qu'ils ont une dimension canonique $2$. En revanche, telle que nous avons établi la dualité jusqu'ici, à basse énergie, les mésons magnétiques sont des champs fondamentaux et ils ont une dimension $1$. Dans les expressions de $K$ et $W$, nous avons considéré $M$ comme le champ électrique, et il faut donc faire intervenir l'échelle dynamique (\ref{Lambda}).

Les trois échelles $\Lambda$, $\widehat \Lambda$ et $\Lambda_m$ sont reliées entre elles par
\be
\Lambda^{3N_c - N_f} {\Lambda_m}^{3(N_f-N_c)-N_f} \ = \ (-1)^{N_f - N_c} {\widehat \Lambda}^{N_f} \quad .
\label{Lambdas}
\ee
Cette expression traduit le fait que lorsque la théorie originelle devient plus fortement couplée, la théorie duale devient plus faiblement couplée. La phase $(-1)^{N_f - N_c}$ permet de trouver l'habituel signe $-1$ présent lorsqu'on applique une dualité électrique-magnétique.

Les paramètres $\beta$ et $\widehat \Lambda$ ne sont pas déterminés de manière unique. En effet, nous avons la liberté de redéfinir $q , \, \widetilde q \rightarrow aq, \, a\widetilde q$, ce qui modifie les deux contantes en conséquence. Ce changement affecte également $\Lambda_m$ car il induit une anomalie. Enfin, les baryons à haute énergie sont modifiés en $\rl B , \, \widetilde B \rr \rightarrow \rl a^{N_f - N_c} B , \, a^{N_f - N_c} \widetilde B\rr$. En revanche, $M$ ne peut pas être redimensionné car implicitement nous avons considéré que le terme de masse $m$ dans (\ref{Wmagn}) était directement celui de la théorie électrique.

Nous pouvons donc fixer $q$ et $\widetilde q$ de sorte que $\beta = 1$. Mais ce faisant, il est devenu impossible de calculer explicitement $\widehat \Lambda$ et $\Lambda_m$ en fonction des quantités électriques. Inversement, nous pouvons fixer les baryons de sorte que $B = q^{N_f - N_c}$, et $\widetilde B = {\widetilde q}^{N_f - N_c}$, mais nous ne pouvons alors plus déterminer $\beta$. Puisque nous n'allons pas nous intéresser aux baryons, nous choisissons la première option : $\beta = 1$.

Au prochain paragraphe, nous étudions le système suivant
\begin{eqnarray}
&&K \ = \ \tr  |q|^2 \ + \ \tr  |{\widetilde q}|^2 \ + \  \tr  |\Phi|^2 \quad , \nonumber \\
&&W \ = \ h \ \tr ({\widetilde q} \Phi q) - h \mu^2 \ \tr \Phi \quad ,
\label{KWiss} 
\end{eqnarray}
avec $\Phi$ les mésons fondamentaux de la théorie duale. Ils ont donc une dimension $1$, et les mêmes nombres quantiques que $M$ (voir table (\ref{nbquantmagn})).

Le dictionnaire déduit en comparant (\ref{Wmagn}) et (\ref{Kmagn}) avec (\ref{KWiss}) est :
\begin{eqnarray}
\Phi = \frac{M}{\sqrt{\alpha}\Lambda} \quad , \quad h = \frac{\sqrt{\alpha}\Lambda}{\widehat \Lambda} \quad , \quad \mu^2 = - m \widehat \Lambda \quad .
\label{dicco}
\end{eqnarray}
Nous nous servirons de ces équivalences lorsqu'il s'agira de retrouver les vides supersymétriques (\ref{Mij}).


\section{Le modèle Intriligator-Seiberg-Shih\label{ISS}}


On s'intéresse désormais à la théorie magnétique dont le groupe de jauge est $SU(N)$, avec $N = N_f - N_c$ dans le domaine $N_f \geqslant 3N$ (ce qui correspond à $N_f \leqslant 3N_c /2$).


\subsection{Résultats au niveau global\label{ungauged}}


Pour discuter des propriétés du modèle \cite{Intriligator:2006dd}, il est plus aisé de se placer d'abord dans le cas non-jaugé. Le groupe $SU(N)$ fait partie des symétries globales ; les quarks $q = \rl \ffi, \chi \rr$ et les anti-quarks $\widetilde q = \rl \ffitil , \widetilde \chi \rr$ sont dans les représentations $N$ et $\overline{N}$, les mésons $\Phi$ sont singlets. La théorie est décrite par le potentiel de K\"ahler canonique et le superpotentiel (\ref{KWiss}). Le terme $\mu$ brise explicitement la symétrie chirale $SU(N_f)\times SU(N_f)$ en $SU(N_f)$ diagonal.

Les champ auxiliaires $F$ déduits des équations du mouvement sont
\begin{eqnarray}
&&F_{q} = h\ffitil \Phi \quad , \nonumber \\
&&F_{\widetilde q} =  h \Phi \ffi \quad , \label{FISS} \\
&&F_{\Phi} = h \ffitil \ffi - h\mu^2 \unit_{N_f} \quad . \nonumber
\end{eqnarray}
Les champs $\ffi^T$ et $\ffitil$ sont des matrices $N_f \times N$, et le champ scalaire $\Phi$ est une matrice $N_f \times N_f$.
Puisque $N_f \geqslant N$, les $F_{\Phi}$ ne peuvent pas être tous nuls : la supersymétrie est spontanément brisée à la O'Raifeartaigh.

Le potentiel scalaire 
\be V_F \ = \ \al h \ar^2 \left[\, \tr \al \ffitil \Phi \ar^2 + \tr \al \Phi \ffi \ar^2 + \tr \al \ffitil  \ffi - \mu^2 \unit_{N_f} \ar^2 \,\right]
\label{VISS} 
\ee
est minimisé par
\be
\Phi = \begin{pmatrix} 0 & 0 \cr 0 & \Phi_0 \end{pmatrix} \quad , \quad \ffi = \begin{pmatrix} \ffi_0 \cr 0 \end{pmatrix} \quad , \quad \ffitil^T = \begin{pmatrix} \ffitil_0  \cr  0 \end{pmatrix} \quad , \nonumber
\ee
avec $\ffi_0 \ffitil_0 = \mu^2 \unit_{N}$, où $\ffi_0$ et $\ffitil_0$ sont carrées de rang $N$, et $\Phi_0$ est carrée de rang $\rl N_f - N\rr$. Ces vides possèdent une énergie
\be
V_{min} \ = \ \al h^2 \mu^4 \ar \rl N_f - N \rr \quad .
\label{Vmin}
\ee

Les symétries globales sont plus ou moins brisées selon ce que valent $\ffi_0$ et $\ffitil_0$. On fait le choix
\be \Phi_0 = 0 \quad , \quad \ffi_0 = \ffitil_0 = \mu \unit_{N} \quad ,
\label{videISS}
\ee
qui préserve un groupe global maximal
\be
SU(N)_D\times SU(N_f - N)\times U(1)'_B\times U(1)_R \quad . \nonumber
\ee

Pour connaître le spectre de la théorie, développons les champs autour de ces vides selon la paramétrisation suivante
\be
\Phi = \begin{pmatrix} \delta Y & \delta Z \cr \delta \widetilde Z & \ \delta \widehat \Phi \end{pmatrix} \quad , \label{paramVideISS1}
\ee
\be
\ffi = \begin{pmatrix} \mu + \frac{1}{\sqrt{2}}\rl \delta \sigma_+ + \delta \sigma_-\rr \cr \frac{1}{\sqrt{2}}\rl \delta \rho_+ + \delta \rho_-\rr  \end{pmatrix} \quad , \quad \ffitil^T = \begin{pmatrix} \mu + \frac{1}{\sqrt{2}}\rl \delta \sigma_+ - \delta \sigma_-\rr \cr \frac{1}{\sqrt{2}}\rl \delta \rho_+ - \delta \rho_-\rr  \end{pmatrix} \quad ,
\label{paramVideISS2}
\ee
qui traduit le fait que les vides (\ref{videISS}) sont invariants par conjugaison de charge entre $\ffi$ et $\ffitil^T$.

Au niveau classique, le développement du potentiel (\ref{VISS}) implique les champs $\delta Y$ et $\delta \sigma_+$ ont une masse $\sqrt{2}\al h \mu \ar$, tandis que $\delta Z$, $\delta \widetilde Z$, $\text{Im} \rl \mu^* \delta \rho_+ \rr / \al \mu \ar$ et $\text{Re} \rl \mu^* \delta \rho_- \rr / \al \mu \ar$ ont une masse $\al h \mu \ar$.

Les champs non-massifs sont, pour certains d'entre eux, les bosons de Goldstone des symétries globales brisées par (\ref{videISS}),
\be
\frac{\mu^* \delta \sigma_- - {\rm h.c.}}{\al \mu \ar} \quad , \quad  \frac{\text{Re} \rl \mu^*\delta \rho_+ \rr}{\al \mu \ar} \quad , \quad  \frac{\text{Im} \rl \mu^*\delta \rho_-  \rr}{\al \mu \ar} \quad . \nonumber
\ee

Les autres champs non-massifs au niveau des arbres acquièrent une masse à une boucle grâce à leurs interactions avec les champs massifs. Il s'agit de $\delta \widehat \Phi$ et $\delta \widehat \sigma = \rl \mu^* \delta \sigma_- + {\rm h.c.}\rr / \al \mu \ar$. Le potentiel effectif à une boucle de Coleman-Weinberg \cite{Coleman:1973jx} est donné par
\be
V^{(1)}_{eff} = \frac{1}{64 \pi^2} \rl \tr \, m_B^4 \ln \frac{m_B^2}{\Lambda^2} - \tr \, m_F^4 \ln \frac{m_F^2}{\Lambda^2} \rr \quad ,
\label{Veff1}
\ee
où $m_B$ et $m_F$ sont les matrices de masses bosonique et fermionique, exprimées en fonction des champs. Autrement dit, ce que l'on appelle masses pour les bosons, ce sont en réalité les dérivées secondes $\displaystyle{\frac{\dd^2 V}{\dd z^i \dd \zb_j}}$ exprimées en fonction des champs scalaires $z^i$. Pour les fermions, il s'agit des dérivées $\displaystyle{\al \frac{\dd W}{\dd z^i}\ar^2}$. Les termes quadratiques de ces dérivées, qui donnent les masses à une boucle, proviennent des couplages quartiques à l'ordre des arbres\footnote{La dépendance de (\ref{Veff1}) en $\Lambda$ peut être réabsorbée dans $h$.}.

On trouve que le potentiel (\ref{Veff1}) a la forme
\be
V^{(1)}_{eff} = h^4 \al \mu^2 \ar \rl \frac{a}{2} \tr \, \delta {\widehat \sigma}^2 + b \tr \, \delta {\widehat \Phi}^{\dagger} \delta {\widehat \Phi} \rr \quad , \nonumber
\ee
avec $$a = \frac{\ln 4 - 1}{8 \pi^2}(N_f - N) \quad \text{et} \quad b= \frac{\ln 4 - 1}{8 \pi^2}N$$ tous deux positifs.

Les vides (\ref{videISS}) sont donc stables. En effet, nous n'avons trouvé aucune direction tachyonique à une boucle, et puisque la théorie est perturbative, les contributions à deux boucles n'excèdent pas celles à une boucle.

Le spectre autour de ces vides se compose de particules de masses $\sim h \al \mu \ar$ au niveau classique, de particules qui acquièrent une masse $\sim h^2 \al \mu \ar$ à une boucle, et de particules sans masse. Ces dernières sont les bosons de Goldstone des symétries globales qui ont été brisées, mais notons aussi la présence d'un Goldstino provenant de la brisure spontanée de la supersymétrie.

La question que nous nous posons dans la suite est de savoir comment ces vides survivent lorsque l'on jauge la théorie.


\subsection{Restauration dynamique de la supersymétrie\label{gauged}}


On rajoute le secteur de jauge $\rl \lambda^A, A_{\mu}^A \rr$, avec $A=1,\ldots,N^2-1$.

Le potentiel scalaire devient $V = V_F + V_D$, où $V_F$ est le potentiel (\ref{VISS}), et $V_D$ est donné par
\be
V_D \ = \ \frac{1}{2}g^{\prime \, 2} \sum_A \left[\ \tr \, \varphi^{\dagger} T^{A} \varphi \ - \ \tr \, {\widetilde \varphi} T^{A} {\widetilde \varphi}^{\dagger} \ \right]^2 \quad .
\label{VDISS}
\ee

Il est clair que ce potentiel est nul dans les vides non-supersymétriques (\ref{videISS}). La nouveauté réside dans le fait que les valeurs moyennes non-nulles des squarks brisent intégralement le groupe de jauge. Le mécanisme de super-Higgs a lieu et les bosons de jauge acquièrent une masse $g\mu$ en absorbant les scalaires $\text{Im}' \rl \mu^* \delta \sigma_- \rr / \al \mu \ar$ où le prime signifie la partie sans trace. Grâce au potentiel (\ref{VDISS}), les scalaires $\text{Re}'\rl \mu^* \delta \sigma_- \rr / \al \mu \ar $, qui représentent la partie sans trace de $\delta \widehat \sigma$, acquièrent une masse $g\mu$ au niveau des arbres. Les champs $\delta \widehat \Phi$ et $\tr \, \delta \widehat \sigma$ demeurent non-massifs au niveau classique.

Ces vides sont toujours stables à une boucle car le secteur de jauge n'est pas affecté par la brisure de supersymétrie. En effet, les valeurs moyennes (\ref{videISS}) des champs $\ffi$ et $\ffitil$, qui donnent leur masse aux bosons de jauge, ne couplent à aucun des termes $F$ non-nuls, et comme nous l'avons vu, les termes $D^A$ du potentiel (\ref{VDISS}) sont nuls dans ces vides. Il s'ensuit que la supertrace apparaissant dans le potentiel effectif (\ref{Veff1}) s'annule par supersymétrie pour le secteur de jauge.

Le calcul de l'indice de Witten, section \ref{indice}, implique que $N$ vides supersymétriques sont présents dans la théorie. Supposons qu'il existe une région de l'espace des champs où les mésons $\Phi$ ont des valeurs moyennes non-nulles dans le vide. D'après le potentiel (\ref{VISS}), les quarks $\ffi$ et les anti-quarks $\ffitil$, c'est-à-dire la matière, ont une masse $\langle h \Phi \rangle$. En dessous de cette échelle, ils découplent de la théorie qui devient alors super-Yang-Mills pure avec des mésons singlets de jauge. Elle est asymptotiquement libre et admet une échelle dynamique
\be
\widetilde \Lambda \ = \ E \, \exp \rl -\frac{8 \pi^2}{3N g^{\prime \, 2}(E)} \rr \quad .
\label{Lambdatilde}
\ee

Le couplage de jauge $g'$ doit être continu lorsque les quarks découplent. En utilisant (\ref{Lambdam}) à l'échelle $\langle h \Phi \rangle$, ceci implique
\be
{\widetilde \Lambda}^{3N} \ = \ \frac{h^{N_f} \det \Phi}{\Lambda_m^{N_f-3N}} \quad . \nonumber
\ee

À basse énergie, les gluinos $SU(N)$ condensent, et l'on trouve un superpotentiel engendré dynamiquement
\be
W_{dyn} \ = \ N \rl \frac{h^{N_f} \det \Phi}{\Lambda_m^{N_f-3N}} \rr - h\mu^2 \tr \, \Phi \quad ,
\label{Wdynmagn}
\ee
où nous avons utilisé la relation (\ref{Lambdatilde}). Ce résultat est très similaire à celui que nous avions trouvé dans la théorie électrique (\ref{Wdyn}). La différence de signe, encore une fois, est essentielle pour nous permettre de retrouver le bon signe en appliquant la dualité. La minimisation du potentiel (\ref{Wdynmagn}) donne $N$ vides supersymétriques
\be
\langle h \Phi \rangle = \Lambda_m \epsilon^{2N/(N_f-N)} \unit_{N_f} \quad , \quad \text{avec} \quad \epsilon = \frac{\mu}{\Lambda_m} \quad .
\label{hphi}
\ee
Pour $\al \epsilon \ar \ll 1$, on trouve une hiérarchie
\be
\al \mu \ar \ll \langle h \Phi \rangle \ll \Lambda_m \quad , \nonumber
\ee
ce qui implique que l'analyse faite ci-dessus est valide puisque nous sommes bien en-dessous du pôle de Landau.
Notons que le résultat (\ref{hphi}) peut aussi s'écrire
\be
\langle h \Phi \rangle = \mu \epsilon^{-(N_f-3N)/(N_f-N)} \unit_{N_f} \quad . \nonumber
\ee

En utilisant les relations établies dans le paragraphe précédent ainsi que la relation liant les échelles des deux théories, équations (\ref{Lambdas}) et (\ref{dicco}), et bien sûr le fait que $N = N_f - N_c$, on retrouve le résultat (\ref{Mij})
\be
\langle M \rangle = \rl \Lambda^{3N_c - N_f} \det m \rr^{1/N_c} m^{-1} \unit_{N_f} \nonumber
\ee
dans le cas d'une matrice de masse diagonale.

\begin{figure}[ht!]
\begin{center}
\includegraphics[scale=0.5]{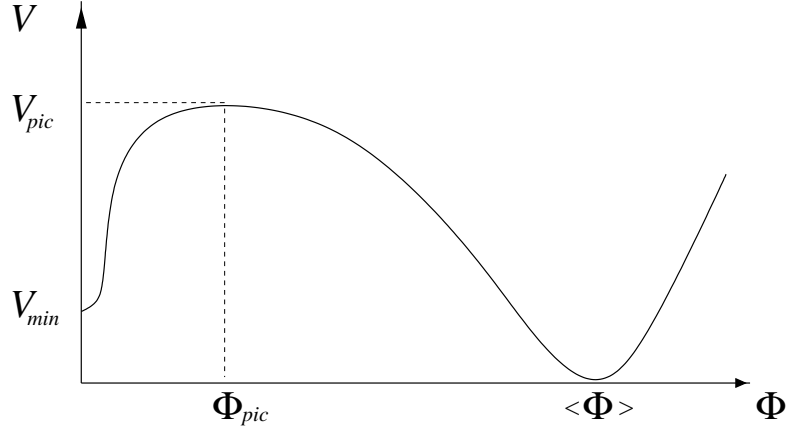}
\caption{Comportement du potentiel total $V = V_F + V_D$ dans la direction des mésons. La contribution $V_F$ est donnée par (\ref{VISS}), et celle de $V_D$ est donnée par (\ref{VDISS}).}
\label{metastable}
\end{center}
\end{figure}

Les résultats que nous venons de dériver semblent indiquer qu'il y a une restauration dynamique de la supersymétrie alors qu'elle est brisée spontanément au niveau classique dans une autre région de l'espace des champs. Ceci a pour conséquence principale de rendre les vides (\ref{videISS}) métastables c'est-à-dire localement stables. En effet, nous avons montré que ces derniers sont stables en ce sens que toutes les masses y sont réelles, mais ils ont une énergie positive puisqu'ils brisent la supersymétrie et la théorie peut aller dans les vides (\ref{hphi}) par effet tunnel. Dans la direction $\Phi$, il existe une barrière de potentiel entre l'origine et la valeur $\langle \Phi \rangle$ comme représenté schématiquement sur la Fig. \ref{metastable}.
L'éloignement entre les deux minima nous empêche de calculer la durée de vie des vides métastables par une approximation ``thin-wall''. En modélisant la barrière par un triangle de hauteur $V_{min}$ et de base $\langle \Phi \rangle$, on obtient l'action ``bounce''
\be
S_{bounce} \ \sim \ \frac{\langle \Phi \rangle^4}{V_{min}} \ \sim \ \al \epsilon \ar^{\frac{4(N_f - 3N)}{N_f-N}} \quad ,
\label{Sbounce}
\ee
qui est reliée à la probabilité $\Gamma$ de transition des vides non-supersymétriques vers les vides (\ref{hphi}) par
\be
\Gamma \ \sim \ e^{-S_{bounce}} \quad . \nonumber
\ee

Plus $\epsilon$ est petit devant $1$ et plus les vides métastables ont une durée de vie longue. Faire tendre $\epsilon \rightarrow 0$ signifie faire tendre $\Lambda_m \rightarrow \infty$ en gardant $\mu$ fixé. Ceci envoie les solutions $\langle \Phi \rangle$ vers l'infini.

Comment se traduisent ces résultats en termes des paramètres de la théorie électrique ? En utilisant les égalités (\ref{dicco}), l'équivalent de $\epsilon$ dans la théorie à haute énergie est $\sqrt{m/\Lambda}$. De même, allonger la durée de vie des vides métastables revient à faire tendre $m \rightarrow 0$ en gardant $\Lambda$ fixé.


\chapter{Les champs de modules et leur stabilisation\label{modules}}


Dans ce chapitre, nous abordons la phénoménologie des théories de cordes. Selon le type de cordes que l'on considère, et les conditions aux limites qu'on lui applique, on obtient différentes actions effectives à dix dimensions. L'espace-temps est un produit $\mathcal M \times I_6$ où $\mathcal M$ est l'espace-temps minskowskien à quatre dimensions, et $I_6$ est l'espace compact à six dimensions. De manière générale, $I_6$ est une variété de Calabi-Yau qui a pour propriété principale de préserver une supersymétrie $\mathcal N = 1$ à quatre dimensions. Lorsque l'on réduit la théorie (dans notre cas, la corde de type IIB) de dix à quatre dimensions, on peut obtenir un très grand nombre de théories différentes selon les symétries que l'on impose aux champs. À la suite de ce processus, des champs supplémentaires, appelés \textit{modules}, sont présents dans la théorie effective (voir aussi le chapitre \ref{dimsup}). Généralement, les modules n'ont pas de potentiel, ils correspondent à des directions plates. Le but de ce chapitre est de présenter certains mécanismes développés ces quatre dernières années pour leur générer un potentiel avec minimum : c'est la stabilisation des modules.


\section{Réduction dimensionnelle et identification des modules\label{dimreduc}}


Dans cette section, nous effectuons la compactification (minimale) des six dimensions supplémentaires pour identifier les modules présents dans la théorie quadri-dimensionnelle.

Le secteur gravitationnel comprend le tenseur métrique $g_{MN}$, le tenseur antisymétrique $B_{MN}$ et le scalaire réel $\phi$, appelé dilaton. Nous n'expliciterons pas le secteur fermionique du spectre car il s'obtient par supersymétrie. On désigne par $M,N = 0, \ldots, 9$ les indices de l'ensemble des dix dimensions. Les indices $m, n = 0, \ldots, 3$ correspondent à notre espace-temps, et $I, J = 4, \ldots, 9$ sont les indices des six dimensions compactes. L'action du secteur gravitationnel est
\be
\mathcal S = \frac{M_S^8}{2} \int d^{10}x \sqrt{g_S^{(10)}}  e^{-2\phi}\pl R_S^{(10)}+4 \rl \nabla_M\phi\rr^2 - \frac{1}{12} H_{MNP}H^{MNP}\pr \ ,
\label{LgravString}
\ee
où $H_{MNP} = \dd_{M} B_{NP}$, avec les indices antisymétrisés, est le tenseur champ associé à $B$, et $M_S^8$ est l'échelle de masse des cordes.

Dans l'expression (\ref{LgravString}), $g_{S}^{(10)}$ est la métrique dans le référentiel des cordes, et l'action $\sqrt{g}R$ porte un facteur dépendant du dilaton. Nous devons effectuer une transformation de Weyl sur la métrique pour retrouver le référentiel d'Einstein de métrique $g_E$, dans lequel l'action d'Einstein-Hilbert est canonique. Sous la transformation
\be
\rl g_S^{(10)}\rr_{MN} \ = \ e^{\phi/2} \rl g_E^{(10)}\rr_{MN} \quad ,
\label{1eretransfWeyl}
\ee
le scalaire de Ricci devient
\be
R_S^{(10)} \ = \ e^{-\phi/2} \left[R_E^{(10)} - \frac{9}{2} \rl \nabla_M \phi \rr^2 + \ldots \right] \quad ,
\nonumber
\ee
où $\ldots$ désigne des termes s'annulant après intégration par partie dans l'action (\ref{LgravString}). Le déterminant de la métrique est $\sqrt{g_S^{(10)}} = e^{5\phi/2}\sqrt{g_E^{(10)}}$. Par ailleurs, il faut aussi transformer les métriques inverses intervenant dans le produit $H_{(3)}\cdot H_{(3)}$ et dans le terme cinétique du dilaton.

L'action dans le référentiel d'Einstein est donc
\be
\mathcal S = \frac{1}{2\kappa_{10}^2} \int d^{10}x \sqrt{g_E^{(10)}} \pl R_E^{(10)}-\frac{1}{2} \rl \dd_M\phi\rr^2 - \frac{e^{-\phi}}{12} H_{MNP}H^{MNP}\pr \quad ,
\label{LgravEinstein}
\ee
où $\kappa_{10}$ est l'échelle de Planck à dix dimensions.

Les six dimensions compactes forment une variété de Calabi-Yau dont on ne connaît pas la forme exacte. Pour pouvoir réduire le nombre de dimensions à quatre, il faut faire des hypothèses sur le comportement des champs \cite{Witten:1985xb}. Les hypothèses minimales que l'on fait sont que les champs qui survivent à la réduction dimensionnelle sont invariants par translation et par certaines rotations des coordonnées supplémentaires $y^I$. La première condition préserve une supergravité $\mathcal N = 4$. La deuxième condition permet d'éliminer trois des supersymétries pour obtenir le cas $\mathcal N =1$ qui nous intéresse. En effet, le groupe de rotations des six coordonnées $y^I$ est $O(6)$, qui est isomorphe à $SU(4)$. Un sous-groupe de $SU(4)$ est $SU(3)$ ; nous imposons que les champs soient invariants sous les rotations dans ce $SU(3)$ des $y^I$. Clairement, les champs $g^{(10)}_{mI}$ et $B_{mI}$ ne sont pas invariants. La métrique $g^{(10)}_{IJ}$ doit s'écrire $e^{2u}\delta_{IJ}$ où $u$ est un champ scalaire réel. Le tenseur antisymétrique, de même, prend la forme $B_{IJ} = a \epsilon_{IJ}$ avec $a$ un pseudo-scalaire. Le reste des champs, $g^{(10)}_{mn}$, $B_{mn}$ et le dilaton sont des scalaires du point de vue quadri-dimensionnel, et survivent donc à la projection.

L'opération décrite ci-dessus implique que le référentiel d'Einstein à dix dimensions ne correspond pas à celui à quatre dimensions. Nous devons donc modifier la métrique $g^{(10)}_{mn}$ de sorte à retrouver un terme d'Einstein-Hilbert canonique. Ceci constitue une nouvelle transformation de Weyl
\be
g^{(10)}_{IJ} \ = \ e^{2u}\delta_{IJ} \quad \quad , \quad \quad g^{(10)}_{mn} = e^{-6 u} g^{(4)}_{mn} \quad ,
\label{2emetransfWeyl}
\ee
Cette transformation fait apparaître un terme cinétique pour le champ $u$. Finalement, l'action à quatre dimensions contient
\begin{eqnarray}
\mathcal S &=& \frac{1}{2\kappa_{4}^2} \int d^{4}x \sqrt{g_E^{(4)}} \pl R_E^{(4)}-24 \rl \dd_m u \rr^2 - \frac{1}{2} \rl \dd_m\phi\rr^2 \right. \nonumber \\
&&\left. \quad \quad \quad \quad - \frac{e^{-\phi}e^{12 u}}{12} H_{mnp}H^{mnp} - \frac{3e^{-\phi}e^{-4u}}{2} \rl \dd_m a \rr^2 \pr \ ,
\label{LgravEinstein4D}
\end{eqnarray}
où $\kappa_4^2 = \kappa_{10}^2 / V_6$ est la masse de Planck et $V_6$ le volume de l'espace compact.

Nous nous intéressons maintenant à la réduction dimensionnelle de la théorie IIB, qui contient des cordes fermées et inclut la chiralité \cite{Ferrara:1986qn, Giddings:2001yu}. Elle induit à basse énergie la présence de formes de rang pair $C_{(0)}, \, C_{(2)}, \, C_{(4)}, \, C_{(6)}, \, C_{(8)}, \, C_{(10)}$ où nous utilisons le langage des formes différentielles. La forme $C_{(0)}$ est un axion. Les formes $C_{(2n)}$ et $C_{(8-2n)}$ sont reliées par dualité et donc on ne considère généralement que les trois premières de ces formes. En particulier le tenseur champ de jauge $F_{(5)} = dC_{(4)}$ est auto-dual : $F_{(5)} = \widetilde F_{(5)}$.

Pour l'instant, nous incluons seulement le terme cinétique de l'axion
\be
\mathcal S =  \frac{M_S^8}{2} \int d^{10}x \sqrt{g_S^{(10)}} \pl - \frac{1}{2} \rl \dd_M C_{(0)} \rr^2 \pr \quad ,
\nonumber
\ee
qui devient
\be
\mathcal S =  \frac{1}{2\kappa_{10}^2} \int d^{10}x \sqrt{g_E^{(10)}} \pl - \frac{e^{2\phi}}{2} \rl \dd_M C_{(0)} \rr^2 \pr
\label{LC0Einstein}
\ee
dans le référentiel d'Einstein. L'axion survit à la réduction dimensionnelle, et de plus, son terme cinétique (\ref{LC0Einstein}) n'est pas modifié par la transformation (\ref{2emetransfWeyl}). À quatre dimensions, on a simplement
\be
\mathcal S =  \frac{1}{2\kappa_{4}^2} \int d^{4}x \sqrt{g_E^{(4)}} \pl - \frac{e^{2\phi}}{2} \rl \dd_m C_{(0)} \rr^2 \pr \quad ,
\label{LC0Einstein4D}
\ee
qui vient s'ajouter à l'action (\ref{LgravEinstein4D}).

La théorie étant supersymétrique, les termes cinétiques doivent s'écrire $-\sqrt{\det g} \, K_{i \bar j}\dd_m z^i \dd^m \zb^{\bar j}$ (voir section \ref{sugra}). Une partie de ces champs scalaires complexes $z^i$ est formée par les pseudo-scalaires (parties imaginaires) et les scalaires (parties réelles) issus de la réduction dimensionnelle ; il s'agit des modules que nous avons évoqués.

Le dilaton complexe $S$, aussi appelé axion-dilaton, est déterminé à partir de sa partie imaginaire $C_{(0)}$. On voit d'après (\ref{LC0Einstein4D}) que la métrique de K\"ahler dans la direction $S$ est
\be
K_{S \overline{S}} \ = \ \frac{e^{2\phi}}{4} \quad , \label{KSS}
\ee
et ce doit aussi être le facteur du terme cinétique de $\text{Re}\, S$. Cette expression suggère que celle-ci a la forme $\text{Re} \, S = a_1 e^{\alpha_1 \phi}$. En comparant avec le terme $\rl \dd_m \phi \rr^2$ dans (\ref{LgravEinstein4D}), on trouve $\alpha_1 = -1$ et $a_1^2 \alpha_1^2 = 1$. Nous en concluons que le module $S$ est
\be
S \ = \  e^{-\phi} + i C_{(0)} \quad  .
\label{dilaton}
\ee
La métrique de K\"ahler (\ref{KSS}) se réécrit
\be
K_{S \overline{S}} \ = \ \frac{1}{4\rl \, \text{Re}\, S \, \rr^2} \ = \ \frac{1}{\rl \, S + \overline{S} \, \rr^2} \quad , \nonumber
\ee
ce qui permet de remonter au potentiel de K\"ahler
\be
K(S, \overline{S}) \ = \ -\ln \rl \, S + \overline{S} \, \rr \quad ,
\label{Kdilaton}
\ee
où $S$ désigne ici le superchamp chiral associé au scalaire (\ref{dilaton}).

Le module de K\"ahler $T$ a pour partie réelle une fonction du deuxième scalaire à notre disposition, $u$, qui est le poids de Weyl de la transformation (\ref{2emetransfWeyl}). Le moyen le plus simple de voir la forme de ce module est d'utiliser les branes. Les $Dp$-branes sont des objets étendus vivant à $p+1$ dimensions d'espace-temps et qui supportent les extrêmités des cordes ouvertes. Ce sont des objets non-perturbatifs (solitons) dans la description supergravité\footnote{Les branes sont des objets perturbatifs du point de vue des cordes.}, qui possèdent une tension positive et une charge pouvant être positive ou négative, ce dernier cas correspondant aux anti-$Dp$-branes. Le couplage d'une théorie de jauge pure sur une $D7$-brane (qui contient $4$ dimensions compactes) s'écrit
\be
\mathcal S \ = \ \int_{D7} d^8 x \sqrt{\gamma} F_{MN} F^{MN} \quad , \nonumber
\ee
où $\gamma$ est la métrique induite sur la brane dans le référentiel d'Einstein, et les indices $M, N$ prennent les valeurs $0, \ldots, 7$. Sous la transformation (\ref{2emetransfWeyl}), on obtient l'action effective
\be
\mathcal S \ = \ \int d^4 x \sqrt{g_E^{(4)}} e^{4u} F_{mn} F^{mn} \quad . \nonumber
\ee
Le couplage d'une théorie de jauge sur une $D7$-brane est donc $g^2 = e^{-4u}$.

L'action étant hermitienne, si on définit le module $T$ par sa partie réelle $e^{4u}= \text{Re}\, T$, alors sa partie imaginaire doit coupler au terme $F_{mn} \widetilde F^{mn}$ avec $\widetilde F^{mn} = \epsilon^{mnpq}F_{pq}$ le dual du tenseur champ de jauge. Le couplage général de ces termes se déduit de l'action de Wess-Zumino sur une $D7$-brane
\be
\mathcal S_{WZ} \ = \ \int_{D7} C \wedge e^F \quad , \label{SWZ}
\ee
où l'écriture $C\wedge e^F$ représente la somme
\be
C\wedge e^F = \sum C_{(2n)}\wedge F \ldots \wedge F  = \sum \epsilon^{n_1 \ldots n_{2l}pq\ldots rs}C_{n_1 \ldots n_{2l}} \ F_{pq} \ldots \ F_{rs} \quad ,
\label{CwedgeexpF}
\ee
de sorte que le nombre d'indices du développement vaut $8$, ce qui fixe le rang $2l$. Le terme qui nous intéresse est $C_{(4)} \wedge F \wedge F$. Plus exactement, pour retrouver le terme $F_{mn} \widetilde F^{mn}$ à quatre dimensions, nous nous intéressons seulement à\footnote{Pour plus de simplicité dans ce qui suit, nous ne tenons pas compte des facteurs numériques.}
\be
C_{(4)} \wedge F \wedge F \supset \epsilon^{I_1I_2I_3I_4m_1m_2m_3m_4}C_{I_1I_2I_3I_4} F_{m_1m_2} F_{m_3 m_4} \quad . \nonumber
\ee

Nous pouvons toujours écrire $F_{m_1m_2} F_{m_3 m_4} = \dd_{m_1} \omega_{m_2m_3m_4}$ puisque les indices sont antisymétrisés par le tenseur $\epsilon$ en facteur, donc l'action qui nous intéresse est
\begin{eqnarray}
\mathcal S &=& \int_{D7} d^{8} x \sqrt{\gamma} \ \epsilon^{I_1I_2I_3I_4m_1m_2m_3m_4}\ C_{I_1I_2I_3I_4} \  \dd_{m_1} \omega_{m_2m_3m_4} \nonumber \\
&=& -  \int_{D7} d^{8} x \sqrt{\gamma}\ \epsilon^{I_1I_2I_3I_4m_1m_2m_3m_4} \rl \dd_{m_1}C_{I_1I_2I_3I_4} \rr \ \omega_{m_2m_3m_4} \quad , \nonumber 
\end{eqnarray}
après intégration par parties. En utilisant l'auto-dualité de $dC_{(4)}$, on trouve
\be
\mathcal S = - \int_{D7} d^{8} x \sqrt{\gamma}\ \epsilon^{I_1I_2I_3I_4m_1m_2m_3m_4}\ \epsilon_{m_1I_1I_2I_3I_4}^{n_1n_2n_3I_5I_6}\ F_{n_1n_2n_3I_5I_6}\ \omega_{m_2m_3m_4} \quad .\nonumber
\ee

Finalement, en réarrangeant les tenseurs antisymétriques, cette expression contient
\be
\mathcal S = - \int_{D7} d^{8} x \sqrt{\gamma}\ \delta^{m_2n_1}\delta^{m_3n_2}\delta^{m_4n_3}\delta^{i \bar j}\ F_{n_1n_2n_3i \bar j}\ \omega_{m_2m_3m_4} \quad ,\nonumber
\ee
où nous nous sommes servi des coordonnées complexes $z_i = \frac{1}{\sqrt{2}}\rl Y_I + i Y_{I+1}\rr$. Nous remarquons que $F_{n_1n_2n_3i \bar j}= \epsilon_{n_1 n_2 n_3 p}\dd^p b_{i \bar j}$. Pour effectuer la réduction dimensionnelle, il faut que $b_{i \bar j} = b \, \delta_{i \bar j}$ avec $b$ pseudo-scalaire. On obtient l'action
\be
\mathcal S = \int d^{4} x \sqrt{g_E^{(4)}}\ b \ \epsilon_{mnpq} \ F^{m n} F^{p q} \quad .\nonumber
\ee

Nous avons donc identifié le module de K\"ahler
\be
T \ = \ e^{4u} + i b \quad .
\label{moduleT}
\ee
Pour trouver le potentiel de K\"ahler $K(T, \overline{T})$, nous procédons comme pour le dilaton, en utilisant le terme cinétique de $u$ dans (\ref{LgravEinstein4D}), et le terme cinétique de $b$, contenu dans l'action
\be
\mathcal S = - \frac{1}{2\kappa_{10}^2} \int d^{10}x \sqrt{g_E^{(10)}} \ \frac{F_{(5)}^2 }{4} \quad . \nonumber
\ee
Ce terme est invariant de Weyl, c'est pourquoi nous l'avons directement exprimé dans le référentiel d'Einstein. Nous ne considérons que $F_{mnpi \bar j}F^{mnpi \bar j}$ car il donne directement le terme cinétique de $b$. En effectuant la deuxième transformation de Weyl (\ref{2emetransfWeyl}) sur ce dernier, on trouve
\be
\mathcal S = - \frac{1}{2\kappa_{4}^2} \int d^{4}x \sqrt{g_E^{(4)}} \ \frac{3}{2} e^{-8u} \rl \dd_m b \rr^2 \quad . \nonumber
\ee
Par ailleurs, en utilisant l'action (\ref{LgravEinstein4D}), le terme cinétique du scalaire $u$ peut se réécrire
\be
\mathcal S = - \frac{1}{2\kappa_{4}^2} \int d^{4}x \sqrt{g_E^{(4)}} \ \frac{3}{2} e^{-8u} \rl \dd_m e^{4u} \rr^2 \quad . \nonumber
\ee
Ceci nous prouve que l'expression (\ref{moduleT}) du module de K\"ahler\footnote{Le module $T$ est souvent appelé ``volume modulus'' car sa partie réelle est directement reliée au volume des dimensions supplémentaires.} est la bonne. La métrique associée se déduit des deux actions ci-dessus
\be
K_{T \overline{T}} \ = \ \frac{3e^{-8u}}{4} \ = \ \frac{3}{\rl \, T + \overline{T} \, \rr^2} \quad , \nonumber
\ee
et le potentiel de K\"ahler est donc
\be
K(T, \overline{T}) \ = \ -3\, \ln \rl \, T + \overline{T} \, \rr \quad .
\label{KmoduleT}
\ee

Nous avons montré que toute théorie de jauge vivant sur une $D7$-brane a un couplage $\text{Re}\, T$ qui ne dépend pas du dilaton. Montrons que l'autre cas existe aussi. Pour cela, considérons une théorie de jauge sur une $D3$-brane, c'est-à-dire une brane pouvant décrire notre espace-temps. Le couplage s'écrit
\be
\mathcal S_{WZ} \ = \ \int_{D3} C \wedge e^F \quad . \nonumber
\ee
En développant le produit $C\wedge e^F$, on trouve que le facteur du terme $F \widetilde F$ est $C_{(0)}$. Il en résulte que le terme cinétique de jauge $F\cdot F$ aura pour facteur la partie réelle associée à $C_{(0)}$, soit $\text{Re}\, S$. Le couplage d'une théorie de jauge vivant sur une $D3$-brane est donc $g^2 = e^{\phi}$.

La réduction dimensionnelle que vous venons d'effectuer est minimale dans le sens des conditions d'invariance par translation et rotation que nous avons imposées. Mais, si l'espace interne est plat, les modules sont des directions plates du potentiel effectif à quatre dimensions. C'est un réel problème non-seulement parce que le spectre contient des modes non-massif, mais parce que les couplages de jauge vivant sur les branes ne seront physiques que si les modules ont des valeurs moyennes, de préférence ni nulles ni infinies.


\section{Superpotentiel non-perturbatif\label{stabilisationTparjauginos}}


Dans ce paragraphe, nous allons utiliser les phénomènes non-perturbatifs étudiés dans la section \ref{SUSYQCD} pour générer un potentiel au module de K\"ahler $T$ \cite{Dine:1985rz}.

Pour cela, considérons une théorie de jauge SUSY-QCD avec $N_f$ saveurs et $N_c$ couleurs située sur une $D7$-brane. Le couplage de jauge à la masse de Planck est $g^2(M_P)=1/\rl \text{Re}\, T\rr$. Nous avons vu qu'à basse énergie, le superpotentiel (\ref{Wdyn}) est généré dynamiquement par la condensation des gluinos. Sans perte de généralité, nous redéfinissons les champs $(\psi, {\widetilde \psi}, M)$ en $(i\psi, i{\widetilde \psi}, -M)$ de façon à changer le signe global du superpotentiel, et nous omettons la phase venant de la valeur moyenne des jauginos :
\be
W_{dyn} \ = \ \rl N_c - N_f \rr \ \rl \frac{\Lambda^{3N_c - N_f}}{\det M}\rr^{\frac{1}{N_c-N_f}} + \tr \ mM \quad .
\label{Wdyn+}
\ee

Dans cette expression, l'échelle dynamique $\Lambda$ est directement reliée au module de K\"ahler $T$ à travers l'expression (\ref{Lambda}) pour $\mu = M_P$
\be
\Lambda \ = \ M_P \exp \rl -\frac{8 \pi^2 \, \text{Re}\, T}{(3N_c - N_f)} \rr \quad . \nonumber
\ee
Le superpotentiel est une fonction holomorphe des superchamps, or la partie réelle de $T$ ne l'est pas. Nous exprimons donc $\Lambda$ directement en fonction de $T$
\be
\Lambda \ = \ M_P \exp \rl -\frac{8 \pi^2 \, T}{(3N_c - N_f)} \rr \quad . \nonumber
\ee
Le superpotentiel (\ref{Wdyn+}) devient
\be
W_{dyn} \ = \ \rl N_c - N_f \rr \ \rl \frac{e^{-8 \pi^2 \, T}}{\det M}\rr^{\frac{1}{N_c-N_f}} + \tr \ mM \quad .
\label{WdynenfctdeT}
\ee

Si nous découplons toutes les saveurs de mésons, c'est-à-dire que nous supposons que toutes les masses $m_i$ sont très grandes, nous obtenons un superpotentiel qui ne dépend plus que du module
\be
W(T) \ = \ N_c \  \exp \rl \frac{-8 \pi^2 \, T}{N_c} \rr \quad .
\label{WdeT}
\ee

Ce résultat peut être obtenu directement en reprenant le superpotentiel (\ref{Wan}) sans saveurs, $N_f=0$ et $\det M = 1$, et en condensant les jauginos, $U \sim \Lambda^3$ ; on retrouve bien l'expression (\ref{WdeT}).

La théorie SUSY-QCD sur la $D7$-brane agit comme un secteur caché. La valeur moyenne dans le vide de $T$ donne le couplage de jauge du secteur caché. Malheureusement, le superpotentiel (\ref{WdeT}) et le potentiel de K\"ahler (\ref{KmoduleT}) ne suffisent pas à stabiliser le module car ils n'engendrent pas de potentiel avec un minimum.


\section{Stabilisation des modules par les flux}


Les flux sont les composantes dans les directions compactes des tenseurs champs de jauge. Par exemple, pour $H_{MNP} = \dd_M B_{NP}$, il s'agira des composantes $H_{IJK}$ avec $I, J, K = 4, \ldots, 9$. L'intégrale de ces champs sur un sous-espace à trois dimensions est quantifiée (quantification de Dirac).


\subsection{Un exemple simple}


L'action effective de la théorie IIB dans le référentiel des cordes contient le terme
\be
\mathcal S = \frac{M_S^8}{2} \int d^{10}x \sqrt{g_S^{(10)}}  \rl - \frac{1}{12} G_{(3)}\overline{G}_{(3)}\rr \quad ,
\label{exemplesimpleString}
\ee
où $G_{(3)} = F_{(3)} + i S H_{(3)}$, avec $F_{(3)}=dC_{(2)}$ le tenseur champ de jauge de la forme de rang $2$, et $H_{(3)}=dB_{(2)}$ que nous connaissons déjà. Le dilaton $S$ présent devant $H_{(3)}$ prend en compte le facteur $e^{-2\phi}$ dans l'action (\ref{LgravString}). Sous la première transformation de Weyl (\ref{1eretransfWeyl}) qui permet de retrouver le référentiel d'Einstein, l'action (\ref{exemplesimpleString}) se réécrit
\begin{eqnarray}
\mathcal S &=& - \frac{1}{2\kappa_{10}^2} \int d^{10}x \sqrt{g_E^{(10)}} \, \frac{e^{\phi}}{12} G_{(3)}\overline{G}_{(3)} \nonumber \\
&=& - \frac{1}{2\kappa_{10}^2} \int d^{10}x \sqrt{g_E^{(10)}}  \, \frac{1}{6\rl S + \overline{S}\rr} G_{(3)}\overline{G}_{(3)} \quad .
\label{exemplesimpleEinstein}
\end{eqnarray}
Nous considérons maintenant les termes contenant les flux $G_{IJK}\overline{G}^{IJK}$. En appliquant la seconde transformation (\ref{2emetransfWeyl}), et en effectuant la réduction dimensionnelle, on obtient l'action
\begin{eqnarray}
\mathcal S &=& - \frac{1}{2\kappa_{4}^2} \int d^{4}x \sqrt{g_E^{(4)}} \, \frac{e^{-12u}}{6\rl S + \overline{S}\rr} G_{IJK}\overline{G}^{IJK} \nonumber \\
&=& - \frac{1}{2\kappa_{4}^2} \int d^{4}x \sqrt{g_E^{(4)}}  \, \frac{4}{3\rl S + \overline{S}\rr\, \rl T + \overline{T}\rr^3} \al m +inS \ar^2 \quad ,
\label{exemplesimple4D}
\end{eqnarray}
où $m$ et $n$ sont des entiers. Nous reconnaissons la métrique de K\"ahler $e^K$ avec
\be
K(S, \overline{S}, T, \overline{T}) \ = \ - \ln \rl S+\overline{S}\rr - 3 \ln \rl T+\overline{T}\rr \nonumber
\ee
en facteur. D'après la section \ref{potentielsugra}, le potentiel scalaire s'écrit
\be
V = e^{K} \ccl K^{i \, \bar j} \,  D_i W  \Db_{\bar j} \, \Wb  \ - \ 3 \al W \ar^2 \ccr \quad ,
\ee
avec $D_i W = W_i + K_i W$, et les indices $i$ portent sur $S$ et $T$. L'argument des crochets dans cette expression doit être identifié avec $\al m + inS \ar^2$ dans l'action (\ref{exemplesimple4D}). Il semble donc que le superpotentiel ne dépende pas de $T$. Dans ce cas, en utilisant (\ref{KmoduleT}),
\be
K^{T \, \overline{T}} \,  D_T W  \Db_{\overline{T}} \, \Wb  \ - \ 3 \al W \ar^2 \ = \ 0 \quad . \nonumber
\ee
On trouve aisément le superpotentiel
\be
W(S) \ = \ m + inS \quad .
\label{WdeS}
\ee

Nous avons ainsi pu générer un potentiel pour le module $S$ en utilisant des flux non-nuls. Dans cet exemple, seul l'un des deux champs $C_{(0)}$ ou $\phi$ aura une valeur moyenne non-nulle, et on voudrait que ce soit le cas pour les deux. De manière un peu plus générale, d'autres modules sont présents après la réduction dimensionnelle : ceux-ci décrivent la structure complexe de l'espace compact $I_6$ (rapport des rayons de compactification, angles, ...). En les désignant par $U_i$, le superpotentiel général que l'on peut obtenir grâce aux flux s'écrit $W = f(U_i) + g(U_i)\, S$, avec $f$ et $g$ des fonctions complexes. Si l'on suppose que ces modules sont stabilisés et prennent des valeurs moyennes, alors on engendre un superpotentiel linéaire du type (\ref{WdeS}) qui donne une valeur moyenne non-nulle aux deux composantes de $S$.

Malheureusement, cet exemple brise la supersymétrie, $\langle D_T W \rangle \neq 0$. Nous ne voulons pas que cette brisure ait lieu lors de la stabilisation du dilaton pour des raisons de stabilité du minimum : la supersymétrie (théorèmes de non-renormalisation) assure que si le potentiel a un minimum stable au niveau classique, alors il le reste à tous les ordres supérieurs. Une solution pour remédier à ce problème est d'imposer la conservation de la supersymétrie, non pas à travers $D_S W = 0$, mais en séparant $\dd_S W = 0 = W$. Ceci assure que $D_T W = 0$ puisque, par ailleurs, aucun terme dépendant de $T$ n'est engendré par les flux dans le superpotentiel.

Le deuxième inconvénient majeur est que l'on ne peut pas combiner aisément le superpotentiel (\ref{WdeS}) avec un superpotentiel de la forme (\ref{WdeT}) pour le dilaton. En effet, il y a un problème de magnitude entre $m,\, n$ qui sont des entiers, donc d'ordre $1$, et l'argument de l'exponentielle, qui est très grand puisque les condensats de jauginos ont lieu à une énergie intermédiaire $\langle \lambda \lambda \rangle \sim \Lambda^3 \ll M_P^3$. Ce problème peut également être levé si l'on impose la conservation de la supersymétrie.


\subsection{Le modèle KKLT\label{KKLT}}


En 2003, Kachru, Kallosh, Linde et Trivedi (KKLT, \cite{Kachru:2003aw}) proposent un modèle de stabilisation de l'ensemble des modules par des flux et par des condensats de jauginos. L'intérêt en particulier pour le dilaton est qu'en utilisant plusieurs flux, celui-ci est stabilisé sans que la supersymétrie ne soit brisée. Dans le scénario KKLT, le module de K\"ahler est traité différemment du reste des modules pour les raisons énoncées ci-dessus : les flux ne générent aucun terme dépendant de $T$ dans le superpotentiel.

Dans la suite, nous considérons que tous les modules excepté $T$ sont stabilisés. Leur interaction avec ce dernier induit une constante $W_0$ qui s'additionne au superpotentiel non-perturbatif (\ref{WdeT}). Cette constante est générée perturbativement et supposée petite (en pratique, ceci est difficile à obtenir). Le modèle est donc
\begin{eqnarray}
&& K(T, \overline{T}) \ = \ -3 \ln \rl T + \overline{T} \rr \quad , \nonumber \\
&& W(T) \ = \ W_0 + a \, e^{-b T} \quad . \label{KWkklt}
\end{eqnarray}

Le potentiel scalaire est calculé en appliquant la formule (\ref{VsugrasansD})
\be
V_{KKLT}(T, \overline{T}) \ = \ \frac{1}{\rl T + \overline{T} \rr^3} \ccl 3 \al \frac{ab\rl T + \overline{T}\rr}{3}e^{-bT} + W \ar^2 - 3 \al W \ar^2  \ccr \quad , \nonumber
\ee
qui se réécrit plus simplement si l'on suppose que $T$ est réel
\be
V_{KKLT}(t) \ = \ \frac{abe^{-bt}}{2t^2} \ccl W_0 + ae^{-bt} \rl 1 + \frac{bt}{3}\rr \ccr \quad ,
\label{Vkklt}
\ee
avec $t = \text{Re}\, T$.

Ce potentiel possède un minimum, et le module $T$ a donc effectivement été stabilisé. Nous demandons que le minimum du potentiel soit supersymétrique, $\left. D_t W \ar_{t=t_0}= 0$, ce qui implique
\be
W_0 \ = \ -ae^{-bt_0} \rl 1 + \frac{2bt_0}{3} \rr \quad . \nonumber
\ee
L'énergie du minimum est
\be
\langle V_{KKLT} \rangle \ = \ - \frac{\rl a b e^{-bt_0}\rr^2}{6 t_0} \ < \ 0 \quad . \nonumber
\ee
Ceci correspond à un vide anti-de Sitter. Puisque l'on sait que la constante cosmologique $\Lambda$ est nulle ou positive très petite, il faut trouver un mécanisme permettant de relever le minimum du potentiel vers des valeurs acceptables. Ce processus s'appelle \textit{uplift}.

Dans le modèle KKLT originel, les auteurs ont supposé l'existence d'une anti-$D3$-brane très loin de là où se situe la matière, dans le sens des dimensions compactes. L'introduction de cette anti-brane a pour effet d'ajouter un potentiel
\be
V_{uplift} (T) \ = \ \frac{c}{t^2} \quad ,
\label{VupliftantiD3brane}
\ee
où $T$ désigne ici le champ scalaire (\ref{moduleT}).

\begin{figure}[ht!]
\begin{center}
\includegraphics[scale=1.]{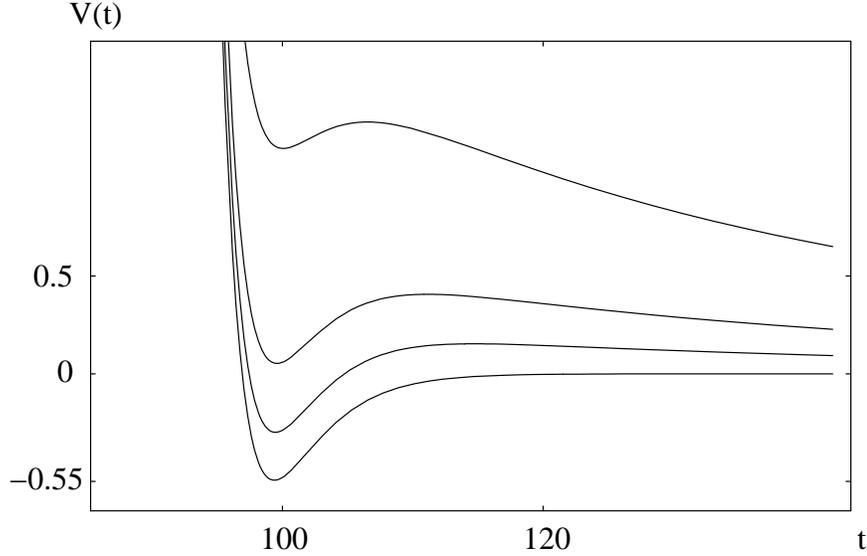}
\caption{Potentiel ($\times 10^{30}$) du module de K\"ahler généré par condensation de jauginos et par les flux. L'augmentation des courbes correspond à une augmentation de la constante $c$. Lorsque celle-ci est trop élevée, le minimum est détruit, et on retrouve le même comportement ``runaway'' que dans le cas du potentiel (\ref{WdeT}).}
\label{VKKLTplusieurs}
\end{center}
\end{figure}

Selon la valeur de $c$, le potentiel total $V = V_{KKLT}+V_{uplift}$ est relevé de manière différente, comme le montre la Fig. \ref{VKKLTplusieurs}.
Dans le modèle considéré, nous avons pris $a=1$, $b=0.3$, $W_0=-10^{-12}$, et la constante $c$ varie entre $4\times 10^{-25}M_P^4$ et $3\times 10^{-24}M_P^4$. La courbe la plus basse correspond au potentiel KKLT (\ref{Vkklt}), c'est-à-dire pour $c=0$.

L'emplacement $t_0 \sim 100$ du minimum n'est pas modifiée par l'ajout du potentiel d'uplift, et elle est telle que $bt_0 \gg 1$ comme nous nous y attendions puisque ce terme résulte de phénomènes non-perturbatifs. La courbure du potentiel autour du minimum ne varie pas non plus lorsqu'on réalise l'uplift. Ceci signifie que si les dimensions supplémentaires sont larges dans le vide anti-de Sitter, elles le restent dans le vide de Sitter. En outre, la masse du module reste très grande devant la constante cosmologique.

L'ajustement entre l'énergie du vide anti-de Sitter $\langle V_{KKLT}\rangle$ et la constante $c$ peut paraître trop fin ($\Lambda = 0$ pour $c \sim 10^{-24}M_P^4$). Ceci s'explique par la courbure des dimensions compactes et par le fait que la matière est très éloignée de l'anti-$D3$-brane. Jusqu'à maintenant, nous avons négligé la courbure des dimensions compactes dûe à la présence de flux non-nuls (voir aussi la section \ref{warpedxdim}). En réalité, la métrique s'écrit
\be
ds^2 \ = \ e^{2A(y)} g_{mn} dx^m dx^n \ + \ e^{-2A(y)} \delta_{IJ} dy^I dy^J \quad , \nonumber
\ee
où $A$, le warp factor, ne dépend que des coordonnées compactes $y^I$.

Pour autant, l'uplift par le potentiel (\ref{VupliftantiD3brane}) n'est pas satisfaisant. En effet, la présence de l'anti-$D3$-brane implique que la supersymétrie est réalisée de manière non-linéaire et donc que l'action effective ne peut plus être mise sous la forme d'une supergravité quadri-dimensionnelle habituelle\footnote{Il est clair que le potentiel (\ref{VupliftantiD3brane}) brise explicitement la supersymétrie.}. De ce fait, les interactions à basse énergie ne peuvent plus être contrôlées comme dans le cadre d'une théorie pleinement supersymétrique. Néanmoins, les contraintes phénoménologiques ($\Lambda \gtrsim 0$, warp factor) forcent l'anti-brane à être située loin de la brane contenant le Modèle Standard ou le MSSM. De ce fait, si le spectre de la matière ordinaire (à basse énergie) est supersymétrique au niveau des arbres, alors le potentiel (\ref{VupliftantiD3brane}) peut n'induire qu'une brisure douce au niveau quantique, donc une brisure relativement contrôlable. Ces aspects du modèle KKLT ont été étudiés en détail et ont donné lieu à une phénoménologie étonnamment riche \cite{Falkowski:2005ck}.

%
%


\section[Modèles de stabilisation avec énergie positive]{Stabilisation des modules et brisure dynamique de supersymétrie}


Depuis quatre ans, une recherche intensive a été menée afin de trouver des mécanismes permettant de relever le potentiel (\ref{Vkklt}) du modèle KKLT à l'aide d'un potentiel supersymétrique.

Rappelons que le potentiel scalaire d'une théorie de supergravité $\mathcal N = 1$ est donné par l'équation (\ref{Vsugra})
\be
V \ = \ e^{K} \pl K^{i \, \bar j} \, D_i W   \Db_{\bar j} \, \Wb  \ - \ 3 \Wb W \pr + \frac{1}{2} D^{(a)} D^{(a)} \quad , \nonumber
\ee
qui s'écrit plus simplement
\be
V \ = \ \sum_i \al F^i \ar^2 + \frac{1}{2}D^{(a)} D^{(a)} \ - \ 3 m_{3/2}^2 M_P^4  \quad ,
\label{VsugraFD}
\ee
avec les champs auxiliaires
\begin{eqnarray}
F^i \ &=& \ - e^{K/2}\,  \rl K^{-1} \rr^{i \, \bar j} \,  \Db_{\bar j} \Wb \quad , \nonumber \\
D^{(a)} \ &=& \ -g K_i  \, T^{(a)\, i}_{\ \ \ j} z^j \quad ,
\label{champsauxsugra2}
\end{eqnarray}
et la masse du gravitino
\be
m_{3/2}^2 \ = \ e^{K} \al W \ar^2 \quad .
\label{m3/22}
\ee

Dans une théorie de supergravité (voir section \ref{potentielsugra}), on peut briser spontanément la supersymétrie tout en ayant une énergie du vide quelconque. Il semble donc naturel de coupler le modèle KKLT à un autre secteur qui apporterait à la fois l'uplift requis et la brisure spontanée de supersymétrie. Ce rôle peut à priori être assuré par des composantes $\langle D \rangle \neq 0$, en utilisant un terme de Fayet-Iliopoulos (\ref{LFI}) si le groupe de jauge du nouveau secteur contient un facteur abélien, ou par des composantes $\langle F\rangle \neq 0$ à travers un modèle à la O'Raifeartaigh (voir paragraphe \ref{BrisureSponSusy}). Dans la suite, nous étudions les deux approches.


\subsection{Uplift grâce à un groupe $U(1)$ anormal\label{Dtermuplifting}}


Quelques mois après la parution du scénario KKLT \cite{Kachru:2003aw}, Burgess, Kallosh et Quevedo \cite{Burgess:2003ic} proposent d'utiliser une composante $D$ brisant spontanément la supersymétrie.

La présence d'un facteur $U(1)_X$ anormal est générique dans les théories de supergravité issues de théories des cordes. À quatre dimensions, la théorie effective possède des anomalies mixtes entre la symétrie de jauge $U(1)_X$ et le groupe $SU(N_c)$ de la $D7$-brane qui permet de stabiliser le module $T$. Les champs de matière présents sur la $D7$-brane doivent être chargés et la somme de leur charge est alors fixée par l'annulation des anomalies. Parallèlement, les auteurs utilisent certains flux sur la $D7$-brane pour engendrer un terme de Fayet-Iliopoulos (FI) à la composante $D$ du groupe $U(1)_X$. Par conséquent, le terme FI dépend du module $T$ puisque ce dernier est directement relié au couplage de jauge sur la brane. Ce mécanisme force le module à se transformer sous le groupe abélien selon
\be
\delta T \ = \ \delta_{GS} \Lambda \quad , \label{deltaTU1}
\ee
où $\delta_{GS}$ est une constante reliée au mécanisme de Green-Schwarz d'annulation des anomalies, et $\Lambda$ est le superchamp chiral paramètre de la transformation de jauge $U(1)_X$. Le potentiel de K\"ahler (\ref{KmoduleT}) n'est plus invariant de jauge dans ce cas, et il faut le modifier en
\be
K(T, \overline{T}) \ = \ -3 \ln \rl T + \overline{T} - \delta_{GS} V_X \rr \quad , \nonumber
\ee
avec $V_X$ le vecteur de jauge de $U(1)_X$ qui a la transformation habituelle $\delta V_X = \Lambda + \overline{\Lambda}$.

Le potentiel scalaire induit par la composante $D_X$ a typiquement la forme
\be
V_D \ = \ \frac{1}{2}g^2 D_X^2 \ = \ \frac{2\pi}{\text{Re} \, T} \rl \sum_i q_i K_i \phi^i + \zeta \rr^2 \quad ,
\label{VDuplift}
\ee
où $\phi^i$ sont les champs de matière de charges $q_i$, et $K_i = \dd K / \dd \phi^i$ avec un potentiel de K\"ahler canonique pour les $\phi^i$. La constante $\zeta$ est le terme FI, il est relié à l'annulation des anomalies par
\be
\zeta \ = \ -\frac{\delta_{GS}}{2} \dd_T K \quad . \nonumber
\ee

On peut montrer \cite{Binetruy:1996uv, Dudas:2005vv, Villadoro:2005yq} que ce schéma n'est pas complet car le superpotentiel non-perturbatif (\ref{WdeT}), $W_{NP}=a e^{-bT}$, n'est pas invariant sous (\ref{deltaTU1}). La complétion du modèle est délicate et nécessite l'introduction d'un champ $\phi$ singlet de $SU(N_c)$ et de charge $U(1)_X$ négative\footnote{Plus exactement, la charge du singlet doit être de signe opposé à celui de la somme des charges $q_i$. En effet, l'annulation des anomalies impose $\delta_{GS} \sim \sum q_i$, et par ailleurs $\dd_T K < 0$. Il s'ensuit que le terme $\zeta$ et la somme $\sum q_i$ sont de même signe.}. De manière générale, un tel champ existe toujours dans les constructions de théories des cordes. Le modèle est alors fixé par l'invariance de jauge.

Après avoir résolu les équations du mouvement dans ce type de constructions, on trouve toujours que le minimum est de la forme
\be
\langle V \rangle \ = \ V_F + V_D \ \sim \ m_{3/2}^4 \ - \ m_{3/2}^2 \zeta^2 \quad . \nonumber
\ee

Pour obtenir une constante cosmologique nulle, il faut donc que la masse du gravitino soit du même ordre que le terme de Fayet-Iliopoulos, qui est généralement très grand $\zeta \sim M_P$. Pourtant, on a typiquement $m_{3/2} \sim W_0 / M_P^2 \ll \Lambda$ avec, encore une fois, $W_0$ très petit, et où $\Lambda$ est l'échelle de condensation des jauginos donnant lieu au superpotentiel $e^{-bT}$.


On trouve que la masse du gravitino dans ces modèles est de l'ordre de $10^{16} \ \text{GeV}$, ce qui n'est pas intéressant du point de vue phénoménologique. On peut modifier le modèle en incluant des corrections non-perturbatives au potentiel de K\"ahler, mais malgré ces tentatives, il semble difficile d'obtenir une masse $m_{3/2}$ inférieure à $10^{12}$-$10^{13}$ GeV.


\subsection{Uplift et brisure de supersymétrie à la O'Raifeartaigh}


Dans la Publication ${\mathcal N}^{\mathrm{o}}\, 3$ (voir aussi \cite{Abe:2006xp}), nous avons réalisé l'uplift du potentiel KKLT d'une nouvelle façon. En couplant le modèle ISS développé dans la section \ref{ISS} avec le modèle KKLT, on obtient l'uplift souhaité et la brisure dynamique de supersymétrie par le terme $F_{\Phi}\neq 0$ (voir équations (\ref{FISS})). Le point majeur est que le gravitino possède naturellement une masse petite $m_{3/2} \sim \mathcal O(\text{TeV})$.

\begin{figure}[ht!]
\begin{center}
\includegraphics[scale=0.55]{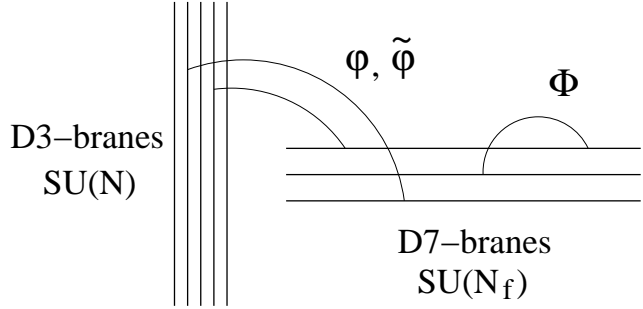}
\caption{La construction faite dans la Publication ${\mathcal N}^{\mathrm{o}}\, 3$.}
\label{BraneISSKKLT}
\end{center}
\end{figure}

La construction des différents secteurs est assez simple à réaliser et est illustrée sur la Fig. \ref{BraneISSKKLT}.

L'espace-temps dix-dimensionnel contient des $D3$ et des $D7$-branes. Rappelons que les mésons $\phi$ sont chargés sous le groupe global $SU(N_f)$, tandis que les quarks $\ffi$ et anti-quarks $\ffitil$ portent à la fois des indices de couleur et de saveur. Les mésons doivent être relativement découplés du module de K\"ahler
\begin{eqnarray}
&& K \ = \ K(T, \overline{T}) \ + \ K(\Phi, \Phi^{\dagger}) \quad , \nonumber \\
&& W \ = \ W(T) \ + \ W(\Phi) \quad , \label{sepaKWphiT}
\end{eqnarray}
de manière à pouvoir induire facilement l'uplift et la brisure de supersymétrie.

Aussi peuvent-ils appartenir à un ensemble de $N_f$ $D7$-branes car ils sont singlets de jauge. Nous les interprétons comme les positions des branes, ce qui garantit la forme (\ref{sepaKWphiT}). En effet, l'action des positions $X^I$ des $D7$-branes dans le référentiel d'Einstein est
\be
\mathcal S \ = \ \int_{D7} d^{8} x \sqrt{\gamma} \, e^{\phi} \, \gamma^{MN} g_{IJ}\dd_M X^I \dd_N X^J \quad , \nonumber
\ee
où le facteur $e^{\phi}$ vient de la tension de la brane, et $M, N = 0, \ldots, 7$. Lorsqu'on réduit cette action en utilisant (\ref{2emetransfWeyl}), on trouve qu'il n'y a aucune dépendance explicite en $T$ dans les termes cinétiques $\dd_m X^I \dd^m X^J$ à quatre dimensions. Par contre, ils dépendent du dilaton selon le facteur $e^{\phi}= 1/ \text{Re}\, S$.

De la même façon, on ne veut pas que l'échelle dynamique $\Lambda$ du secteur ISS dépende du module $T$. Il faut donc que le couplage de jauge vive sur un ensemble de $N$ $D3$-branes, pour lesquelles nous avons montré à la fin de la section \ref{dimreduc} que le couplage ne dépend que de $S$. Les quarks et anti-quarks sont décrits par des cordes ouvertes dont les extrêmités lient les $D7$-branes et les $D3$-branes. À priori, ils sont couplés de manière non-triviale aux modules $T$ et $S$, mais puisqu'ils ne contribuent pas à la brisure de supersymétrie, leur potentiel de K\"ahler (supposé canonique pour plus de simplicité dans la Publication ${\mathcal N}^{\mathrm{o}}\, 3$) est moins important que celui des mésons.

Le couplage direct entre le secteur ISS et le secteur KKLT produit un terme $C/ t^3$, avec $t=\text{Re}\, T$, dans le potentiel scalaire. Ce terme joue le rôle de potentiel d'uplift, et comme sur la Fig. \ref{VKKLTplusieurs}, ne modifie pas la valeur moyenne $t_0$ dans le vide.

Dans le modèle, l'uplift et la brisure dynamique de supersymétrie à la ISS dépendent du paramètre de masse $\mu^2$ des mésons (équations (\ref{KWiss})). De ce fait, l'annulation de la constante cosmologique demande un ajustement entre $W_0$ et $\mu^2$
\be
3 \al W_0 \ar^2 \ \sim \ \al h^2 \mu^4 \ar \, \rl N_f - N \rr \quad . \nonumber
\ee
Nous avons vu que $W_0$ est très petit $\sim 10^{-12}$, et il faut donc que $\mu^2$ le soit aussi. Ceci se produit naturellement car $\mu^2$ est généré dynamiquement dans la théorie électrique \cite{Intriligator:2006dd}. On obtient un gravitino de masse
\be
m_{3/2} \ \sim \ \frac{\al W_0\ar}{\rl T_0+\overline{T_0}\rr^{3/2}} \quad . \nonumber
\ee
Puisque $t_0 \sim 100$, on a $m_{3/2} \sim 10^{-15} M_P \sim 10 \ \text{TeV}$.

Notons que l'uplift du potentiel (\ref{Vkklt}) par des composantes $F$ peut être réalisé grâce à n'importe quel modèle à la O'Raifeartaigh. Cela a été brièvement montré à la fin de la Publication ${\mathcal N}^{\mathrm{o}}\, 3$. L'avantage de coupler un modèle de type ISS avec le module est que la présence de ce dernier rallonge la durée de vie des vides métastables. Pour finir, la phénoménologie de ces modèles a été récemment développée \cite{Lebedev:2006qc}.

\end{document}